\documentclass[preprint]{aastex}
\usepackage{epstopdf}
 \usepackage{psfig}
 \usepackage{epsf}
\usepackage{graphics}
\usepackage{natbib}
\usepackage{hyperref}
\newcommand{\msun}{\mbox{$M_{\odot}$}~}

\newcommand{\lsun}{\mbox{$L_{\odot}$}~}
\newcommand{\kms}{\mbox{km s$^{-1}$}}

\newcommand{\etal}[1]{{ et al.}~}
\def\kms{\ifmmode \hbox{km~s}^{-1}\else km~s$^{-1}$\fi}
\def\etal {{\it et al.}}
\def\deg      {{\ifmmode^\circ\else$^\circ$\fi} } %%% Overwrites TeX \deg

\def\h2     {H$_2$}
\def\arcdeg{\hbox{$^\circ$}}
\def\arcmin{\hbox{$^\prime$}}
\def\arcsec{\hbox{$^{\prime\prime}$}}

\shorttitle{Galaxy Evolution in COSMOS}
\shortauthors{Scoville et al.}

\begin{document}


\title{Evolution of Galaxies and their Environments \\
at z = 0.1 to 3 in COSMOS}

\date{Accepted ApJSuppl 3/26/13}

 \author{ N. Scoville\altaffilmark{1}, 
 S. Arnouts\altaffilmark{2,25},
 H. Aussel\altaffilmark{3},
A. Benson\altaffilmark{1,4},
A. Bongiorno\altaffilmark{29},
K. Bundy\altaffilmark{5},
M. A. A. Calvo\altaffilmark{6},
P. Capak\altaffilmark{7},
%R. S. Ellis\altaffilmark{1},
M. Carollo\altaffilmark{8},
F. Civano\altaffilmark{31},
J. Dunlop\altaffilmark{32},     
M. Elvis\altaffilmark{31},
A. Faisst\altaffilmark{8},
A. Finoguenov\altaffilmark{9},
Hai Fu\altaffilmark{1,10},
M. Giavalisco\altaffilmark{11},
Q. Guo\altaffilmark{9,12},
%L. Guzzo\altaffilmark{6},
O. Ilbert \altaffilmark{13},
A. Iovino \altaffilmark{28},
%G. Li\altaffilmark{1},
M. Kajisawa\altaffilmark{23}, 
J. Kartaltepe\altaffilmark{25},
A. Leauthaud\altaffilmark{5}, %
O. Le Fe`vre\altaffilmark{13}, 
E. LeFloch\altaffilmark{3},
S. J. Lilly\altaffilmark{8},
C. T-C. Liu\altaffilmark{26,27},
S. Manohar\altaffilmark{1},  %
R. Massey\altaffilmark{14},
D. Masters\altaffilmark{16}, 
H. J. McCracken\altaffilmark{15},
B. Mobasher\altaffilmark{16},
Y-J. Peng\altaffilmark{8},
A. Renzini\altaffilmark{30},
J. Rhodes\altaffilmark{1,17},
M. Salvato\altaffilmark{1,9},
D. B. Sanders\altaffilmark{18},
B. D. Sarvestani\altaffilmark{16},
C. Scarlata\altaffilmark{19},
E. Schinnerer\altaffilmark{20},
K. Sheth\altaffilmark{21},
P. L. Shopbell\altaffilmark{1},
V. Smol'cic\altaffilmark{1,22},
Y. Taniguchi\altaffilmark{23}, 
J. E. Taylor\altaffilmark{24},
%D. J. Thompson\altaffilmark{15,18}
S. D. M. White\altaffilmark{9}
and
L. Yan\altaffilmark{7}}


\altaffiltext{$\star$}{Based on observations with the NASA/ESA {\em
Hubble Space Telescope}, obtained at the Space Telescope Science
Institute, which is operated by AURA Inc, under NASA contract NAS
5-26555; and Spitzer Space Telescope, which is operated by the Jet Propulsion Laboratory,
 California Institute of Technology under NASA contract 1407; also based on data collected at : the Subaru Telescope, which is operated by
the National Astronomical Observatory of Japan; the XMM-Newton, an ESA science mission with
instruments and contributions directly funded by ESA Member States and NASA; the European Southern Observatory under Large Program 175.A-01279, Chile; Kitt Peak National Observatory, Cerro Tololo Inter-American
Observatory, and the National Optical Astronomy Observatory, which are
operated by the Association of Universities for Research in Astronomy, Inc.
(AURA) under cooperative agreement with the National Science Foundation;
the National Radio Astronomy Observatory which is a facility of the National Science
Foundation operated under cooperative agreement by Associated Universities, Inc ;
and the Canada-France-Hawaii Telescope with MegaPrime/MegaCam operated as a
joint project by the CFHT Corporation, CEA/DAPNIA, the NRC and CADC of Canada, the CNRS of France, TERAPIX and the Univ. of
Hawaii.}
\altaffiltext{1}{California Institute of Technology, MC 249-17, 1200 East California Boulevard, Pasadena, CA 91125}
\altaffiltext{2}{Canada-France-Hawaii Telescope Corporation, 65-1238 Mamalahoa Hwy, Kamuela, HI 96743, USA}
\altaffiltext{3}{AIM Unit\'e Mixte de Recherche CEA CNRS, Universit\'e Paris VII UMR n158, Paris, France}
\altaffiltext{4}{Carnegie Observatories, Pasadena, CA}
\altaffiltext{5}{Institute for the Physics and Mathematics of the Universe, University of Tokyo, Kashiwa 277-8582, Japan}
\altaffiltext{6}{Dept. of Physics and Astronomy, Johns Hopkins University,  3400 N. Charles St., Baltimore, Md 21218-2686,   USA}
\altaffiltext{7}{Spitzer Science Center, MS 314-6, California Institute of Technology, Pasadena, CA 91125}
\altaffiltext{8}{Institute for Astronomy, ETH Zurich, Wolfgang-Pauli-strasse 27, 8093 Zurich, Switzerland}
\altaffiltext{9}{Max Planck Institut f\"ur Extraterrestrische Physik,  D-85478 Garching, Germany}
\altaffiltext{10}{Department of Physics and Astronomy, University of California, Irvine, CA}
\altaffiltext{11}{Department of Astronomy, University of Massachusetts, Amherst, MA 01003, USA}
\altaffiltext{12}{National Astronomical Observatories, Chinese Academy of Sciences, Beijing, 100012, China}
\altaffiltext{13}{Laboratoire dÕAstrophysique de Marseille, B.P. 8, Traverse du Siphon, 13376 Marseille Cedex 12, France}
\altaffiltext{14}{Institute for Astronomy, Blackford Hill, Edinburgh EH9 3HJ UK}
\altaffiltext{15}{Institut d'Astrophysique de Paris, UMR7095 CNRS, Universit\'e Pierre et Marie Curie, 98 bis Boulevard Arago, 75014 Paris, France}
\altaffiltext{16}{Department of Physics and Astronomy, University of California, Riverside, CA 92521, USA}
\altaffiltext{17}{Jet Propulsion Laboratory, Pasadena, CA 91109}
\altaffiltext{18}{Institute for Astronomy, 2680 Woodlawn Dr., University of Hawaii, Honolulu, Hawaii, 96822}
\altaffiltext{19}{Department of Physics and Astronomy, University of Minnesota, Minneapolis, Minn.}
\altaffiltext{20}{Max Planck Institut f\"ur Astronomie, K\"onigstuhl 17, Heidelberg, D-69117, Germany}
\altaffiltext{21}{National Radio Astronomy Observatory, 520 Edgemont Road, Charlottesville, VA 22903, USA}
\altaffiltext{22}{Argelander Institut for Astronomy, Auf dem H\"ugel 71, Bonn, 53121, Germany}
\altaffiltext{23}{Physics Department, Graduate School of Science, Ehime University, 2-5 Bunkyou, Matsuyama, 790-8577, Japan}
\altaffiltext{24}{Department of Physics and Astronomy, University of Waterloo, Waterloo, Ontario N2L 3G1, Canada}
\altaffiltext{25}{Aix Marseille UniversitŽ, CNRS, LAM (Laboratoire d'Astrophysique de Marseille) UMR 7326, 13388, Marseille, France}
\altaffiltext{26}{Astrophysical Observatory, Department of Engineering Science and Physics, CUNY College of Staten Island, 2800 Victory Blvd, Staten Island, NY  10314}
\altaffiltext{27}{Department of Astrophysics and Hayden Planetarium, American Museum of Natural History, Central Park West at 79th Street, New York, NY 10024}
\altaffiltext{28}{AA(INAF - Osservatorio Astronomico di Brera, via Brera, 28, 20159 Milano, Italy}
\altaffiltext{29}{INAF-Osservatorio Astronomico di Roma, Via di Frascati 33, I-00040, Monteporzio Catone, Rome, Italy}
\altaffiltext{30}{INAF - Osservatorio Astronomico di Padova, Vicolo dellÕOsservatorio 5, I-35122 Padova, Italy}
\altaffiltext{31}{Harvard Smithsonian Center for astrophysics, 60 Garden St., Cambridge, MA 02138, USA}
\altaffiltext{32}{Institute for Astronomy, University of Edinburgh, Royal Observatory, Edinburgh, EH9 3HJ, UK}
%

%

%

%
%\altaffiltext{19}{Dipartimento di Astronomia, Universitˆ di Padova, vicolo dell'Osservatorio 2, I-35122 Padua, Italy}

% had to put the lines below otherwise 'Abstract' appears above afiliations
\altaffiltext{}{}


\begin{abstract}
Large-scale structures (LSS) out to z $< 3.0$ are measured in the Cosmic Evolution Survey (COSMOS)
using extremely accurate photometric redshifts (photoz). The Ks-band selected sample (from Ultra-Vista)
is comprised of 155,954 galaxies. 
 Two techniques -- adaptive smoothing and Voronoi tessellation -- are used to estimate the 
 environmental densities within 127 redshift slices.  
Approximately 250 statistically significant overdense structures are 
identified out to z $= 3.0$ with shapes varying from elongated filamentary structures 
to more circularly symmetric concentrations.
We also compare the densities derived for COSMOS with those based on semi-analytic
predictions for a $\Lambda$CDM simulation and find excellent overall agreement between the 
mean densities as a function of redshift and the range of densities. 
The galaxy properties 
(stellar mass, spectral energy distributions (SEDs) and star formation rates (SFRs)) are strongly 
correlated with environmental 
density and redshift, particularly at z $< 1.0 - 1.2$. 
Classifying the spectral type of each galaxy using the rest-frame b-i color (from the photoz
SED fitting), we find a strong correlation of early type galaxies (E-Sa) with high density 
environments, while the degree of environmental segregation varies systematically with 
redshift out to z $\sim 1.3$. In the highest density regions, 80\% of the galaxies are early types at z=0.2 compared to only 20\% at z = 1.5. 
The SFRs and the star formation timescales  
exhibit clear environmental correlations. At z $> 0.8$, the star formation rate density (SFRD)
is uniformly distributed over all environmental density percentiles, while at lower redshifts the dominant contribution is shifted to galaxies in lower density environments.

\end{abstract}


\section{Introduction}\label{intro}
 
 The cosmic evolution of 
 galaxies and dark matter is strongly linked through both the environmental influences 
 and feedback due to starbursts and AGN. \cite{pen10} have recently shown that the quenching of star formation (SF) activity 
 in low redshift SDSS galaxies is clearly separable into galaxy-mass and environmental-density dependent effects.
A major motivation for the Cosmic Evolution Survey (COSMOS) was to provide a sufficiently large area, 
 to probe the expected range of environments (large-scale structure -- LSS) with high sensitivity to detect large samples 
 of objects at high redshifts, and to minimize the effects of cosmic variance. The COSMOS 2 deg$^2$ survey samples scales of LSS out to  $\sim$ 50 -- 100 Mpc  and detects approximately two million 
 galaxies at z $= 0.1 - 5$ at I $< 26.5$ mag(AB). Initial identifications of LSS 
 galaxy clusters in COSMOS were compared with the total mass densities determined from weak lensing tomography and hot X-ray 
 emitting gas in the virialized clusters/groups of galaxies at z $<1.1$ \citep{sco_lss,mas07,fin07}; 
 here we extend this investigation to higher z and lower density LSS using deeper photometry and high accuracy photometric  redshifts.

The identification of LSS from the observed surface-density of galaxies requires 
separation of galaxies at different distances along the line of 
sight, otherwise, the superposition of LSS at different redshifts would preclude a mapping of the 
structure morphology and the LSS overdensities would  be diluted by foreground and background galaxies. Ideally, redshift (or distance discrimination) precision is desired at the level of the internal 
velocity dispersions of the structures. For LSS mapping, line-of-sight 
discrimination is usually accomplished using:
1) color selection \citep[e.g.~using broadband colors to select red sequence galaxies,][]{gla05};  
2) spectroscopic redshifts  \citep[in COSMOS e.g.,][]{kov10,pen10} and 3) 
photometric redshifts derived from fitting the broadband spectral energy distributions (SEDs) of the galaxies \citep[][]{wey94,pos96, sch98,mar02}. Color selection clearly biases any correlation between environment and galaxy SED or morphological type \citep[][]{dre97,smi05} since the resulting LSS 
are {\it a priori} based on a particular SED type (e.g. early type galaxies). In addition, the red early-type galaxies must become rarer at early epochs, simply due to the short cosmic age, and identification of clustering using the red sequence is bound to become problematic at higher redshifts.  Spectroscopic redshifts are of course 
most desirable and have been used extensively in low z 
studies with relatively bright galaxies (e.g. SDSS); however, they are not presently feasible 
for the samples of hundreds of thousands of high redshift galaxies going fainter than  $I_{AB} = 24$ mag and galaxies lacking strong emission lines. Spectroscopy of such faint galaxies requires the largest telescopes and integration times of a  few hours
\citep{lef05,ger05,men06,cop06,coi06,lil07}.  

In this paper, we identify LSS in the 2 square degree COSMOS 
field using the most recent COSMOS photometric redshifts \citep{ilb09,ilb13} to analyze the galaxy surface 
densities in redshift slices out to z = 3.0, covering cosmic ages from 2.1 to 12 Gyr. These photometric redshifts are extremely accurate (see \S \ref{selection})
since they are based on deep 30-band UV-IR photometry and they cover all galaxy spectral types. The galaxy sample used for this work and the associated stellar mass limits  
are discussed in \S \ref{selection} and a similar sample is generated from a $\Lambda$CDM simulation of size 1/64 of Millennium (\S \ref{millennium}).  We use two independent techniques: adaptive smoothing \citep{sco_lss} and 2-d Voronoi tessellation \citep{ebe93} to measure the local density associated with each galaxy,
and to map and visualize coherent LSS in COSMOS (\S \ref{densities}). Maps of the LSS are presented in \S \ref{lss}. The derived densities are 
compared with predictions from the simulation (\S  \ref{comp_millennium}).
We analyze the evolution of the galaxy population with redshift  and environmental density in \S \ref{evolution} and a simplified schematic model
for the evolution processes is presented in \S \ref{mat}.

Adopted cosmological parameters, used 
throughout, are: H$_0$ = 70 km s$^{-1}$ Mpc$^{-1}$, $\Omega_M$ = 0.3 and  $\Omega_\Lambda$ = 0.7. The AB magnitude system is used throughout. For computing 
stellar masses and star formation rates (SFRs), we adopt a Chabrier  IMF; for the Salpeter IMF, both the mass and SFR estimates 
should be increased by a factor of 1.78.

\begin{figure}[ht]
\epsscale{0.6}
%\plotone{specz-eps-converted-to.pdf}
\plotone{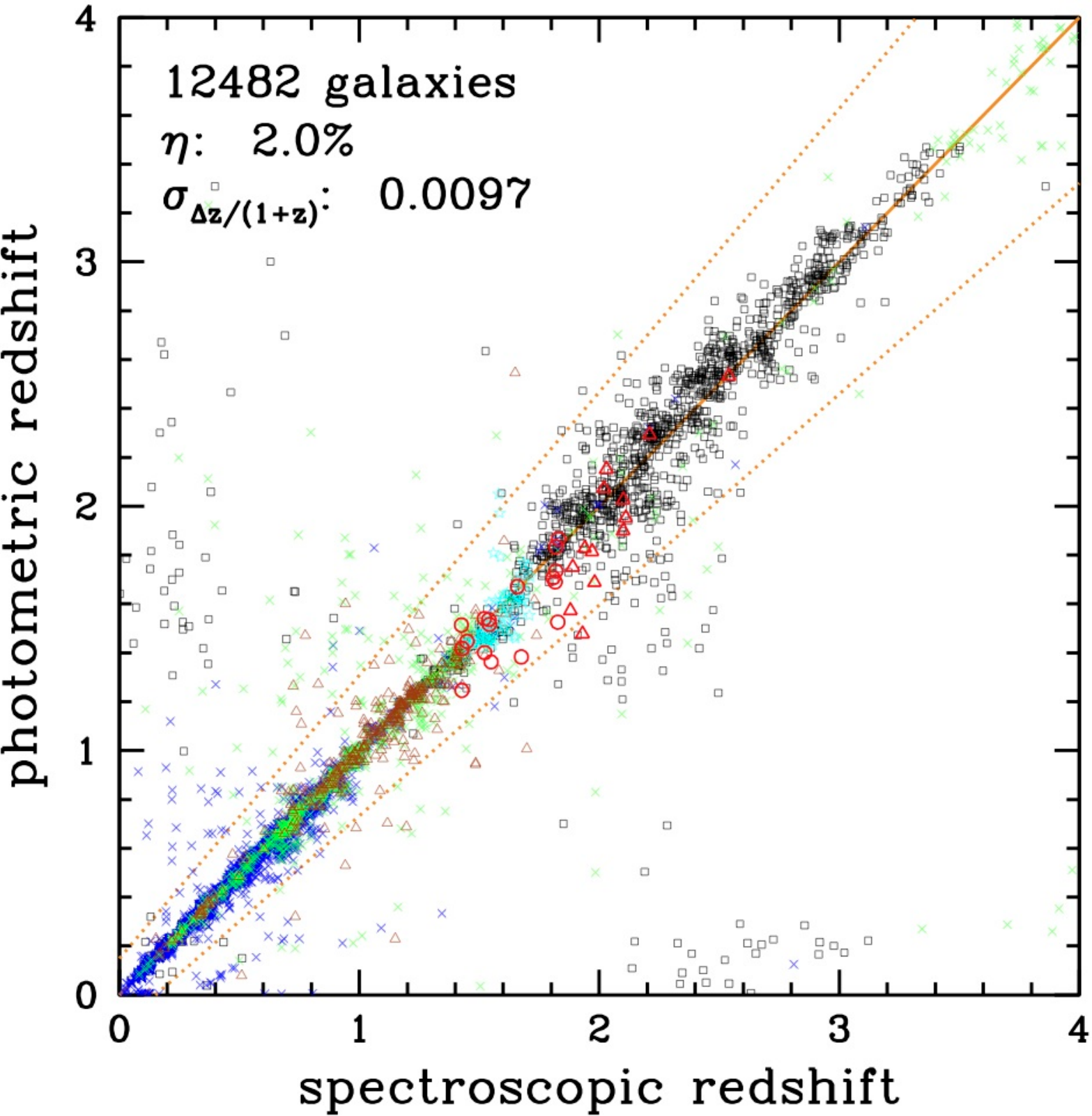}
\caption{Comparison of photometric and spectroscopic redshifts for a sample of 12,482 galaxies down to $I_{AB} \sim 24.5$ mag in COSMOS \citep[figure taken directly from][]{ilb13}. For this sample, the $\sigma_z / (1+z)$ = 0.9\% and the catastrophic failure rate is 2\%. The dotted lines indicate the 2 $\sigma$ dispersions. The COSMOS photometric redshifts are from \cite{ilb13} and the spectroscopic 
redshifts are from: the zCOSMOS VIMOS bright (black squares) and faint (blue x's) surveys \citep{lil07}, the Keck DEIMOS survey (green x's, Capak \etal ~2013 in preparation), the FORS2 survey (brown triangles,  Comparat \etal  ~2013 in preparation), the FMOS survey (cyan stars, ~Silverman \etal 2013 in preparation), the MOIRCS survey (red circles, \cite{ono12}), and the WFC3 grism survey (red triangles, Krogager \etal ~2013 in preparation). Only galaxies with reliable (at least 2 spectral lines) spectroscopic redshifts are used here.}\label{photz_specz}
\end{figure}

\section{Photometric Redshifts and Sample Selection}\label{selection}

%\begin{figure}
%\epsscale{1.}
%\plotone{cosmol_time-eps-converted-to.pdf}
%\caption{The lookback times and cosmological ages at z $< 2.6$ are shown 
%(H$_0$ = 70 km s$^{-1}$ Mpc$^{-1}$, $\Omega_M$ = 0.3 and  $\Omega_\Lambda$ = 0.7).}\label{cosmol_time}
%\end{figure}

%\clearpage

\begin{figure}[ht]
\epsscale{0.6}
\plotone{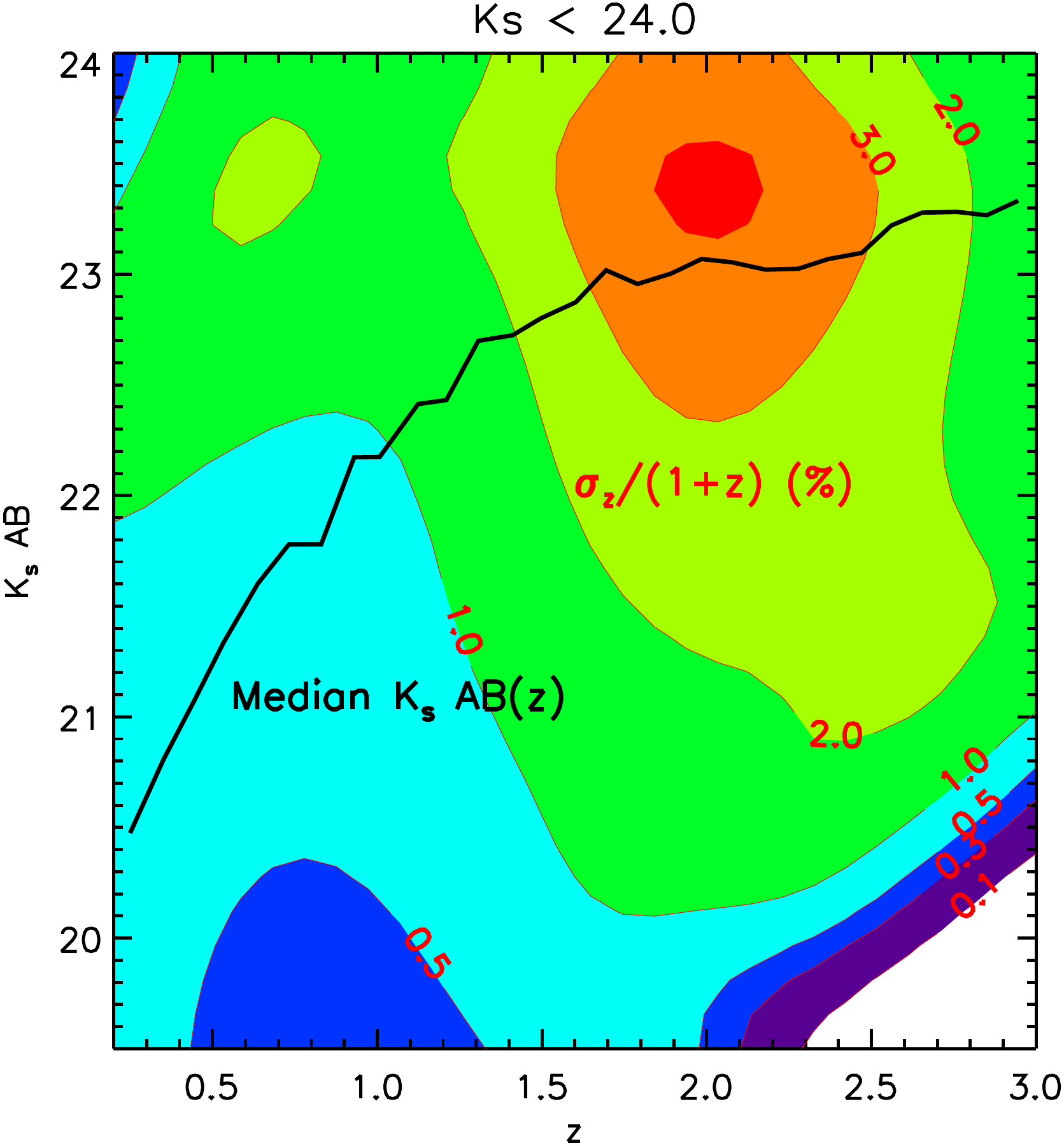}
\caption{The accuracy ($\sigma_z/(1+z)$ in \%) for the  photometric redshifts is shown as a function of observed Ks-band magnitude and redshift. These uncertainty estimates were derived from the dispersion in the photoz probability density distributions (PDF) as discussed in the text. For bright galaxies at z $< 1.2$, $\sigma_z/(1+z)<$ 1\% 
but degrades by a factor of two at higher z and for fainter galaxies. 
Also shown is the median magnitude, Ks$_{AB}$, for galaxies in our 
sample -- the widths of the redshift slices is chosen to approximately follow the 
FWHM (=2$\sigma_z$) of the median galaxy at each z.}\label{sigz}
\end{figure}

The COSMOS photometric catalog is derived from deep ground and space-based 
imaging in 37 broad and intermediate-width bands. This includes  HST-ACS F814W \citep{sco_hst}, Suprime-Cam on Subaru \citep{tan07} and  CFHT-MagaCam/WIRCam  \citep{mcc10};  near infrared imaging (Y, J, H \& K$_s$) 
from NOAO-4m, UH88, UKIRT  \citep{cap07,mcc10}, Ultra-Vista \citep{mcc12}; Spitzer IRAC 3.6-8.5$\mu$m \citep{san07}
and Galex NUV \& FUV  \citep{zam07}. Typical sensitivities (5$\sigma$ in a 3\arcsec aperture) are 26-27 mag (AB) in the 
optical and 25 mag (AB) in the infrared \citep{cap09,mcc12}. 
The original COSMOS photometric catalogs were based on primary source detections in the Subaru and CFHT i-band imaging. 2.1 million objects 
are included at I$_{AB} < 26.5$ mag \citep{cap09}. The new catalog we make use of here is based on primary source detection in the Ultra-Vista Ks band and 
this provides significantly better completeness at $z > 1$, especially for red or passive galaxies. 

The COSMOS photometric redshift (photoz) catalog using primary source selection in Ks band is described in detail in \cite{ilb13}.  For these most recent photometric redshifts the 30 broad, intermediate and narrow bands include : u$^*$, BJ, VJ, r$^+$, i$^+$, z$^+$, IA484, IA527, IA624, IA679, IA738, IA767, IA427, IA464, IA505, IA574, IA709, IA827, NB711, NB816, the four Spitzer IRAC bands and four UltraVISTA bands (Y, J, H, Ks).
Photoz were derived for 218,000 galaxies with 
Ks (AB)$ < 24$ mag using a $\chi^2$ template fitting procedure \citep[Le Phare : ][]{arn02,ilb06}. The fitting presumes 31 basic spectral energy distributions (SEDs) with dust extinctions varying from A$_{V}$ = 0 to 1.5 mag with \cite{cal00} and \cite{pre84} extinction laws. Emission lines are included
in the photoz fitting. 

Spectroscopic redshifts in COSMOS have been obtained from  the VLT-VIMOS zCOSMOS survey \citep{lil07} for approximately 20,000 galaxies (I$_{AB} \leq 24.5$ mag)
and Keck-DEIMOS \citep{cap13} for approximately 3400 galaxies (I$_{AB} \leq 25$ mag). The offsets between the photometric and spectroscopic redshifts
for 12,482 galaxies with high reliability spectroscopic redshifts at z = 0.05 to 4 down to I$_{AB} \leq 24.5$ mag yield $\sigma_z / (1 + z) \simeq$ 0.9\% with a catastrophic ($>2\sigma$) failure rate typically only 2\% as shown 
in Fig. \ref{photz_specz} \citep{ilb13}.

\subsection{Sample Selection}

For this study, we adopt selection criteria requiring that 
the galaxies be detected in the near infrared band (K$_s$ and in most cases they are detected in IRAC1-3.6$\mu$m) -- in order to 
provide more reliable mass estimates from the long wavelength continuum. 
We exclude objects classified as stellar or AGN-dominated as indicated by their measured size in the HST-ACS images or an X-ray detection. We impose the following selection criteria on the Ultra-Vista Ks selected photoz redshift catalog :
\begin{mathletters}
\begin{equation}
z = 0.15 - 3.0
\end{equation}
\begin{equation}
 K_s (AB) \leq 24  ~mag
\end{equation}
\begin{equation}
M_* \geq 10^9 \msun
\end{equation}
\begin{equation}
149.4\deg < \alpha_{2000} < 150.8\deg ~~~ \rm{and}~~~ 1.5\deg < \delta_{2000} < 2.9 \deg.
\end{equation}
\label{selection_eq}
\end{mathletters}

\noindent The stellar mass (see \S \ref{sfr_section}) selection criteria was imposed so that the LSS would be mapped using reasonably massive galaxies; it has impact only at z $<$ 0.5 since such low mass galaxies are not 
detected at the higher redshifts. The last selection by position, was used to provide an approximately square area for imaging LSS, and to minimize the effects of the irregular Ultra-Vista coverage 
at the field edges. These four combined selection criteria 
yield a sample of 155,954 galaxies. [All areas masked for proximity to a bright star were also excluded, as shown in Fig. \ref{mask}.]
A principal goal of this study is to explore the correlation of galaxy properties with environment and redshift. 
Thus, it is important to recognize that we have avoided selection based on a specific type of galaxy, i.e. using colors to select red sequence galaxies. 
%Some earlier studies aimed at locating dense environments or clusters have preselected the red galaxy population \citep[e.g.][]{gla05}.  Although this 
%may be an efficient means of reducing the background shot noise of blue galaxies which are not so concentrated in the densest regions at z $<$ 1, color selection would prejudice investigations of environment correlations 
%with galaxy mass or SED type (see \S \ref{sfr_section}). In addition, the color selection technique is bound to eventually break down at high redshift when the cluster galaxies were actively forming their stellar populations and hence, were intrinsically blue. 

The above selection criteria were arrived at as a compromise between two goals: 1) maintaining high accuracy in redshifts to enable narrow redshift slices for
delineating the LSS and 2) providing large samples of galaxies in each slice so that the LSS can be traced to lower densities. Trials with the simulation mock catalogs (see \S \ref{millennium})
indicated that redshift slice widths of $\Delta z \sim 1-2$\% at z = 1 and $\sim$ 5-10\% at z = 2 are adequate for detecting the LSS without excessive contamination or dilution of the LSS.  

The spectroscopic/
photometric redshift comparison shown in Fig. \ref{photz_specz}  is limited to a  sample of only 12,482 galaxies (mostly 
brighter objects). To extend our understanding of the photoz uncertainties, we used the probability distribution function (PDF) from the 
photometric redshift solutions for a more general assessment of the 
redshift accuracies as a function of both redshift and magnitude. As discussed in \cite{ilb09}, the width of the highest peak in the PDF agrees well with that of the specz-photoz comparison at redshifts and magnitudes where there are sufficient spectroscopic redshifts for a comparison \citep[see Fig. 9;][]{ilb09}. Fig. \ref{sigz} shows 
 $\sigma_z / (1 + z)$ as a function of redshift and galaxy magnitude from the Ks-selected photoz catalog. For sources with spectroscopic redshifts, the PDF yields  
uncertainty estimates in good agreement with the dispersions between the spectroscopic and photometric redshifts shown in Fig. \ref{photz_specz} \citep{ilb13}. 

Fig. \ref{sigz} shows that at Ks (AB)$ < 22.5$ and low z,  $\sigma_z/ (1 + z) < 0.01$ but the accuracy 
degrades significantly at fainter magnitudes and above 
$z \sim 1.1$. The black line in Fig. \ref{sigz} indicates the median 
observed Ks-magnitude of galaxies in our sample as a function of redshift. At z $= 1$, a redshift slice 
of width $\Delta z = 0.02$ ($\sim 2\sigma_z$) is appropriate while at z $= 2$ the width should increase to $\sim0.2$. 
In fact, these variable width bins in redshift result in fairly similar spans in lookback time ($\Delta t_{LB} = 0.17 - 0.42$ Gyr).

%\clearpage

\subsection{Galaxy Classification, Stellar Mass and SFR}\label{sfr_section}

\begin{figure}[ht]
\epsscale{1.0}
\plottwo{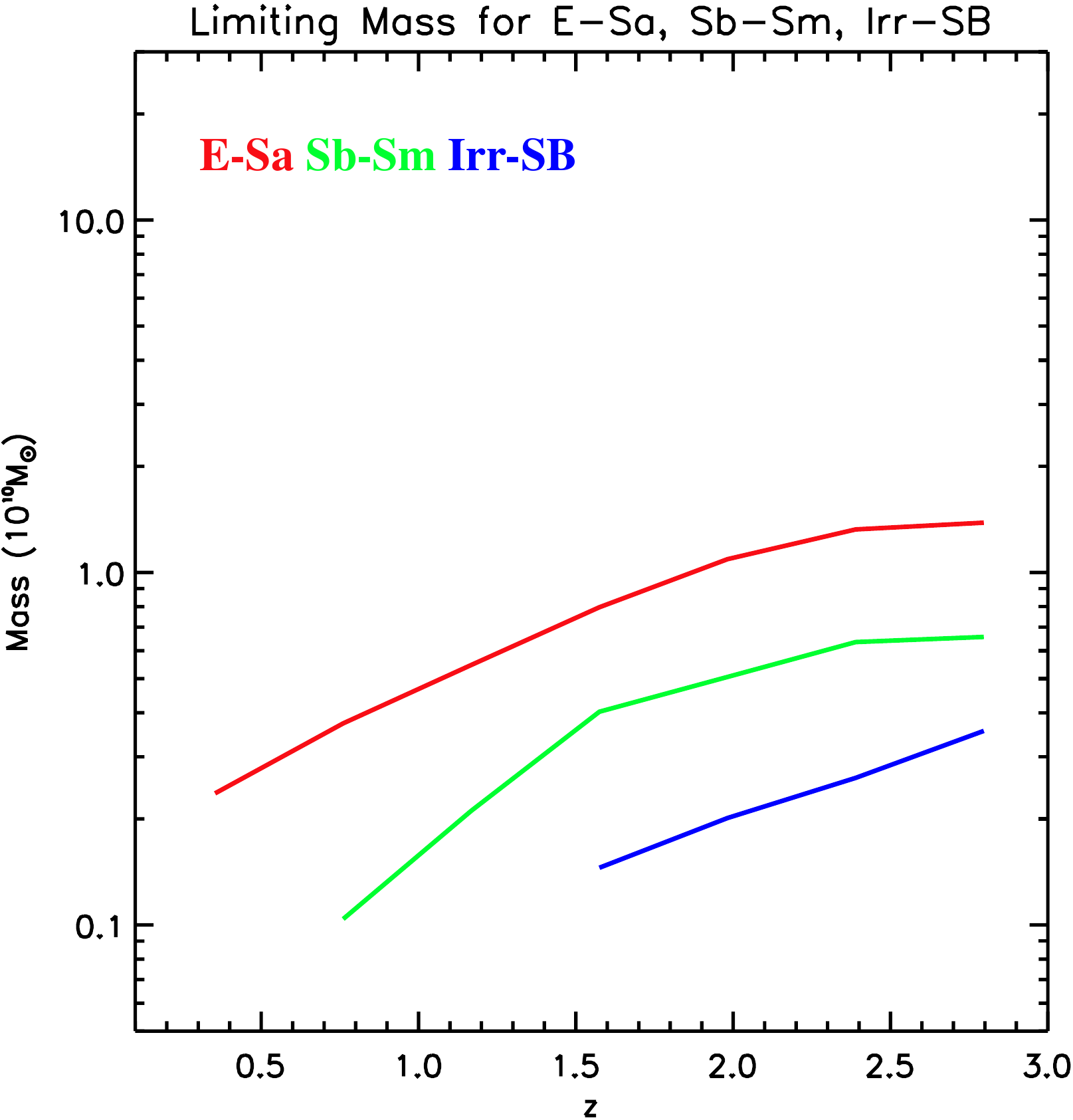}{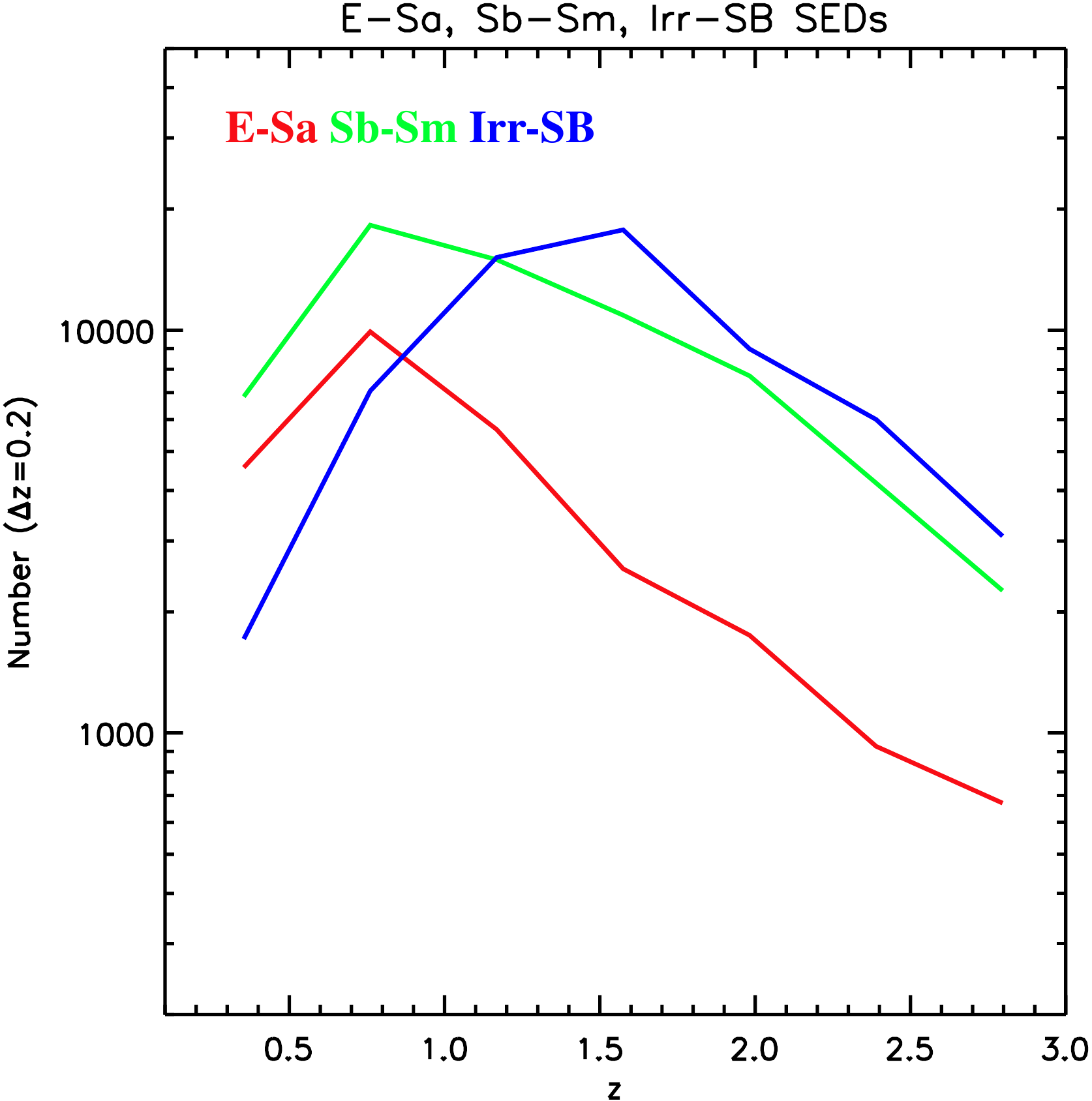}
\caption{For the sample selection with K$_s < 24$, yielding 155,954 galaxies, the stellar mass limits for early, spiral and starburst galaxies are shown as a function of redshift
together with the number counts of each type in bins of width $\Delta z = 0.2$. The lower mass limit for galaxy selection was $10^9$\msun. The small breaks in these curves are a result of the SED features redshifting through the COSMOS photometric bands which have varying sensitivities. The actual sample selection is given by the joint criteria given in Eq. \ref{selection_eq}. The gradual drop in the counts of red galaxies above z = 1.5 (right panel) is due to 
both their decreased numbers at high z and the fact that the red galaxies have lower fluxes rest frame blue which is redshifted to the Ks selection band. The decreases in total number counts at $z < 0.5$ are due simply to the decreased cosmic volume. Extensive discussion of the mass completeness is provided in \cite{ilb13}.}\label{sample}
\end{figure}\label{mass_limits_counts}

In the most recent COSMOS photoz catalog which is used here, stellar masses and SFRs were derived from fitting the template SEDs to BC03 models \citep{bru03} as discussed in \cite{ilb13}.  These models assume a Chabrier stellar initial mass function (IMF, \cite{cha03}). %[As noted above, the SFRs and stellar masses 
%would be scaled up by a factor of 1.78 if a Salpeter IMF was adopted.] 
The SFRs were estimated from both the rest frame UV continuum and the Spitzer 24$\mu$m flux (for galaxies with 24$\mu$m detections). In cases where both IR and UV SFRs were available, we used a SFR given by the extincted UV continuum plus the IR SFR. 
For the IR-based SFRs, the 24$\mu$m fluxes were converted to total L$_{IR}$ using the procedures of \cite{lee10} and using $SFR$(\msun /yr)$= 1.14\times10^{-10}$(L$_{IR}$/\lsun).
For galaxies lacking a 24$\mu$m detection, the SFR was estimated from the extinction-corrected UV continuum derived from the photoz SED fitting, using the relations given in \cite{ken98} and \cite{sch05} scaled to the Chabrier IMF, i.e.  $SFR = 1.0\times10^{-28}L_{\nu}(NUV)$ (cgs).
In order to study the impact of galaxy SED on our results, we assigned a type to each of them according to their rest-frame B-i color (including reddening), using the  types : 'SB1',  'Im',  'SB2',  'Sd',  'Sc',  'Sb', 'Sa', 'S0' and 'E', respectively \citep[similar to the b-i color classes of ][but shifted slightly to account for the different COSMOS filter bandpasses]{arn05}. For the analysis here we define three broad classes with b-i color: $>0.84$ (E-Sa), 0.45 - 0.84 (Sab-Sd) and $<0.45$ (IRR/SB). 

%$>0.1277$ (E-Sa), 0.1277 - 0.45 (Sab-Sd) and $<0.45$ (IRR/SB). 

%Following \cite{arn07}, we classify the SED type of each galaxy using its rest-frame b-i color
%in the photoz best fit SED. The adopted color classifications  :  b-i = 0.193,   0.345,  0.393,  0.50, 0.61,  0.73, 0.837, 1.0  and  1.16 
%for types : 'SB1',    'Im',  'SB2',  'Sd',  'Sc', 'Sb','Sa',  'S0' and  'E', respectively, slightly modifying the data given by \cite{arn07} to account for the different COSMOS filter passbands.

 Although color selection was not used for the sample, the resultant mass limits differ for the red and blue galaxies as a function of redshift.
Figure \ref{sample} shows the stellar mass limits for three characteristic SED types (early, spiral and starburst) resulting from the photometric selection criteria in Eq.~\ref{selection_eq}. To compute the limiting mass curves shown in Fig.  \ref{sample}, we derived the 
mean mass-to-light ratios (using the observed magnitudes in the Ks filter) for all the galaxies of each spectral class as a function of redshift,
and then scaled this ratio by the limiting magnitude (24 AB). (These limits correspond to $\sim75$\% completeness.) The number counts of the three basic SED types with the combined selection criteria are also shown.  The mass limits clearly depend on the SED of the galaxy, but having a catalog with primary source selection in Ks (rather than optical bands) greatly reduces the bias against early types \citep{ilb13}. The starburst galaxies are relatively bright at short wavelength, and therefore 
easier to detect in the observed optical at high redshift, since their UV continua will be redshifted 
to optical bands. In contrast, the early type red galaxies become much more difficult to detect at high redshift  
(i.e. a higher stellar mass is required) since they have relatively weak restframe UV continua. This difficulty is alleviated to some extent by the 
fact that at $z = 1 - 2$ the mass function of passive (red) galaxies appears to have decreased numbers of low mass systems ($< 10^{10.8}$\msun)  compared to higher masses \citep{ilb10,ilb13}; thus the lower mass red galaxies are intrinsically rare above z = 1. The percentage of passive, intrinsically red galaxies, 
is of course also much lower at z $>1$ \citep{ilb13}. 

The Ks band {\it photometric}  selection results in mass detection limits:
0.3, 0.6, 0.8 and 1.5 $\times10^{10}$\msun at z = 0.5, 1, 1.5 and  2.5 for the E-Sa SED types. For the Irr-SB SED types,
the equivalent limits are:  $<$0.08, 0.1, 0.12 and 0.2 $\times10^{10}$\msun at z = 0.5, 1, 1.5 and  2.5. [The explicit mass 
selection in Eq.~\ref{selection_eq} removes all galaxies with mass below $10^{9}$\msun.]
At z = 0.5 to 2 , the knee in the galaxy stellar mass function drops from
$log M_* = 10.9$ to 10.6 \msun, i.e. from 8 to 4$\times10^{10}$ \msun \citep{ilb13} for quiescent galaxies. For the blue galaxy SEDs, our selection reaches 
more than an order of magnitude below these M$_*$ values, even at the highest redshifts. For the 
red SEDs, the mass limit reaches $< 0.2$ M$_*$ all the way to z $\sim$3.
%[It is important to note that the \cite{arn07} relation used to calculate the stellar masses 
%in the COSMOS photoz catalog has only modest evolution in the ratio M$_{stellar}/L_K$ due to stellar population aging over the redshift range 
%z = 0 to 1.5. This mass-to-light ratio decreases by factors of 1.9 and 2.6 for the red and blue galaxies respectively. Stronger stellar population evolution would of course decrease the high-z mass-limits.]

\subsection{$\Lambda$CDM Simulation}\label{millennium}

One of the goals of this study is a comparison of the observed evolution in the COSMOS LSS with current theoretical
models. For this, we make use of mock simulation catalogs generated for an area and volume equivalent to the COSMOS survey. The mock catalogs are based on $\Lambda CDM$ simulations which start at z = 127 evolved down to z = 0 \citep{wan08}.  For comparison with the COSMOS data we make use of their WMAP3YC simulation, which 
adopts cosmological parameters derived from a combination of
third-year WMAP data on large scales, and Cosmic Background Imager and
extended Very Small Array data on small scales \citep{spe07} (with $\Omega_M=0.226$ and $\Omega_{\Lambda}=0.774$).
Their mass and force resolution are the same as used in the
Millennium Simulation \citep{spr05}, while the volume is smaller by a factor of
64.

\begin{figure}[ht]
\epsscale{0.5}
\plotone{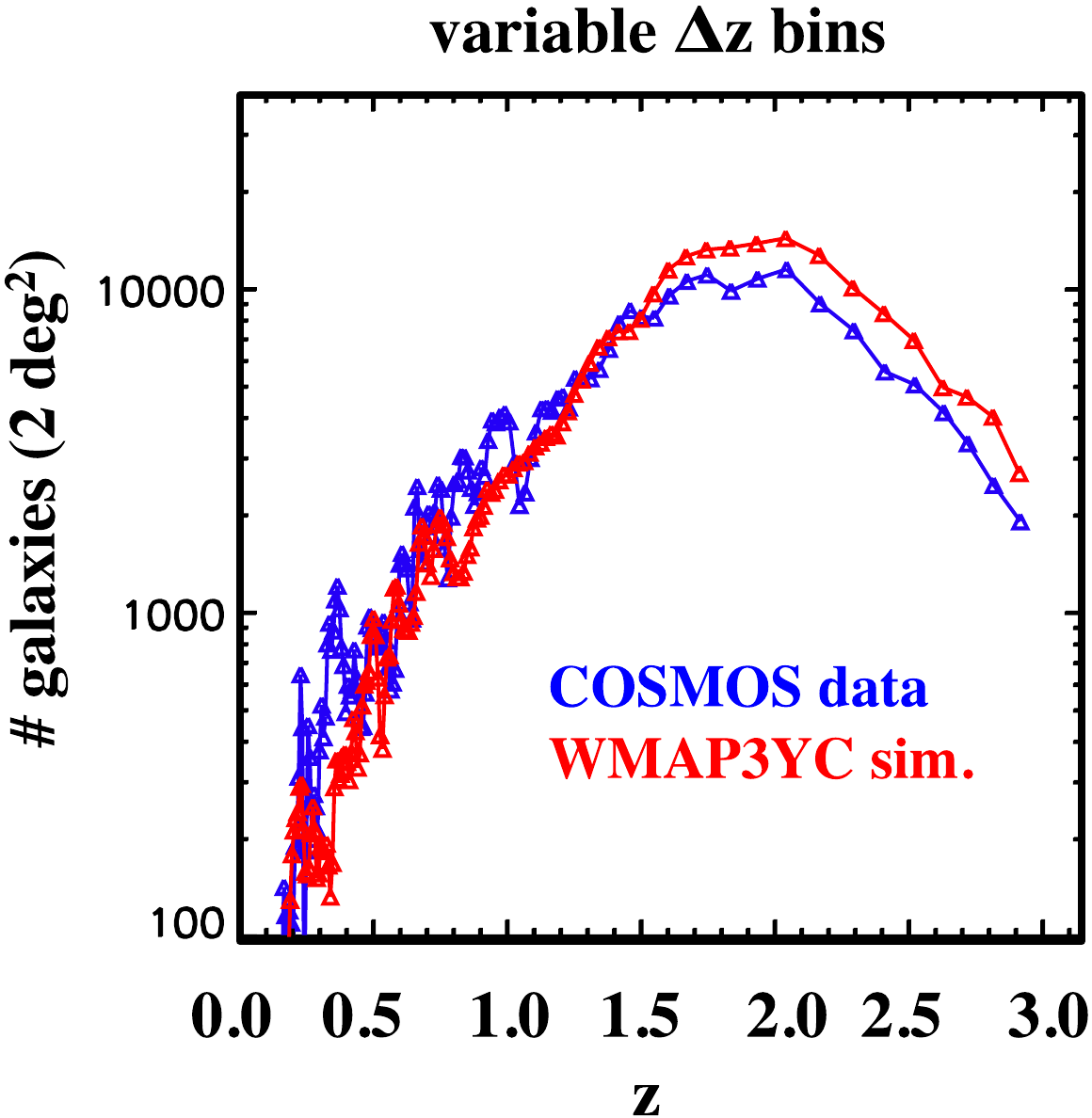}
\caption{The redshift distributions for COSMOS  (blue) and simulation galaxies (red)) 
are shown for the adopted selection criteria (K$_s < 24$ AB mag at z = 0.15 to 3.0). The width $\Delta z$ of the redshift slices varies with redshift 
to match the photoz accuracy (Fig. \ref{sigz}), i.e. larger width bins at high z. These are the same bin widths used in the LSS mapping, giving 1000 to 10000 galaxies in each redshift slice at z $>$0.5. Note that these bins are spaced by half a bin width so the total number counts are half of the sums obtained from this figure.}\label{counts}
\end{figure}

The galaxy formation model of \cite{del07}  was adopted to calculate the galaxy properties. This
model has been able to reproduce many aspects of local galaxy
populations \citep[e.g.][]{cro06,del07}  and
high redshift galaxy properties \citep{kit07,guo09}. For WMAP3, two sets of
parameters are found to reproduce the local observational data \citep[see][]{wan08,del07,cro06,spr05}. The simulations track halo dark matter masses, star formation rates (SFRs) and stellar
masses.

This simulation was extremely valuable for evaluating the effectiveness of our techniques for identifying LSS in the
presence of redshift errors similar to those of the COSMOS photoz,
and for analysis of the scaling between the derived 2-d surface
densities of galaxies and the 3-d volume density of galaxies, 
 for the range of LSS expected to be present at high redshift. % (see Appendix \ref{appendix}).  

 The mock catalog includes photometric magnitudes in the
COSMOS filter passbands from FUV to IRAC1-4 and rest frame absolute
magnitudes, with and without dust extinction. Galaxies were selected from the simulation using the same photometric cuts/limits as 
used for COSMOS (Eq. \ref{selection_eq}). Redshifts from the simulations were also scattered with a dispersion identical to those in the 
COSMOS photometric redshifts, as a function of magnitude and redshift
(see Fig. \ref{sigz}). A known problem with the simulation is an overabundance of low stellar mass galaxies \citep[see Fig. 1 in ][]{guo11}; to alleviate this problem, we imposed a stellar mass limit of $2\times10^9$\msun (instead of $10^9$\msun) to yield similar sample sizes
to the observed galaxy sample. 
In each
mock catalog, two dust extinction curves ('dust1' and 'dust2') were employed to relate the  dust extinction to the surface
density of HI and the metallicity of the ISM. The major difference between
the two dust models is that the 'dust2' has weaker  
dependence on redshift \citep{guo09} to better reproduce the observed counts of Lyman break galaxies \citep{guo09}. %motivated by
%observational
%indications that dust-to-gas ratios are lower at high redshift than
%in the local universe for galaxies with the same bolometric
%luminosity and metallicity  \citep[e.g.][]{ade00}. This
%redshift dependence has been proven necessary to reproduce the
%properties of high redshift Lyman-break galaxies and distant red
%galaxies simultaneously \citep{guo09}. 
In this paper, we
compare observational results with the WMAP3YC model using the magnitudes computed with the 
'dust2' extinction curve. 

The redshifts of the galaxies from the simulation were then dispersed with the same uncertainties as 
for the COSMOS photoz catalog (Fig. \ref{sigz}). The LSS in the simulation was also measured with the 
same routines used for the COSMOS galaxy sample (\S \ref{densities}).  
Fig. \ref{counts} shows a comparison between the
redshift  distributions of galaxies in the mock and  in
the COSMOS sample used here. Overall, there is very good correspondence in the two redshift distributions.

\section{Galaxy Environmental Densities}\label{densities}

The environmental density for each galaxy was derived from the local surface 
density of galaxies within the same redshift slice (\S \ref{slice}) based on the high accuracy COSMOS photometric redshifts for the 155,954 galaxies.
[We note that \cite{kno09,kno12} provide a catalog of galaxy groups and \cite{kov10} the density field, both based on the 
zCOSMOS spectroscopic redshifts for 16,500 galaxies.]
Two techniques were employed here to map the LSS: adaptive spatial smoothing and Voronoi 2-d tessellation (\S \ref{density}).

\subsection{Redshift Slices}\label{slice}

For mapping LSS it is vital that the binning in redshift 
be matched to the accuracy of the redshifts, to provide 
optimum detection of the overdensities associated with LSS.
Using redshift bins that are finer
than the redshift uncertainties distributes the galaxies from a single structure
over multiple redshift slices and thus reduces the signal-to-noise ratios in each slice. Conversely,
 bins of width larger than the redshift uncertainties will increase the shot noise associated
with foreground and background galaxies, relative to the large-scale structure signal, i.e.
galaxies from neighboring redshifts are superposed on the LSS at the redshift of interest.

For the adaptive smoothing algorithm discussed in \cite{sco_lss},
each galaxy is distributed in z according to its photoz PDF (probability density function); for the Voronoi tessellation, each galaxy 
is placed at the maximum likelihood photometric redshift. (Rather than using the minimum chisq photoz, we use the median of the marginalization of the redshift probability distribution.)
If the uncertainties in the galaxy redshifts were a Gaussian distribution, the optimum 
smoothing or binning in redshift would be a Gaussian of FHWM $\simeq 2\sigma_z$ (if there are approximately equal densities of galaxies in LSS, and a uniformly distributed field population). The width of this optimum redshift binning should increase as the number of randomly superposed 'field' galaxies is decreased. In the following, we adopt redshift bin widths of $\Delta z = 2\sigma_z$ where $\sigma_z/(1+z)$ is shown in Fig. \ref{sigz} as the line corresponding to the expected 
uncertainty at the median magnitude of sample galaxies as a function of redshift. 
The adjacent redshift slices are spaced by half of the width of the slices at each redshift.
The result is a total of 127 redshift slices ranging from z $= 0.15$ to 3.0 which 
are analyzed for significant LSS. This results in the bins having galaxy counts as shown in Fig. \ref{counts}.

\subsection{Galaxy Density Measurements}\label{density}

%\clearpage

\begin{figure}[ht]
\epsscale{0.5}
\plotone{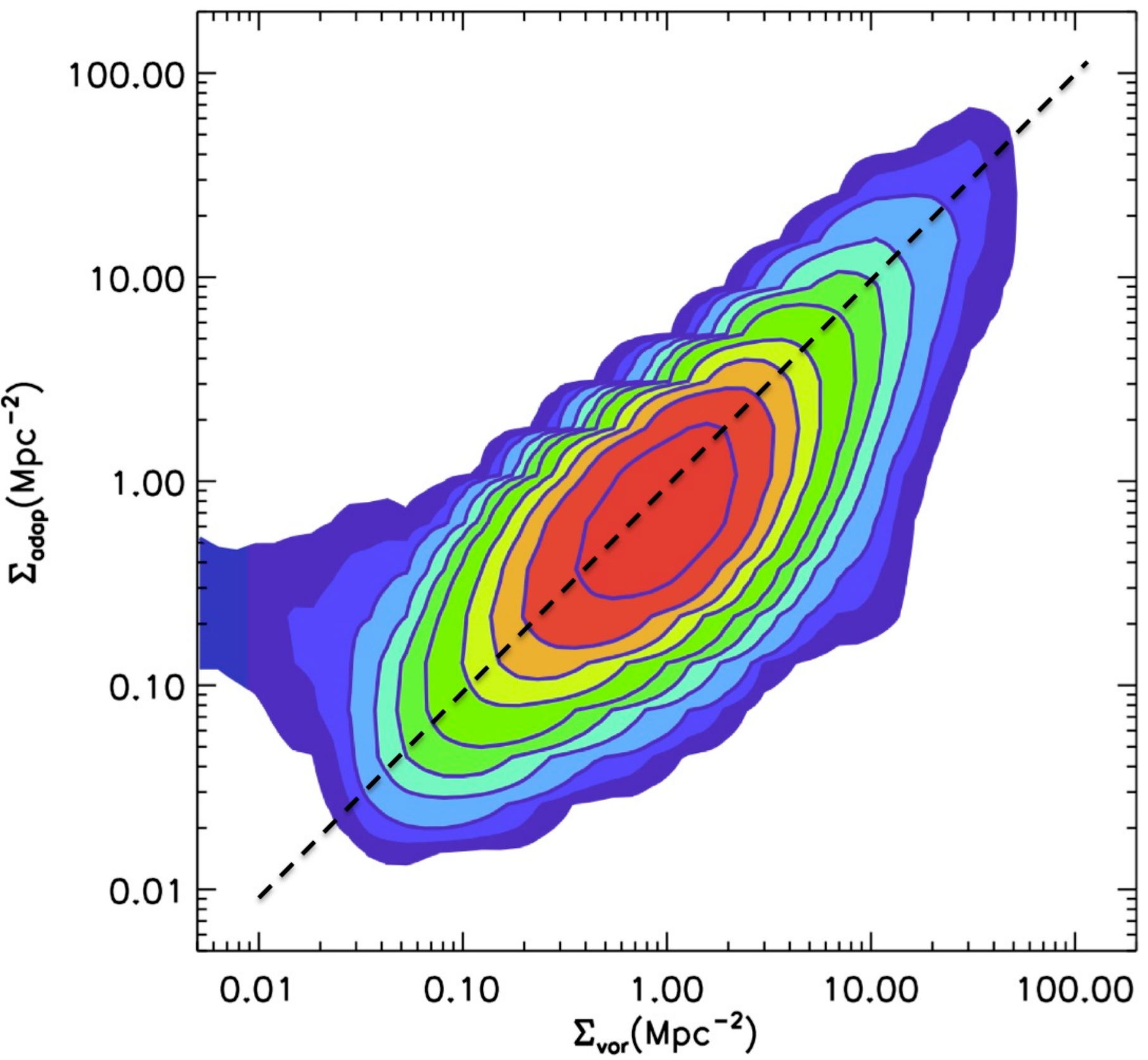}
\caption{ The environmental densities obtained from the adaptive smoothing and Voronoi techniques 
are compared for the sample of 150,852 galaxies.  Contours are at 0.0005, 0.001, 0.002, 0.005, 0.01, 0.02, 0.05, 0.1, 0.2 and 0.5. Over almost 3 orders of magnitude in the surface density, the two techniques give similar results following a line of 1:1 correspondence along the ridge line with the highest number of objects. The deviation seen on the outermost contour on the left side is due to the fact that the adaptive smoothing is designed to detect only statistically significant overdensities in each redshift slice whereas the Voronoi densities are derived everywhere. 
}\label{rho_vs_rho}
\end{figure}

Two techniques are used here to image the LSS environments in the galaxy surface density 
distribution in the 127 redshift slices : adaptive spatial filtering and Voronoi 2-d tessellation. 
The former was developed and tested in our previous analysis of COSMOS LSS
\citep[see][]{sco_lss}; the latter has been used in many earlier investigations of galaxy 
LSS \citep{wey94,ebe93,mar02,ger05}. Both techniques are used here since each has clear advantages and disadvantages
and the generally good agreement in the derived density fields provides confidence in the 
results of both (see Fig. \ref{rho_vs_rho}). The sample numbers shown in the figures below are less than the total sample of 155,954 galaxies, since galaxies at the edge of the survey area do not have closed Voronoi polygons.

The adaptive smoothing procedure has a clearly specified level of significance,
and structures of lower statistical significance are simply not detected. On the other hand, 
the adaptive filtering which makes use of a variable width Gaussian spatial smoothing function
is less appropriate than the Voronoi tessellation for detection of elongated and irregular structures. 
The latter  technique locates the polygon area closest to each galaxy and is therefore not making an assumption of structure shape. For the adaptive smoothing the tests, run on a 'redshift slice', in which 
50\% of the galaxies were in modeled overdense concentrations and 50\% were randomly 
distributed, showed extremely good proportionality between the recovered densities and 
the models, with virtually no spurious features when compared to the input model (see Appendix in 
\cite{sco_lss}. 
\footnote{For the adaptive smoothing, the two adjustable parameters in the algorithm were the same as those used in \cite{sco_lss}. Specifically, at 
a given spatial filter width, the smoothed surface density was required to be detected at a  
significance of 2.5$\sigma_{Poisson}$ and the gradient detection significance  \citep[see][]{sco_lss} was set to 0.5$\sigma_{Poisson}$ (where $\sigma_{Poisson}$ is the Poisson noise level calculated from the mean surface density in the redshift slice).}

A second difference between these techniques arises from the fact that adaptive smoothing searches a defined 
range of angular scales, whereas the Voronoi tessellation is unrestricted. For the former, the data 
is spatially binned in 600$\times$600 pixels (0.2\arcmin) across the 2 deg field and smoothing 
filters from 1 to 60 pixels (FWHM) are searched for significant overdensity. The filtering width 
thus corresponds to 0.2\arcmin ~to 0.2\deg, corresponding to comoving scales
of 200 kpc to 12 Mpc at z = 1. Thus one anticipates that the Voronoi technique can yield higher 
densities on scales smaller than 0.2\arcmin ~or in elongated structures. The Voronoi technique will
also provide a density estimate for all galaxies independent of whether the environmental density is statistically significant. The latter can be an 
advantage or a disadvantage depending on how the density estimates are to be employed, so we feel it is beneficial to have both density fields.  

Both techniques yield the 2-d surface density of galaxies in each redshift slice
rather than the true 3-d volume density of galaxies.  Direct determination of the 3-d volume densities 
would require more precise redshifts and a means of correcting for non-Hubble flow streaming and increased velocity dispersion due to LSS mass concentrations. 
The accuracy of the redshifts would need to be  
a factor of $\sim10$ higher to resolve the cluster velocity dispersions. In the very dense environments, the increased velocity dispersions 
may actually indicate that the 2-d surface densities  provide a more robust measure of the galaxy environment (provided this surface density 
is mostly dominated by the LSS in the slice with little foreground and background contamination). 
In general, one expects proportionality between the derived projected 2-d and true 3-d densities 
as long as the redshift slices are fine enough that there are few galaxies superposed from other redshifts. To test the proportionality, 
we have run both the adaptive smoothing and Voronoi 2-d tessellation algorithms 
on the simulation mock catalog. Since the simulation has 
accurate 3-d positions, we were also able to evaluate 
the 3-d densities using a 3-d tessellation.% In order to understand the smearing effects of the photoz redshift uncertainties,
%we introduced a dispersion on to the simulation redshifts similar to that shown in Fig. \ref{sigz} for the COSMOS photoz. 
%In Appendix \ref{appendix}, 
We found that for the galaxy densities and redshift uncertainties in our samples, the 2-d projected 
densities were monotonically related to the true 3-d volume densities with a $\sim0.67$ power law as expected
for linear structures. % (Fig. \ref{3d2d}). A fit to the data in Fig. \ref{3d2d} yields $\Sigma_{2d} \propto \rho_{3d}^{0.54}$
%but clearly the outer contours are somewhat steeper and similar to the 0.67 power expected
%for linear structures. At higher environmental densities, the slope steepens somewhat; this is also expected since the high density regions in the simulation are likely to be relaxed, i.e. more spherical and less filamentary. 

\subsection{Comparison of Adaptive Smoothing and Voronoi Densities}\label{comparison}

The adaptive smoothing and Voronoi techniques give estimates for the local surface 
densities of galaxies which are in reasonable correspondence, given their very different approaches and assumptions (as discussed above in \S \ref{density}). Figure \ref{rho_vs_rho} shows 
the distribution of the $\Sigma_{adap}$ versus $\Sigma_{vor}$ for the sample of 150,852 
galaxies. This number of galaxies is slightly less than the sample number quoted earlier since 
the Voronoi polygons are not closed at the outer edges of the field and no area and density 
estimate is obtained for those galaxies. Over $\sim$3 orders of magnitude in the surface density the two techniques give similar results with the ridge line for the highest number of objects tilted somewhat, relative to the shown 45 degree equality line. The tilt offset is due to the fact that the adaptive 
smoothing algorithm only recovers densities at a smoothing scale length such that the density is statistically significant, whereas the Voronoi densities do not have this restriction. Deviations can also be seen in the outer 4 contours at level 1/256 of the peak : these are due to the ability of the Voronoi to go to effectively higher resolution at higher densities. The maximum resolution in the adaptive smoothing is set at 1/600 of the field or $\sim$10.8\arcsec). 

In the following, we use the densities derived from the Voronoi tessellation for 
correlating galaxy properties with environmental density. The tessellation provides an
estimate of the environment of all galaxies even if these are not significantly overdense. On the other hand,  the adaptive smoothing  
is more appropriate for the identification of statistically significant large scale structures if that is required (although not the subject of the work here).

The environmental density estimates in the COSMOS field as derived here are available for 
download from the IPAC/IRSA COSMOS archive at http://irsa.ipac.caltech.edu/data/COSMOS/.

\begin{figure}
\epsscale{1.}
\plotone{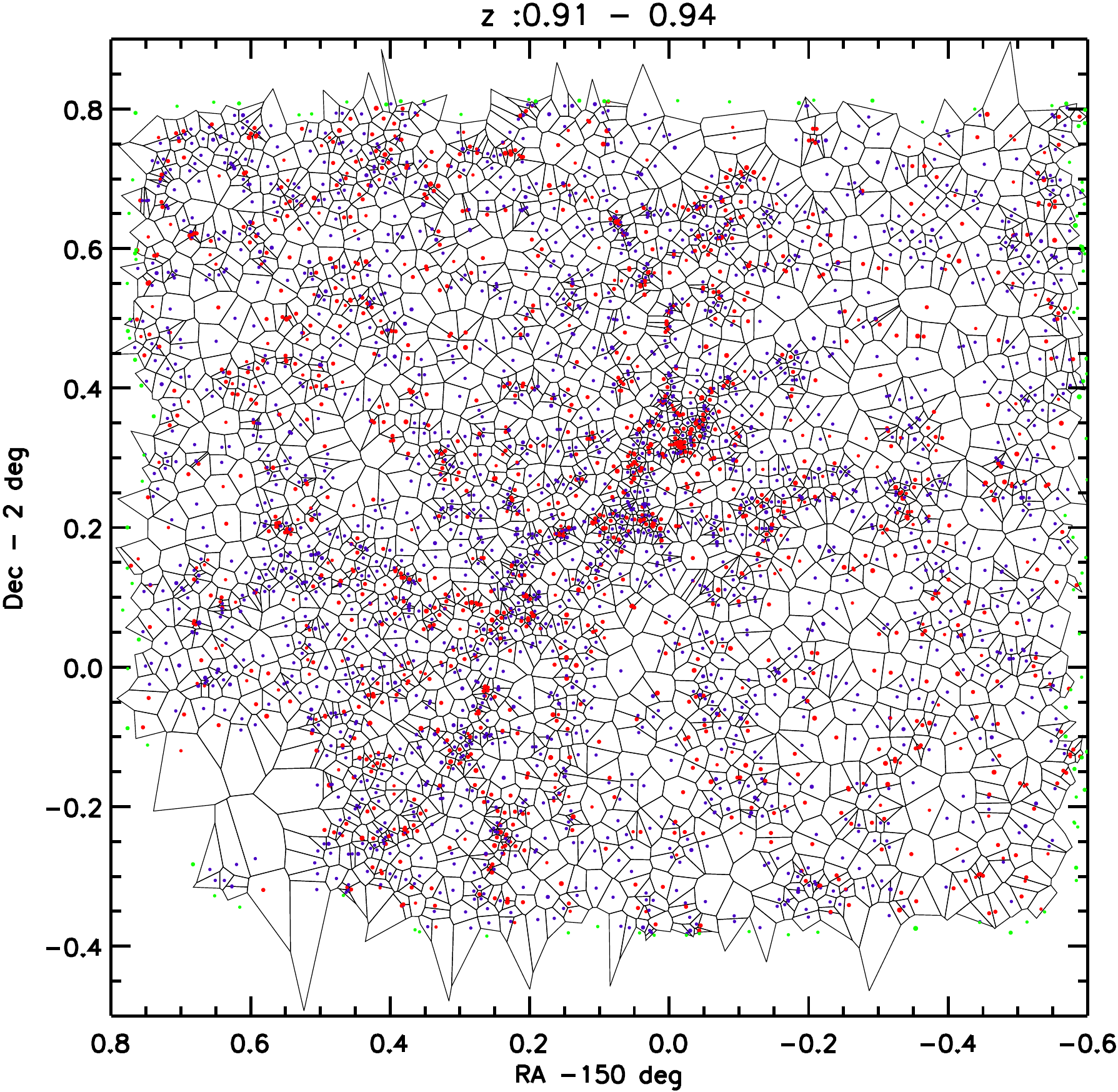}
\caption{The galaxy density field is shown for a sample redshift slice at z $= 0.93$ with the width ($\Delta z = 0.03$) of the slice matching the accuracy of the photometric redshifts for the selected galaxy sample.  The Voronoi 2-d polygons outline the area closest to each individual galaxy. In the tessellation diagrams 
the individual galaxies are shown as red or blue points depending on the SED type of the galaxy (early and late, respectively), clearly showing the correlation of early type galaxies with denser environments 
at these redshifts.  (Green points indicate galaxies on the outside of the area for which the Voronoi polygons are not closed.) %The vertical bar in the lower right indicates the comoving scale length. 
The sparse region in the lower left corner is a masked area where the Ultra-Vista coverage was incomplete (see Fig. \ref{mask}). Animation 1 shows the galaxies distribution which was the basis for the derived densities and 
 full sets of the voronoi plots for the 127 redshift slices are available in animation 2 in the electronic edition of the
{\it Astrophysical Journal}. }\label{vor}
\end{figure}

\begin{figure}
\epsscale{1.}
\plotone{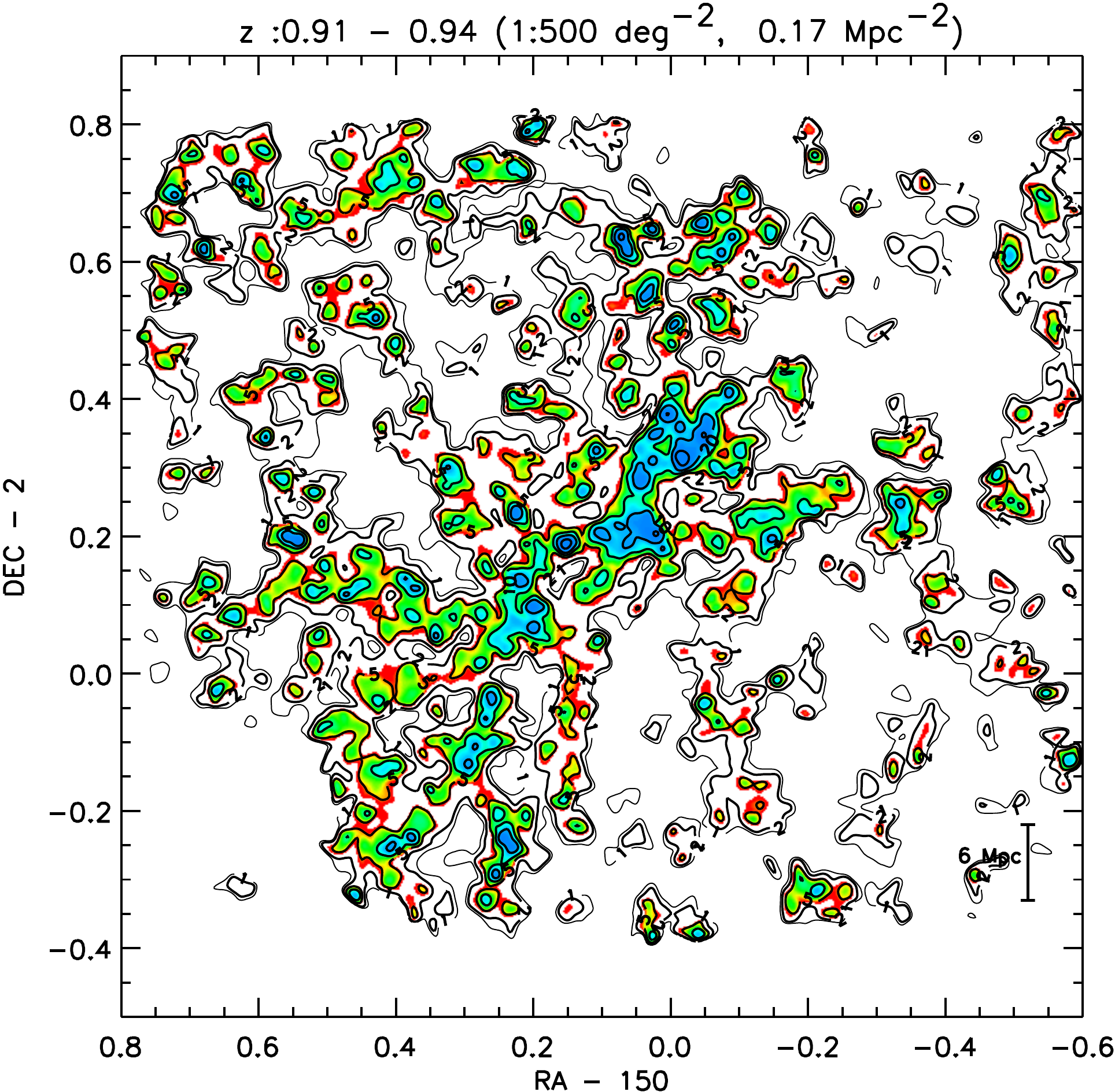}
\caption{The density field estimated from the inverse area of each galaxy shown in Fig. \ref{vor}. The vertical bar in the lower right indicates the comoving scale length of 6 Mpc. 
Contours are at : 1, 2 , 5, 10, 20, 50, 100, 200 and 500 times the density units (per deg$^2$ and per Mpc$^2$) given in the 
upper legend of the plot. Full sets of these plots for the 127 redshift slices are available in animation 3 in the electronic edition of the
{\it Astrophysical Journal}. }\label{vor_den}
\end{figure}

\begin{figure}

\epsscale{1.}
\plotone{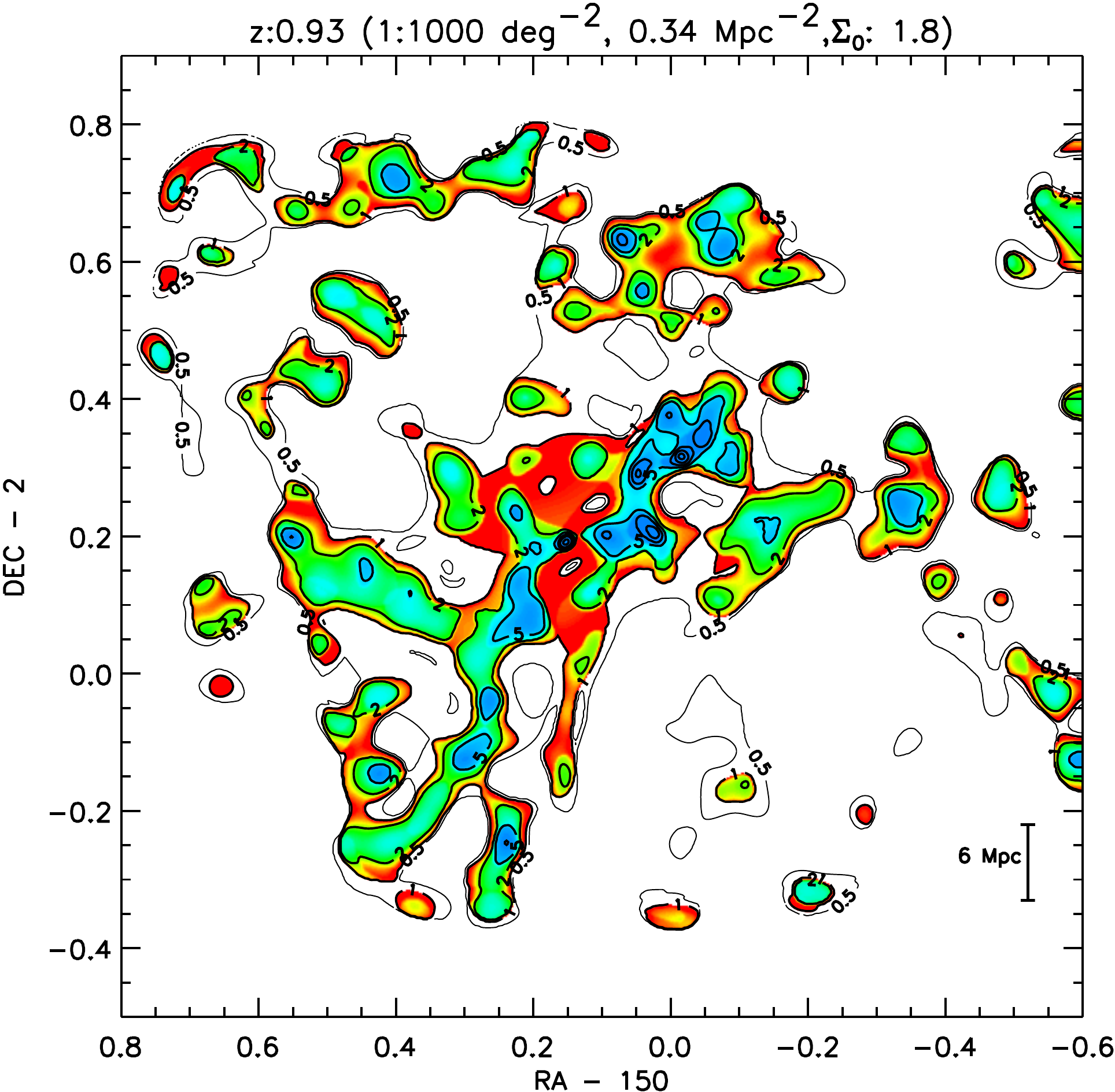}
\caption{Statistically significant overdense regions (relative to the mean density at the same redshift) are
shown from adaptively smoothing the spatial distribution of galaxies in redshift slice at z $= 0.93$ (see Fig \ref{vor}).
The vertical bar in the lower right indicates the comoving scale length of 6 Mpc. 
Contours are at : 0.5, 1, 2 , 5, 10, 20, 50, 100, 200 and 500 times the density units (per deg$^2$ and per comoving Mpc$^2$) given in the 
upper legend of the plot. The mean density given by $\Sigma_0$ in the upper legend has been removed before computing the overdensity. Full sets of these plots for the 127 redshift slices are available in animation 4  in the electronic edition of the
{\it Astrophysical Journal}. }\label{ad}
\end{figure}

\section{COSMOS LSS}\label{lss}

Figures \ref{vor} --  \ref{ad} show the derived density fields of galaxies and overdensities for the selected redshift slices. From the tessellation
analysis, both the Voronoi polygons for each galaxy (color coded red for early-type SEDs 
and blue for late-type or starburst SEDs) and the derived density fields are shown. Statistically significant overdensities are revealed from the adaptively  smoothed surface densities in Fig. \ref{ad}. 
These figures illustrate well the spatial clustering of the galaxies which can be seen in COSMOS using  
accurate photoz to remove foreground and background galaxies for each redshift. In each redshift slice 
many overdense structures are seen -- both dense 'circular clumps' and elongated filamentary 
structures. \emph{ The routine detection of the filamentary structures at most redshifts is a new feature provided by COSMOS -- enabled by the large galaxy samples having high accuracy photometric redshifts.} The very large samples of galaxies available through photometric redshifts 
enable the mapping of structures even at relatively low densities. [Areas masked due to bright stars contaminating the photometry are 
shown in Fig. \ref{mask} and these appear as blank regions in the LSS at all redshifts.]

\begin{figure}[ht]
\epsscale{1.6}
%\plotfiddle{PSFILE}{VSIZE}{ROTANG}{HSCALE}{VSCALE}{HTRANS}{VTRANS}
%\vskip -9in
%\plotfiddle{lss_2d_p19-eps-converted-to.pdf}{9.in}{90.}{600.}{600.}{-100}{600}
\plottwo{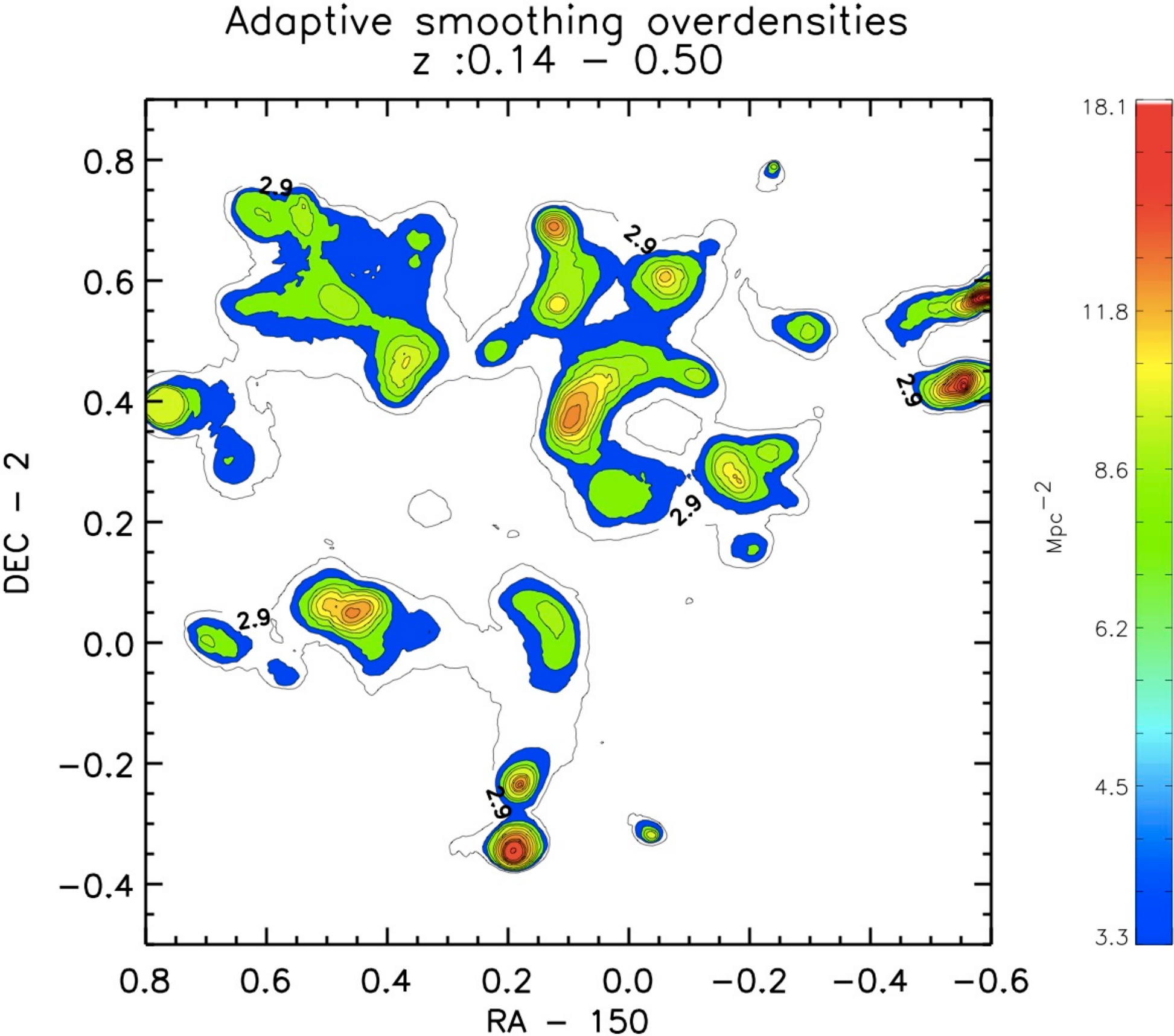}{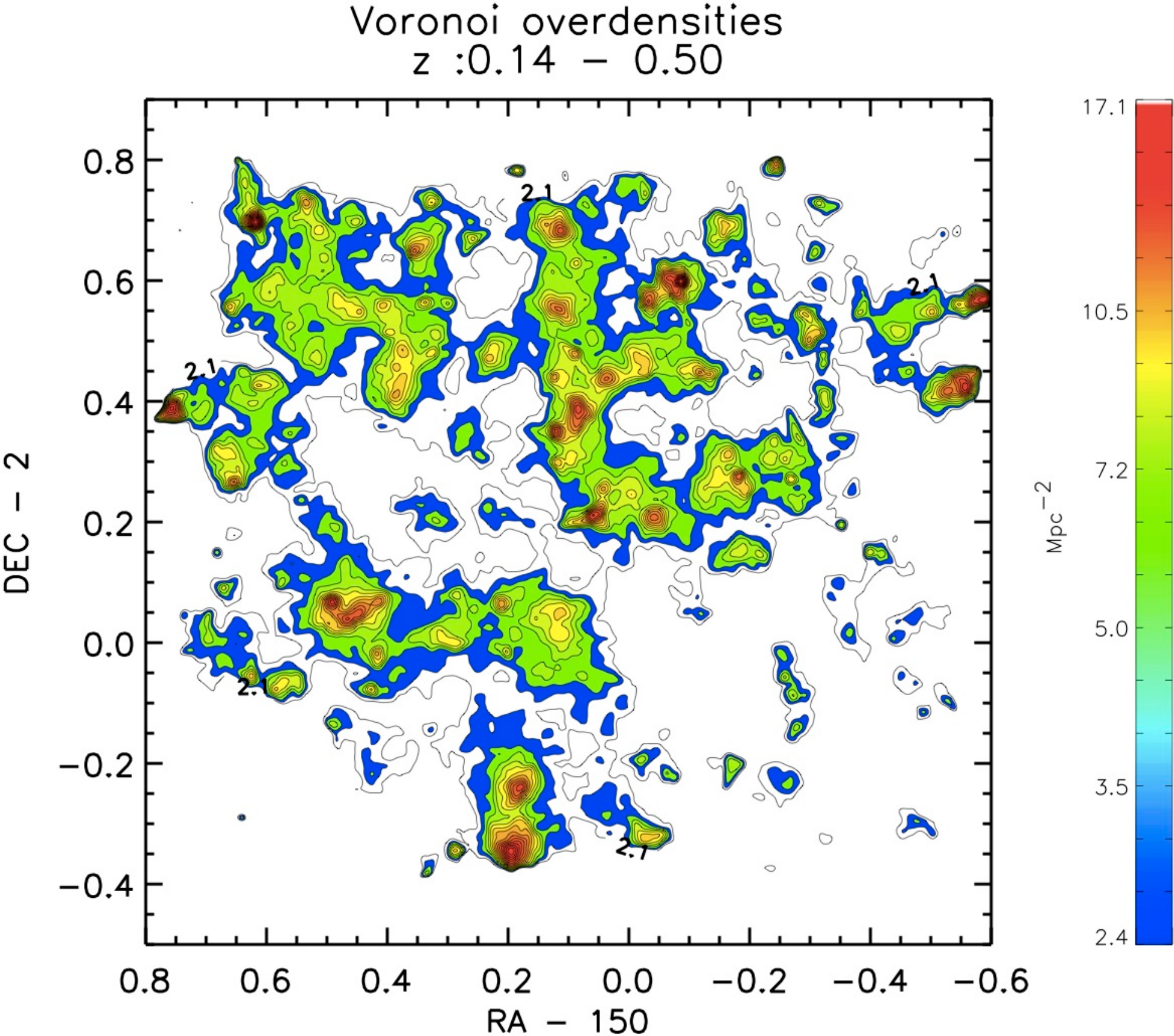}
%\plotone{lss_2d_p20-eps-converted-to.pdf}
\caption{ Overdense regions at z = 0.14 - 0.5 are shown for the adaptive smoothing (Top) and Voronoi (Bottom)
 techniques. The images were made by summing the derived overdensities measured 
from the individual redshift slices. The numbers on the lowest uncolored contour correspond to the projected density in Mpc$^{-2}$. %The regions with no overdensity at any redshift on these figures are often areas masked due to  'bright' stars (see Fig \ref{mask}).
}\label{lss_2d}
\end{figure}

In Figure \ref{lss_2d}, the adaptively smoothed and 
Voronoi projected surface densities are shown for selected ranges of redshift. These images were made by summing the densities over 
the range of redshifts specified  on each plot. In general, there is extremely good correspondence 
between the structures derived using the two techniques after allowing for their different objectives and strengths :
the adaptive smoothing picks up only statistically significant overdensities while the Voronoi 
technique shows all overdensites and is less shape and scale dependent. Approximately 
250 significantly overdense regions are detected with scales 1 to 30 Mpc (comoving). 
We did not attempt to catalog the separate structures -- tracing their full extent 
and deciding whether multiple peaks are really part of a single larger structure becomes 
quite subjective. (Automated delineation of the structures was attempted with only limited success; the parameters 
appropriate to different redshifts must be changed as a function of redshift, due to the varying levels of confusion and thus, the resulting
catalogs are non-uniform in their selection biases.)

\clearpage

\begin{figure}[ht]
\figurenum{\ref{lss_2d} b}
\epsscale{1.6}
%\plotfiddle{PSFILE}{VSIZE}{ROTANG}{HSCALE}{VSCALE}{HTRANS}{VTRANS}
%\vskip -9in
%\plotfiddle{lss_2d_p19-eps-converted-to.pdf}{9.in}{90.}{600.}{600.}{-100}{600}
\plottwo{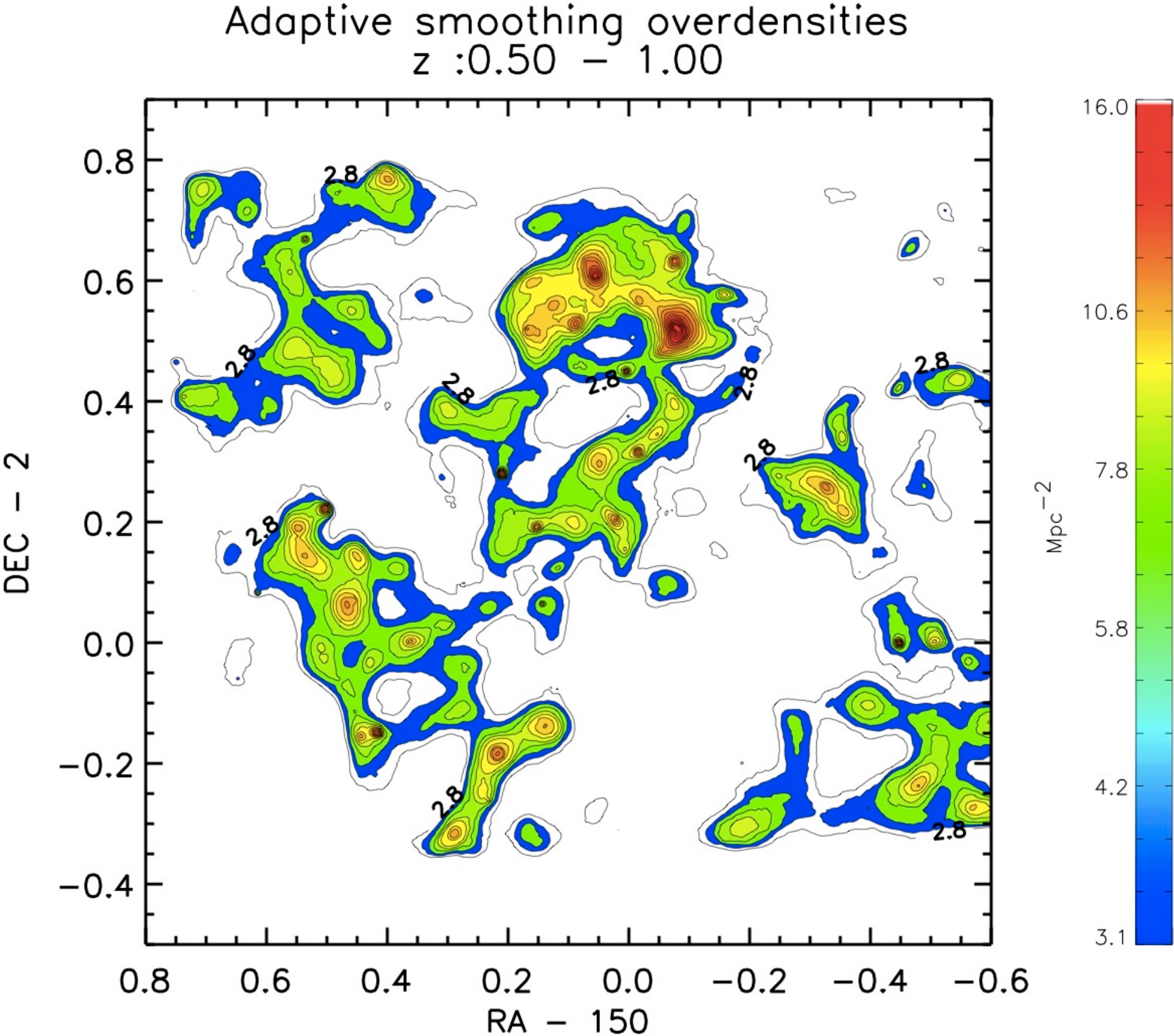}{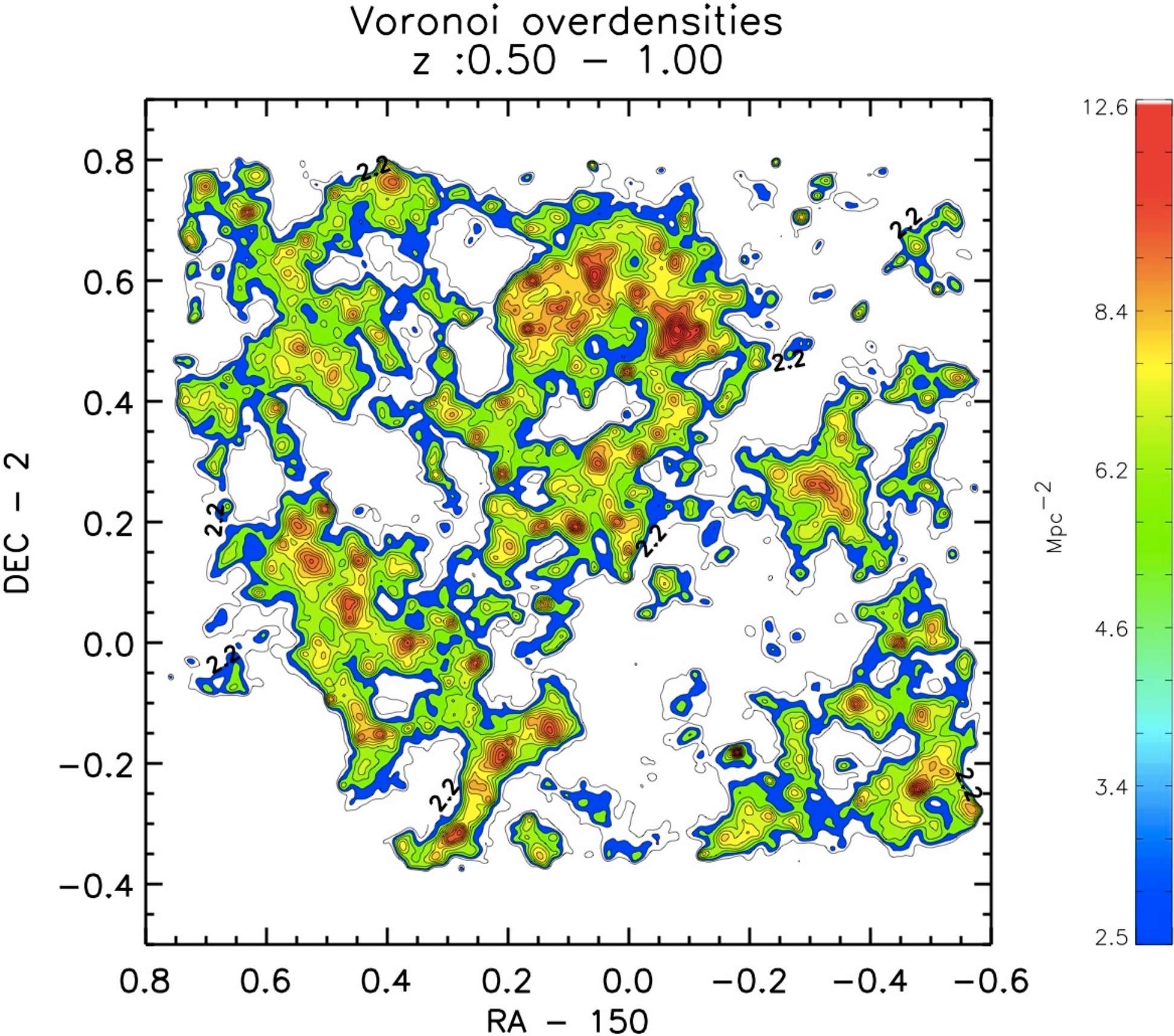}
%\plotone{lss_2d_p20-eps-converted-to.pdf}
\caption{ Overdense regions at z = 0.5 - 1.0 are shown for the adaptive smoothing (Top) and Voronoi (Bottom)
 techniques. The most massive structure in COSMOS at z = 0.73 can be seen here in the top center \citep{guz07,cas07}.
}
\end{figure}

\clearpage

\begin{figure}[ht]
\figurenum{\ref{lss_2d} c}
\epsscale{1.6}
%\plotfiddle{PSFILE}{VSIZE}{ROTANG}{HSCALE}{VSCALE}{HTRANS}{VTRANS}
%\vskip -9in
%\plotfiddle{lss_2d_p19-eps-converted-to.pdf}{9.in}{90.}{600.}{600.}{-100}{600}
\plottwo{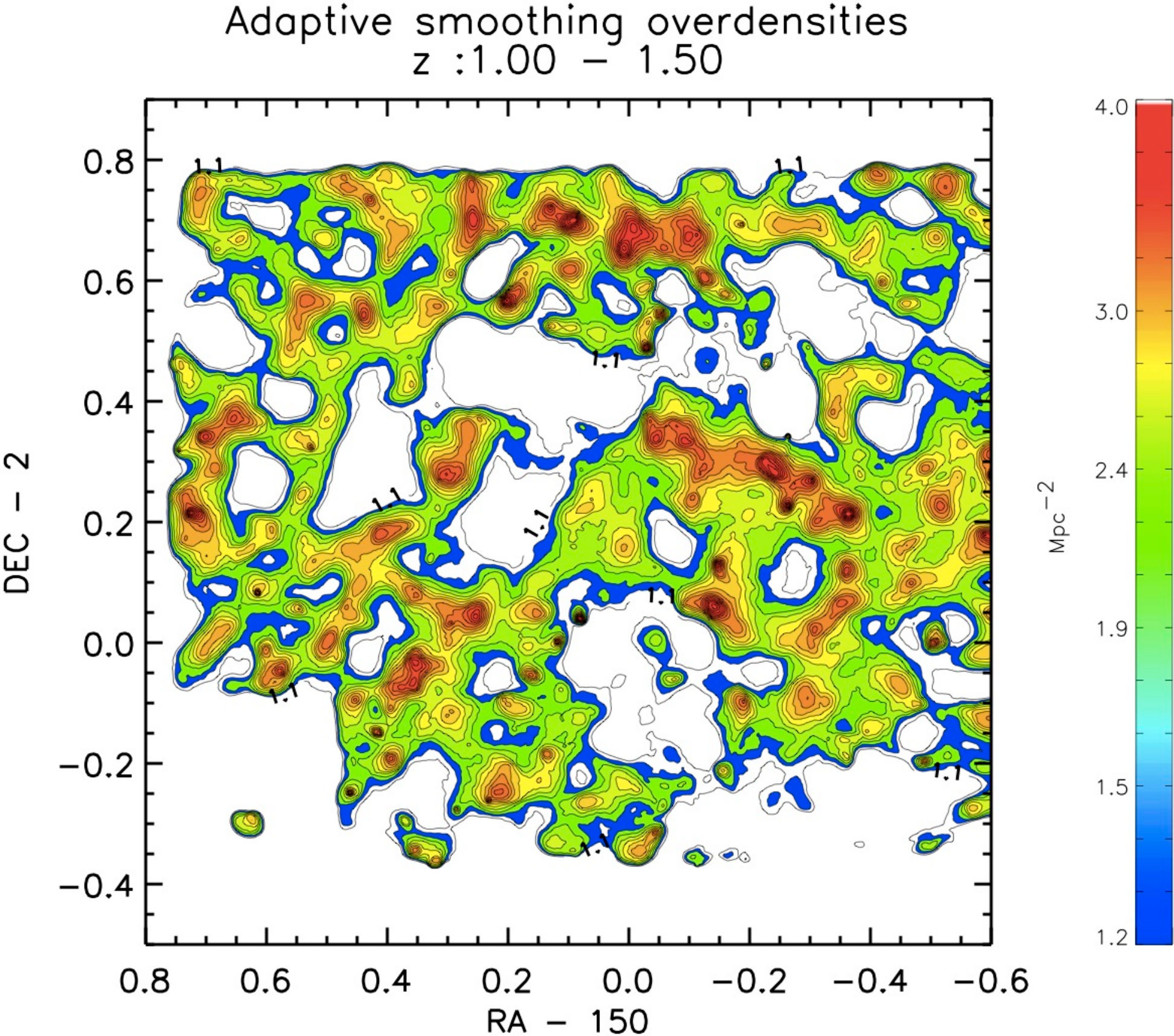}{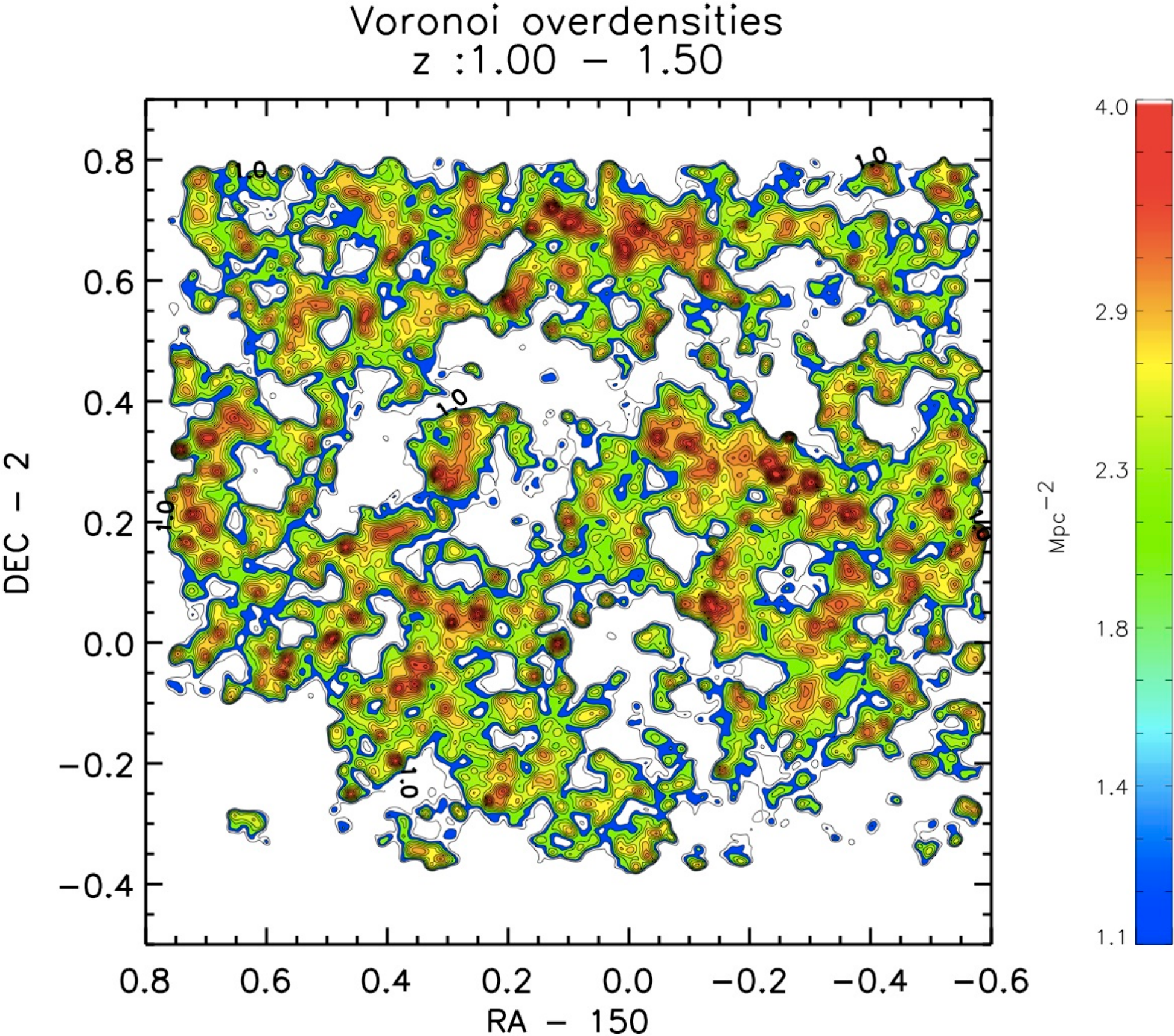}
%\plotone{lss_2d_p20-eps-converted-to.pdf}
\caption{ Overdense regions at z = 1.0 - 1.5 are shown for the adaptive smoothing (Top) and Voronoi (Bottom)
 techniques. 
}
\end{figure}
\clearpage

\begin{figure}[ht]
\figurenum{\ref{lss_2d} d}
\epsscale{1.6}
%\plotfiddle{PSFILE}{VSIZE}{ROTANG}{HSCALE}{VSCALE}{HTRANS}{VTRANS}
%\vskip -9in
%\plotfiddle{lss_2d_p19-eps-converted-to.pdf}{9.in}{90.}{600.}{600.}{-100}{600}
\plottwo{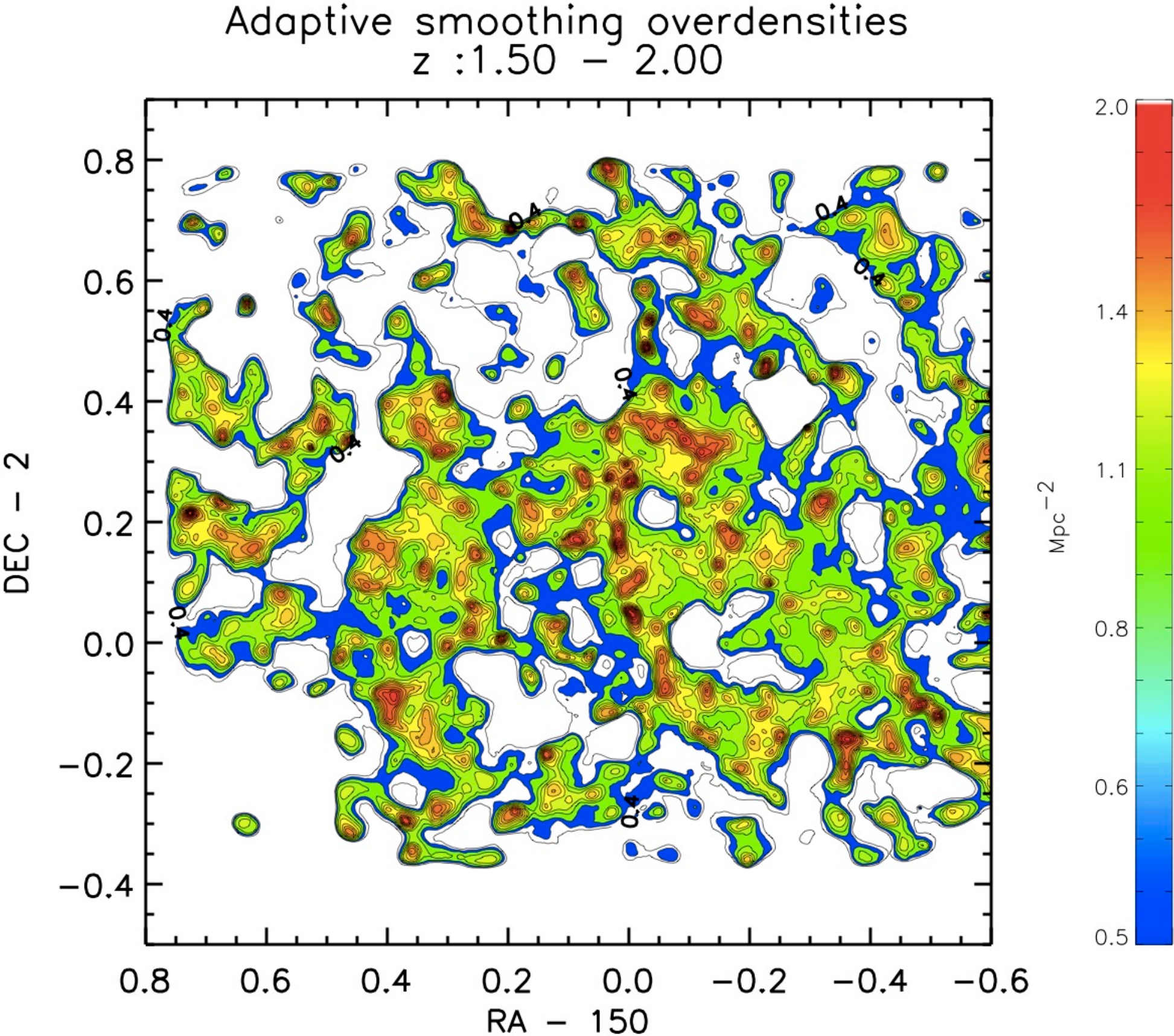}{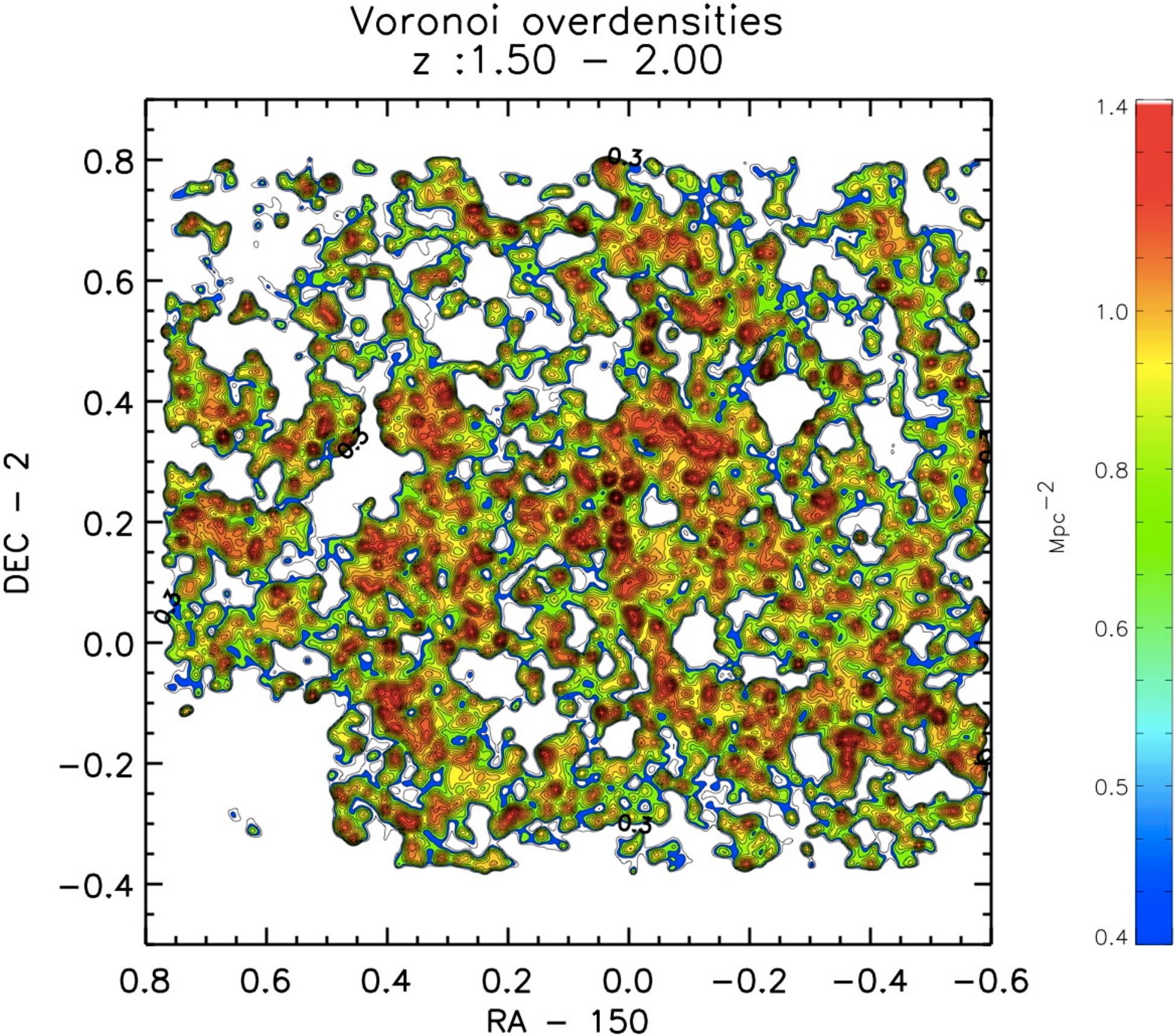}
%\plotone{lss_2d_p20-eps-converted-to.pdf}
\caption{ Overdense regions at z = 1.5 - 2.0 are shown for the adaptive smoothing (Top) and Voronoi (Bottom)
 techniques.
}
\end{figure}
\clearpage

\begin{figure}[ht]
\figurenum{\ref{lss_2d} e}
\epsscale{1.6}
%\plotfiddle{PSFILE}{VSIZE}{ROTANG}{HSCALE}{VSCALE}{HTRANS}{VTRANS}
%\vskip -9in
%\plotfiddle{lss_2d_p19-eps-converted-to.pdf}{9.in}{90.}{600.}{600.}{-100}{600}
\plottwo{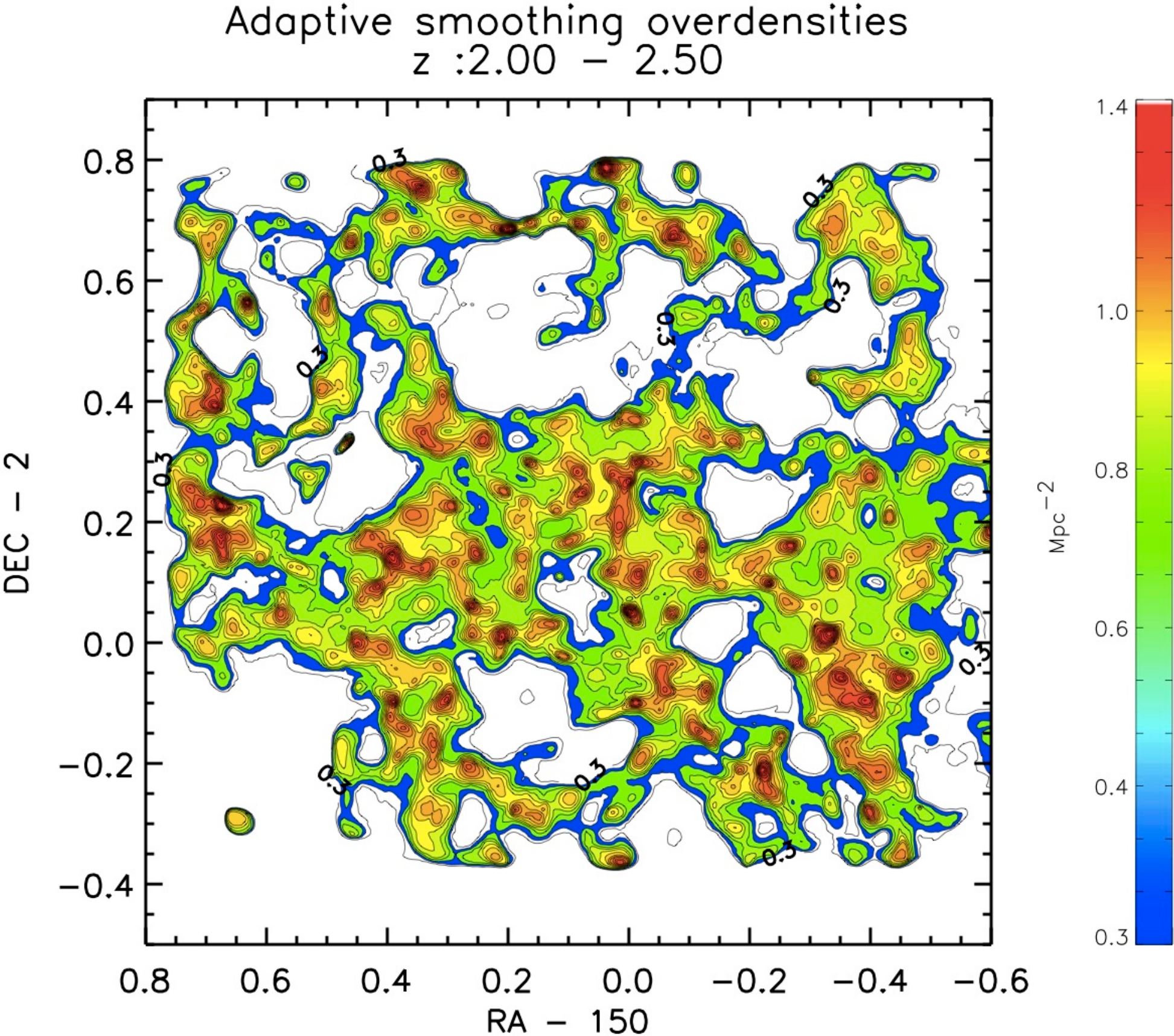}{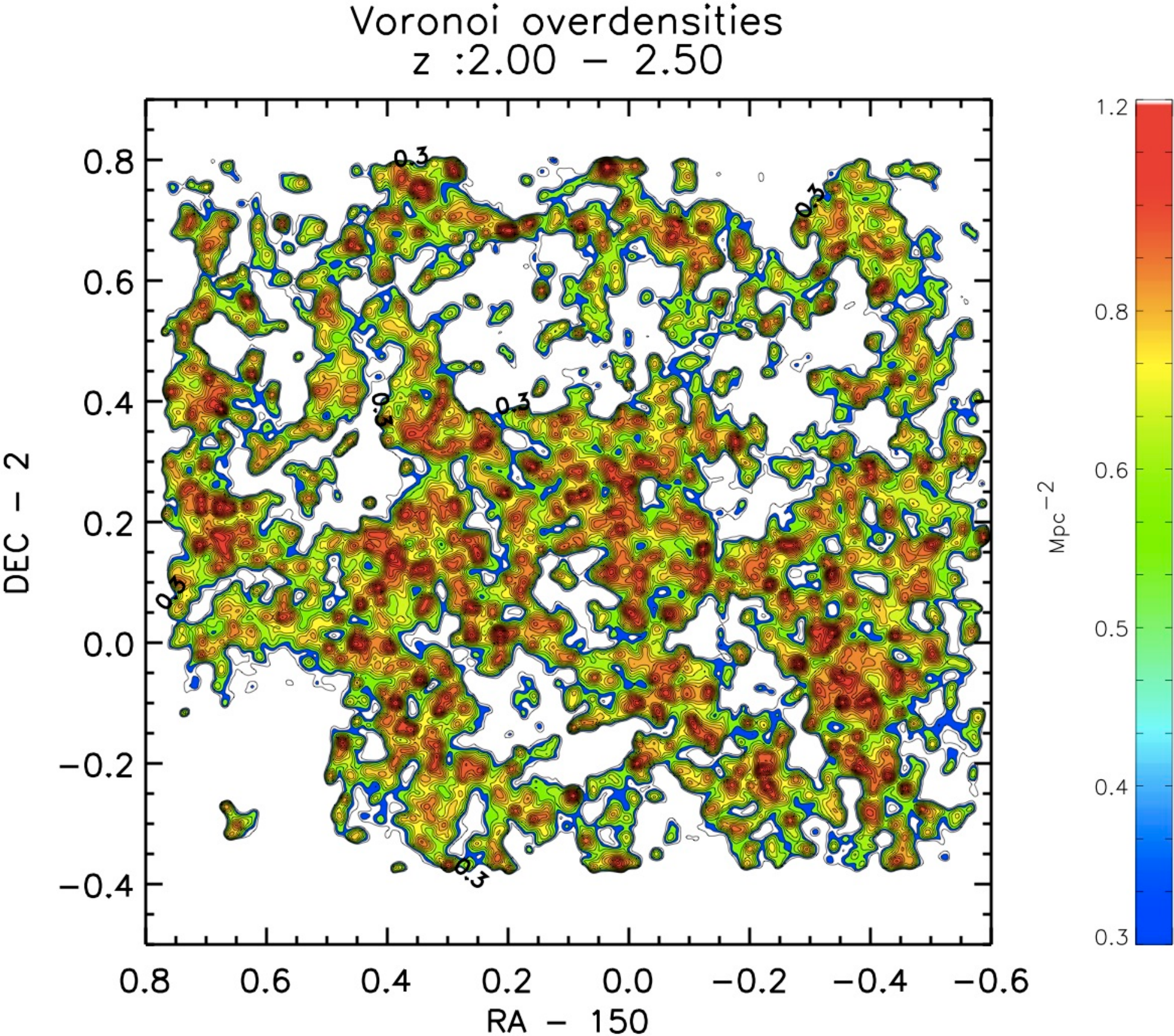}
%\plotone{lss_2d_p20-eps-converted-to.pdf}
\caption{ Overdense regions at z = 2.0 - 2.5 are shown for the adaptive smoothing (Top) and Voronoi (Bottom)
 techniques. 
}
\end{figure}
\clearpage

\begin{figure}[ht]
\figurenum{\ref{lss_2d} f}
\epsscale{1.6}
%\plotfiddle{PSFILE}{VSIZE}{ROTANG}{HSCALE}{VSCALE}{HTRANS}{VTRANS}
%\vskip -9in
%\plotfiddle{lss_2d_p19-eps-converted-to.pdf}{9.in}{90.}{600.}{600.}{-100}{600}
\plottwo{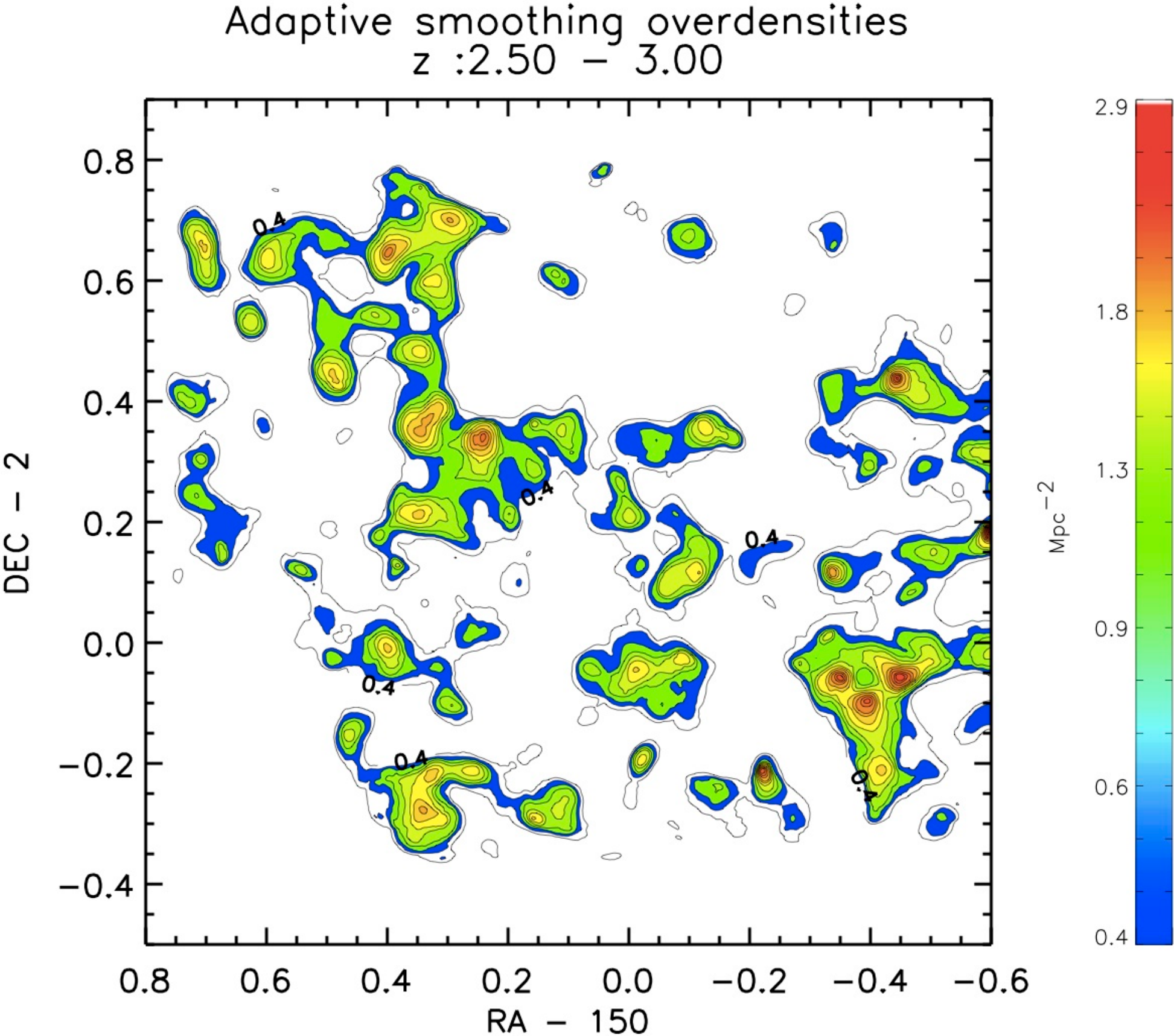}{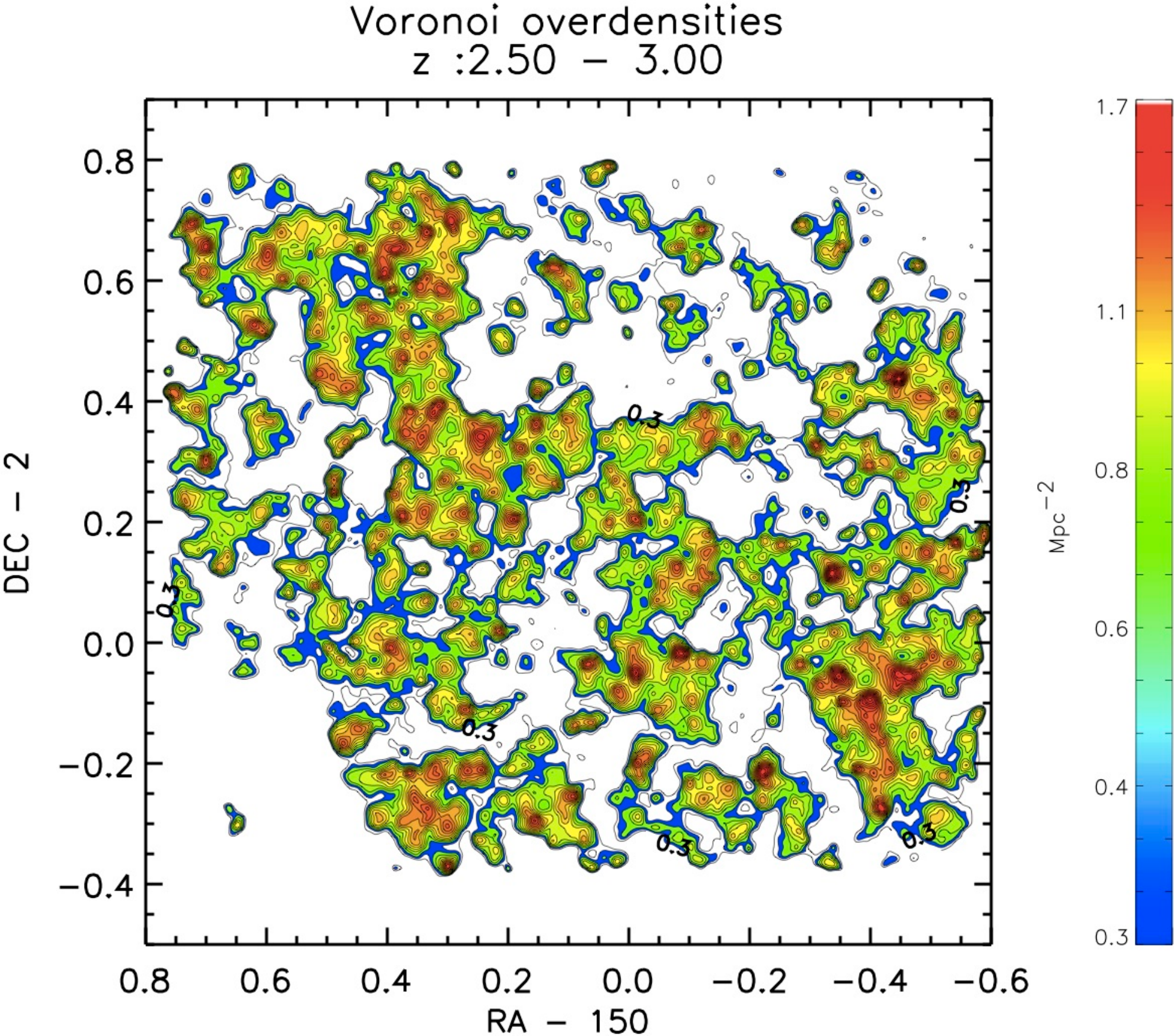}
%\plotone{lss_2d_p20-eps-converted-to.pdf}
\caption{ Overdense regions at z = 2.5 - 3.0 are shown for the adaptive smoothing (Top) and Voronoi (Bottom)
 techniques. 
}
\end{figure}
\clearpage

\begin{figure}[ht]
\figurenum{\ref{lss_2d} g}
\epsscale{1.6}
%\plotfiddle{PSFILE}{VSIZE}{ROTANG}{HSCALE}{VSCALE}{HTRANS}{VTRANS}
%\vskip -9in
%\plotfiddle{lss_2d_p19-eps-converted-to.pdf}{9.in}{90.}{600.}{600.}{-100}{600}
\plottwo{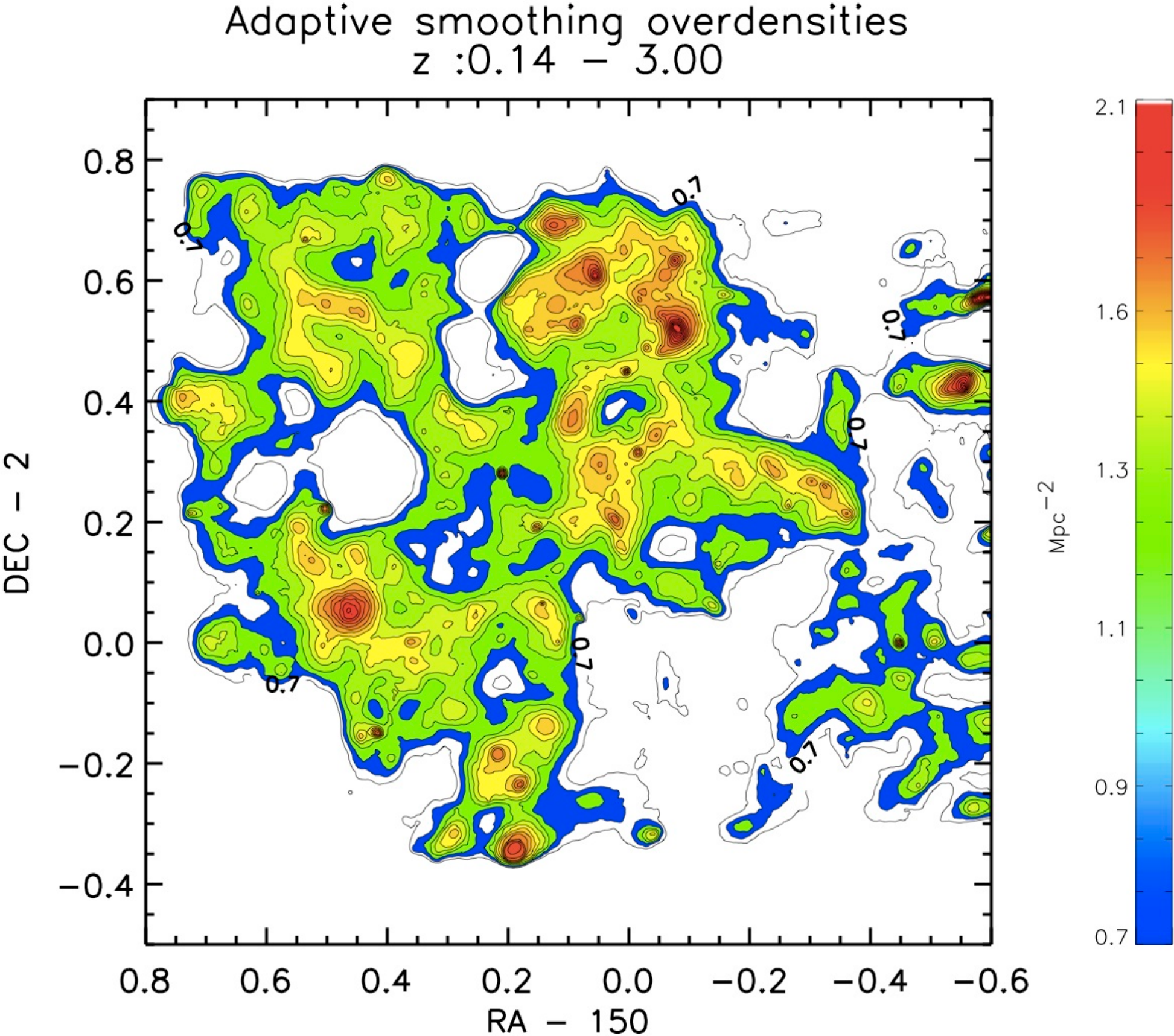}{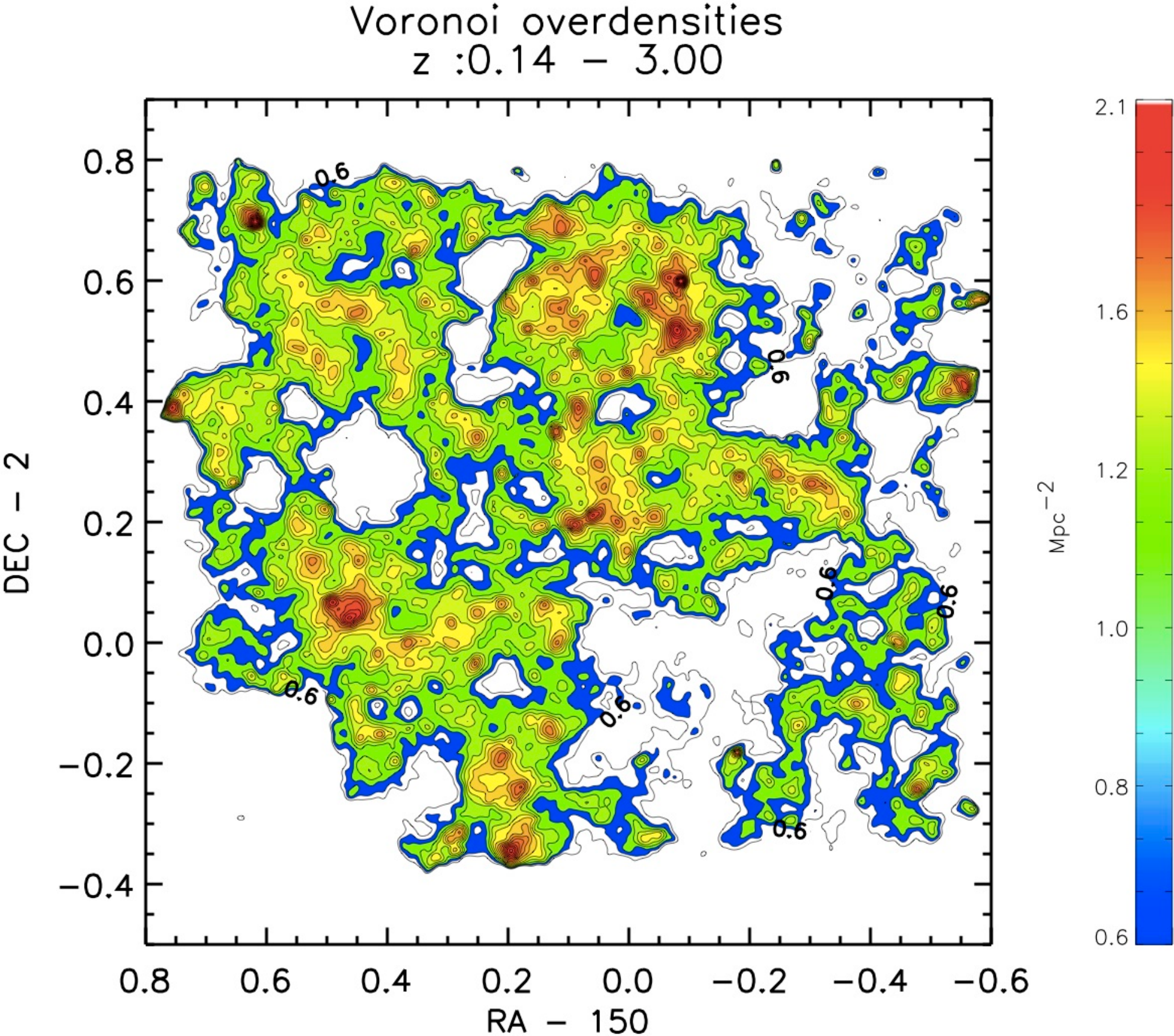}
%\plotone{lss_2d_p20-eps-converted-to.pdf}
\caption{ Overdense regions for the full redshift range at z = 0.15 - 3.0 are shown for the adaptive smoothing (Top) and Voronoi (Bottom)
 techniques.
}
\end{figure}
\clearpage

\section{Evolution of COSMOS LSS and Comparison with the Simulation}\label{comp_millennium}

\begin{figure}[ht]
\epsscale{0.5}
%\plottwo{rho1-eps-converted-to.pdf}{rho2-eps-converted-to.pdf}
%\caption{ The densities measured with the Voronoi technique are shown for  the sample of 150,852 COSMOS galaxies (Left) and 119,293 galaxies from the  WMAP3YC simulation (Right) from z = 0.14 to 3.0. 
%The simulation galaxies were selected using the same photometric selection function as for the COSMOS galaxies and given the same redshift uncertainties as the photoz for COSMOS (see Fig. \ref{sigz}). Excellent correspondence is seen between the COSMOS and simulation LSS environments --  the overall range of densities and their variation with redshift is remarkably similar between the two samples (see Fig. \ref{counts}). 
%At low z, these densities extend over 3 orders of magnitude, typically from 0.1 to over 100 Mpc$^{-2}$ and at the highest z approximately 2 orders of magnitude.  The contours, showing the relative numbers of galaxies at each density and redshift, decrease by a factor of 2 for a full
%range  of 1/4096 at the outermost contour level. At high redshift, the decrease in the number of galaxies at high densities is due to both cosmic evolution of the density field and the decreasing numbers of galaxies at high z due to the rising mass limits for galaxies (see Fig. \ref{sample}). 
\plotone{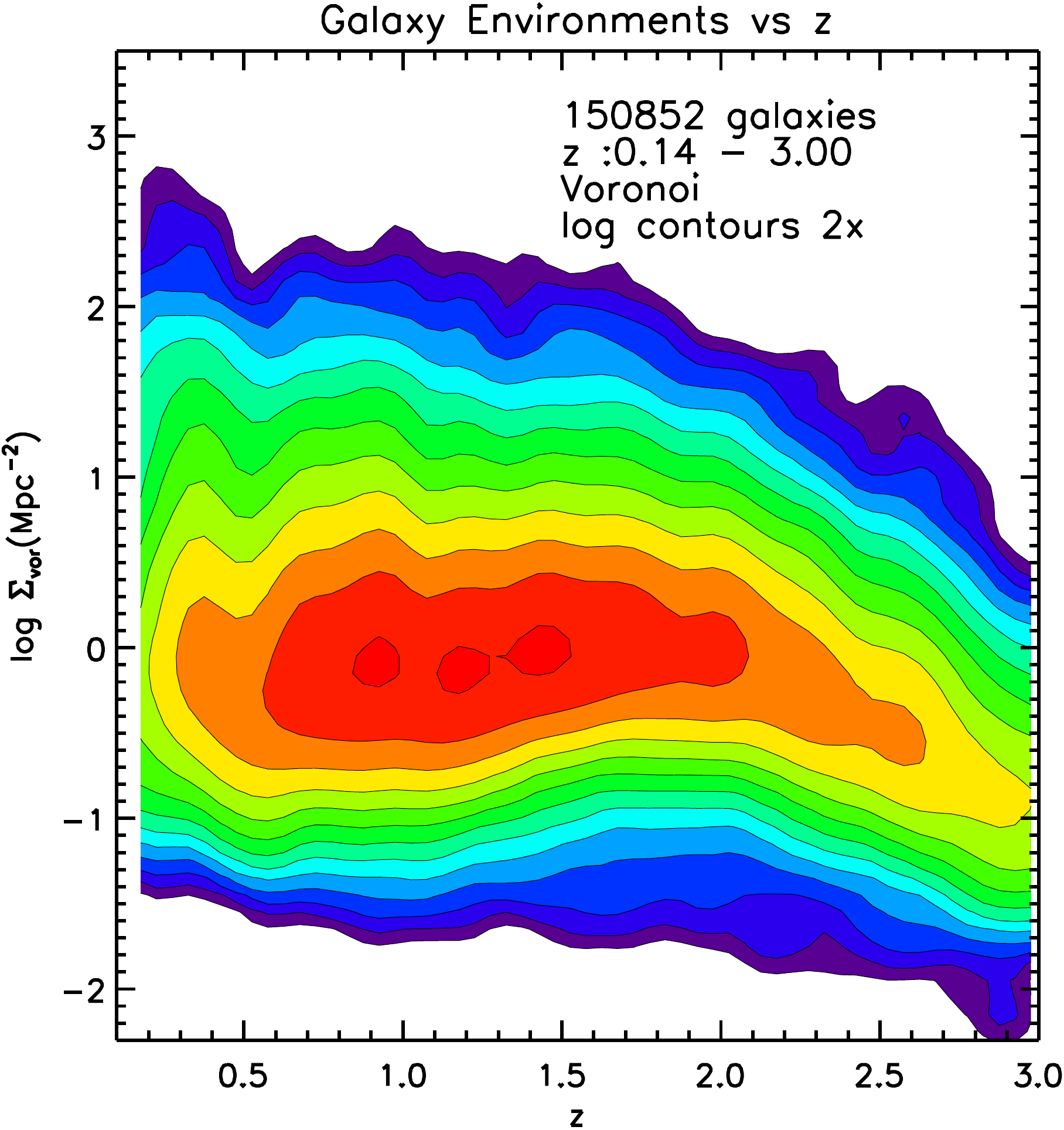}
\caption{ The densities measured with the Voronoi technique are shown for  the sample of 150,852 COSMOS galaxies from z = 0.14 to 3.0. 
At low z, these densities extend over 3 orders of magnitude, typically from 0.1 to over 100 Mpc$^{-2}$ and at the highest z approximately 2 orders of magnitude.  The contours, showing the relative numbers of galaxies at each density and redshift, decrease by a factor of 2 for a full
range  of 1/4096 at the outermost contour level. At high redshift, the decrease in the number of galaxies at high densities is due to cosmic evolution of the density field (see Fig. \ref{sample}). 
}\label{rho_vs_z_cos}
\end{figure}

Figure \ref{rho_vs_z_cos} shows the range of environments for the 
COSMOS sample as a function of redshift. The contours indicate  
relative numbers of galaxies as a function of environmental density and redshift. 
Overall, we find excellent correspondence between the COSMOS sample and that from 
the simulation -- both in the relative number of galaxies 
at different environmental densities and the variation of the structure densities with redshift (see below). 

\begin{figure}[ht]
\epsscale{0.5}
\plotone{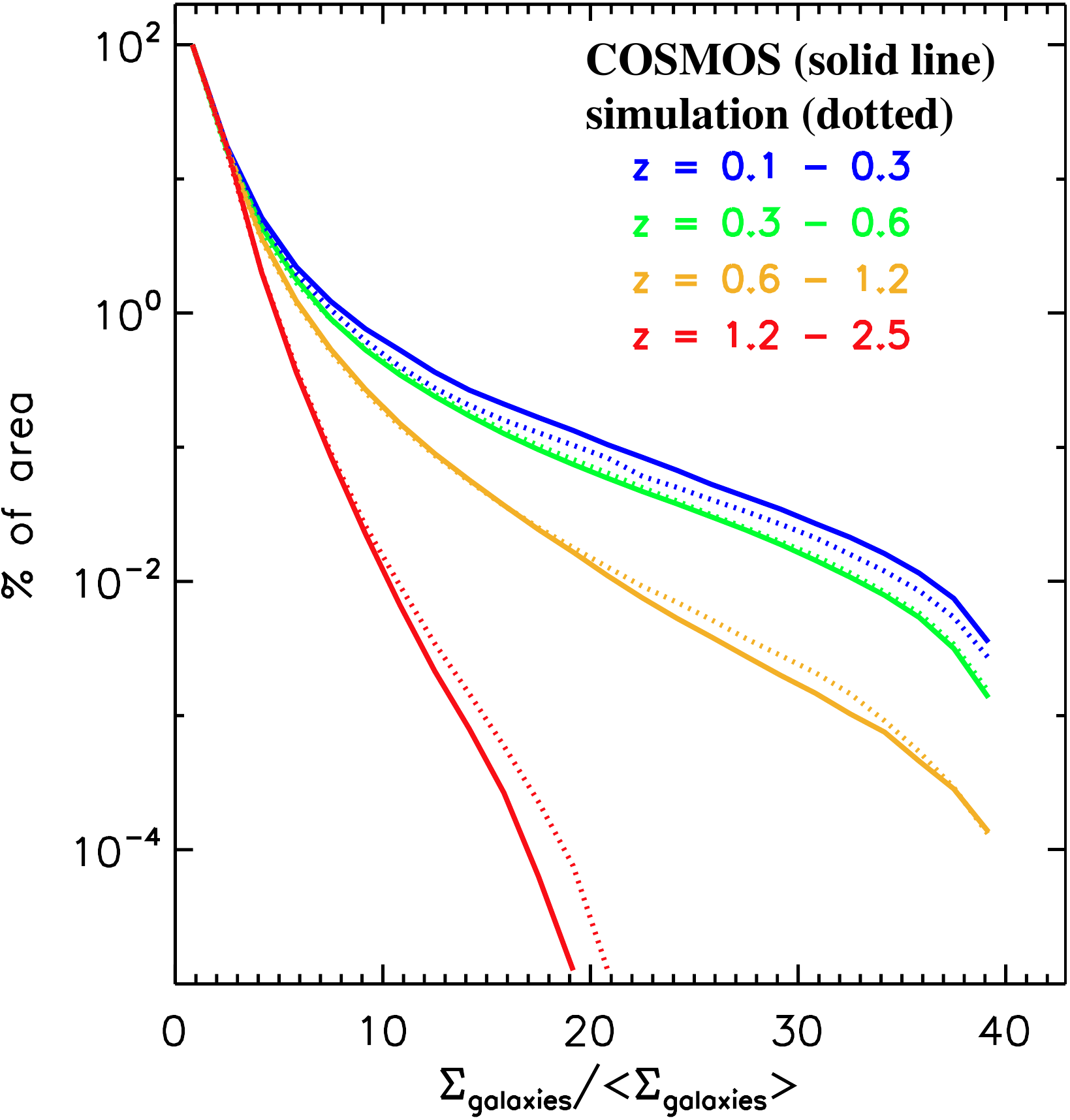}
%\includegraphics[scale=0.5,angle=-90]{fract_area_vor_dens-eps-converted-to.pdf}
%\plotfiddle{fract_area_vor_dens-eps-converted-to.pdf}{0. in}{-90}{4.in}{4.in}{0.in}{0.in}
\caption{The percentage of area on the sky occupied as a function of environmental overdensity for selected redshift ranges is compared between the COSMOS data and WMAP3YC simulation for areas with significant overdensity. The simulation very accurately reproduces the relative amounts of structure as function of both environmental density and the evolution of this structure.}\label{area_filling}
\end{figure}

The LSS seen in COSMOS and those in the $\Lambda$CDM simulation can be compared by measuring the fractional area occupied by environments of varying overdensities. For the simulation, 
the redshifts were given the same dispersion as the COSMOS photoz (see Fig. \ref{sigz}) and the structures were measured using the same techniques as discussed in \S \ref{millennium}
In Fig. \ref{area_filling}, this area filling percentage is shown as a function of overdensity for 4 redshift ranges. This figure clearly illustrates the increasing range of overdensities seen at low redshift compared to higher redshifts. The figure also shows extremely good correspondence in the area filling fractions and their evolution with redshift between COSMOS and the simulation.  This area filling percentage is analogous to a spatial power spectrum, but perhaps more easily visualized.  The relative frequency of a given overdensity at each redshift is more directly apparent than would be the case for a power spectrum.

\section{Correlation of Galaxy Properties with Environment}\label{evolution}

A major motivation of this study is the exploration of the environmental influence on galaxy properties --
their SED types, star formation rates (SFR) and stellar masses. Given the well-known correlation of early type massive galaxies with dense/cluster environments at low redshifts, we can now investigate 
at which redshifts these influences develop, and explore in more detail the dependence on environmental density, using the enormous galaxy samples in COSMOS. And since similar processing has been employed on the simulation, we can compare in detail the observations with 
the semi-analytic model predictions. 

\subsection{Galaxy Colors and SED Types}\label{colors}

\begin{figure}[ht]

\epsscale{1.}
\plottwo{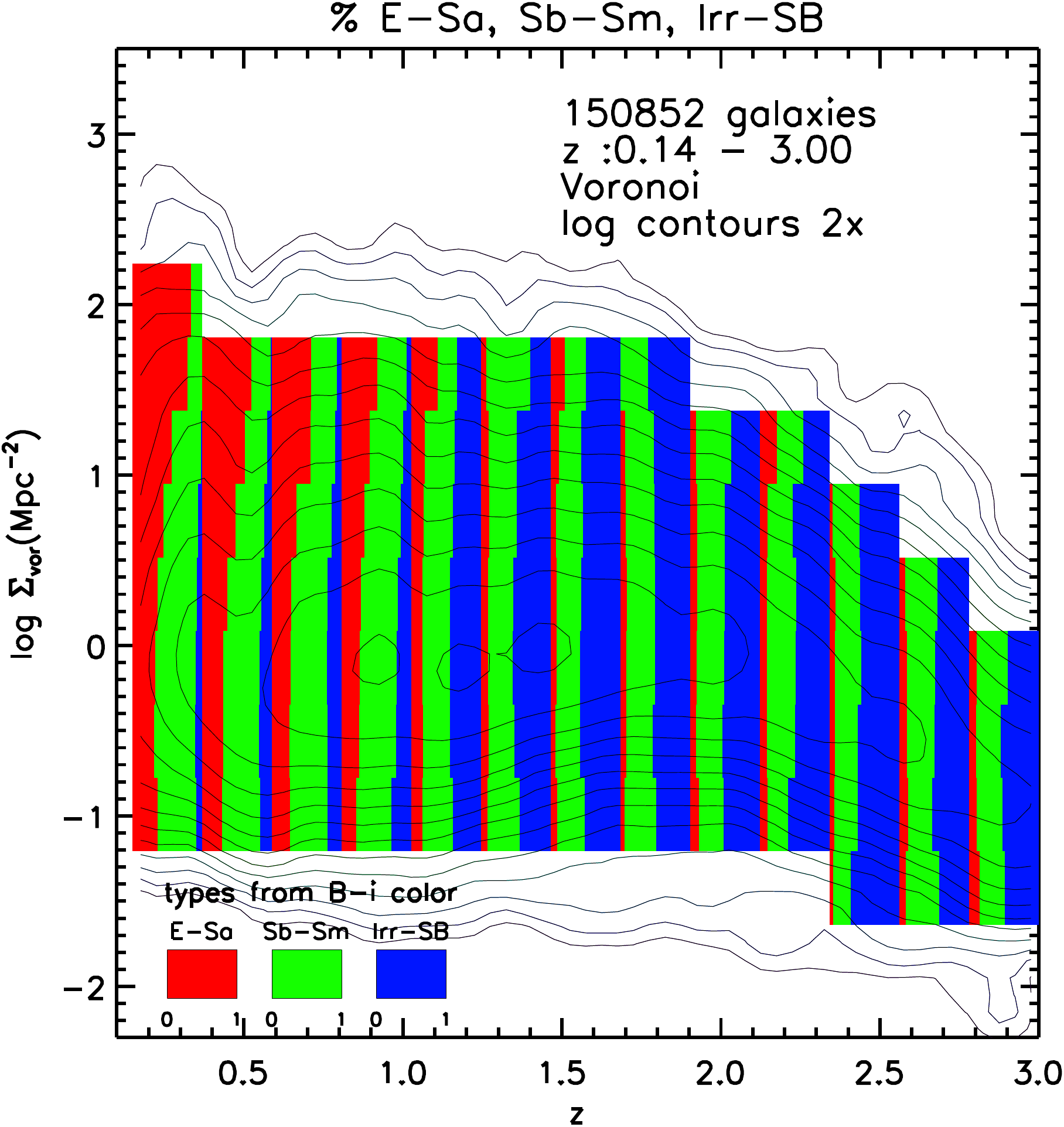}{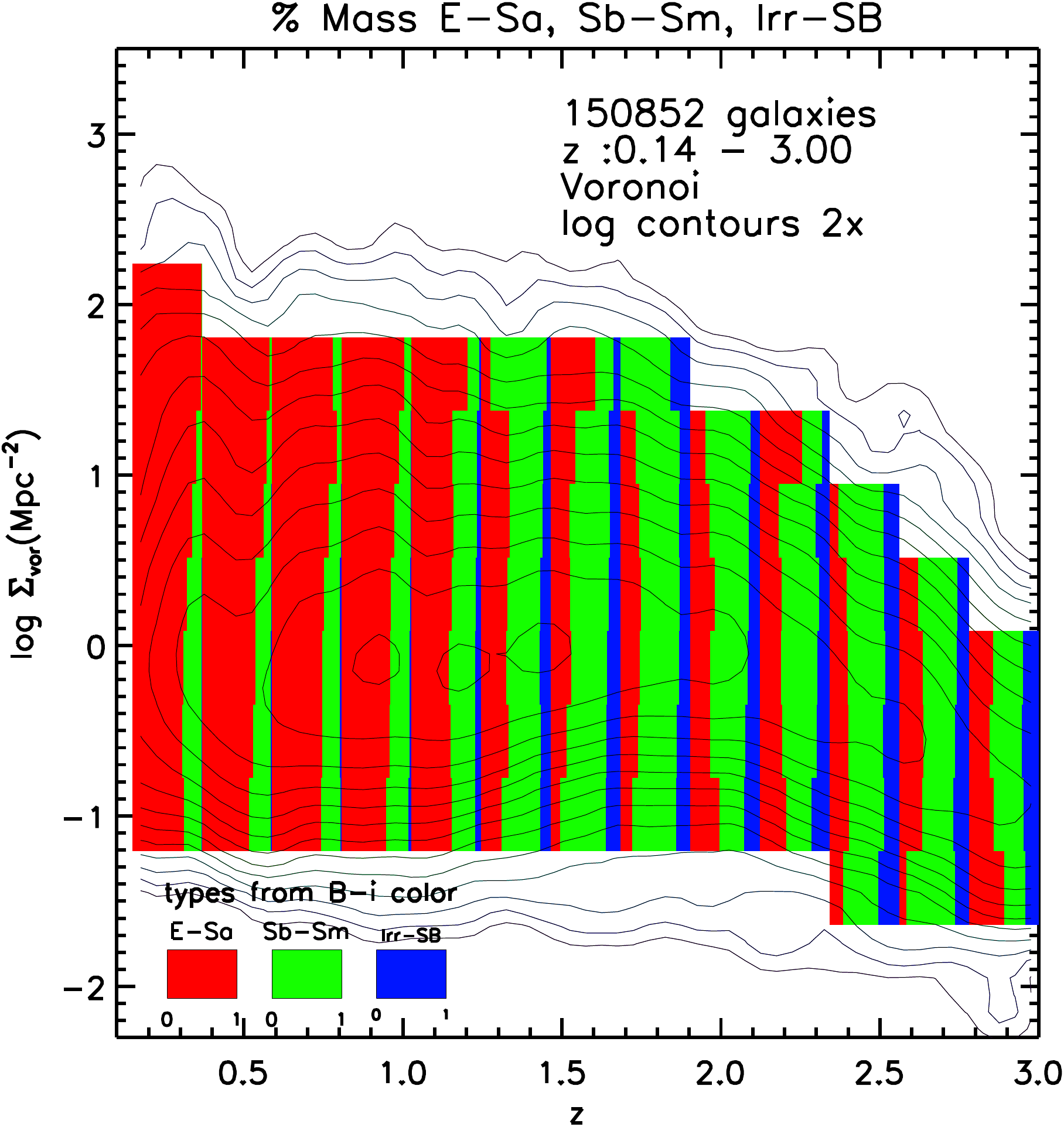}
\caption{Galaxy percentages in COSMOS (classified as early-type (E-Sa), Spiral (Sab-Sd) and IRR/SB on the basis of their 
rest-frame reddened b-i colors) are shown as a function of environmental density and redshift. The left panel shows the percentage by number and the right panel weights each galaxy by its mass.
}\label{color_z}
\end{figure}

In Figure \ref{color_z} the correlations of galaxy SED type (\S \ref{sfr_section}) with density and redshift are 
shown. For each redshift-density cell, the color fractions are proportional to the fraction of each galaxy type. In the left panel, 
the galaxy number fraction is shown and in the right panel each galaxy is weighted by its mass. As noted earlier,  the correspondence between the 
rest-frame b-i color and the galaxy type is taken from \cite{arn07} and the stellar 
mass from the COSMOS photoz catalog was estimated using a color dependent mass-to-light
ratio \citep[see][]{ilb09}.

\begin{figure}[ht]
\epsscale{0.6}
\plotone{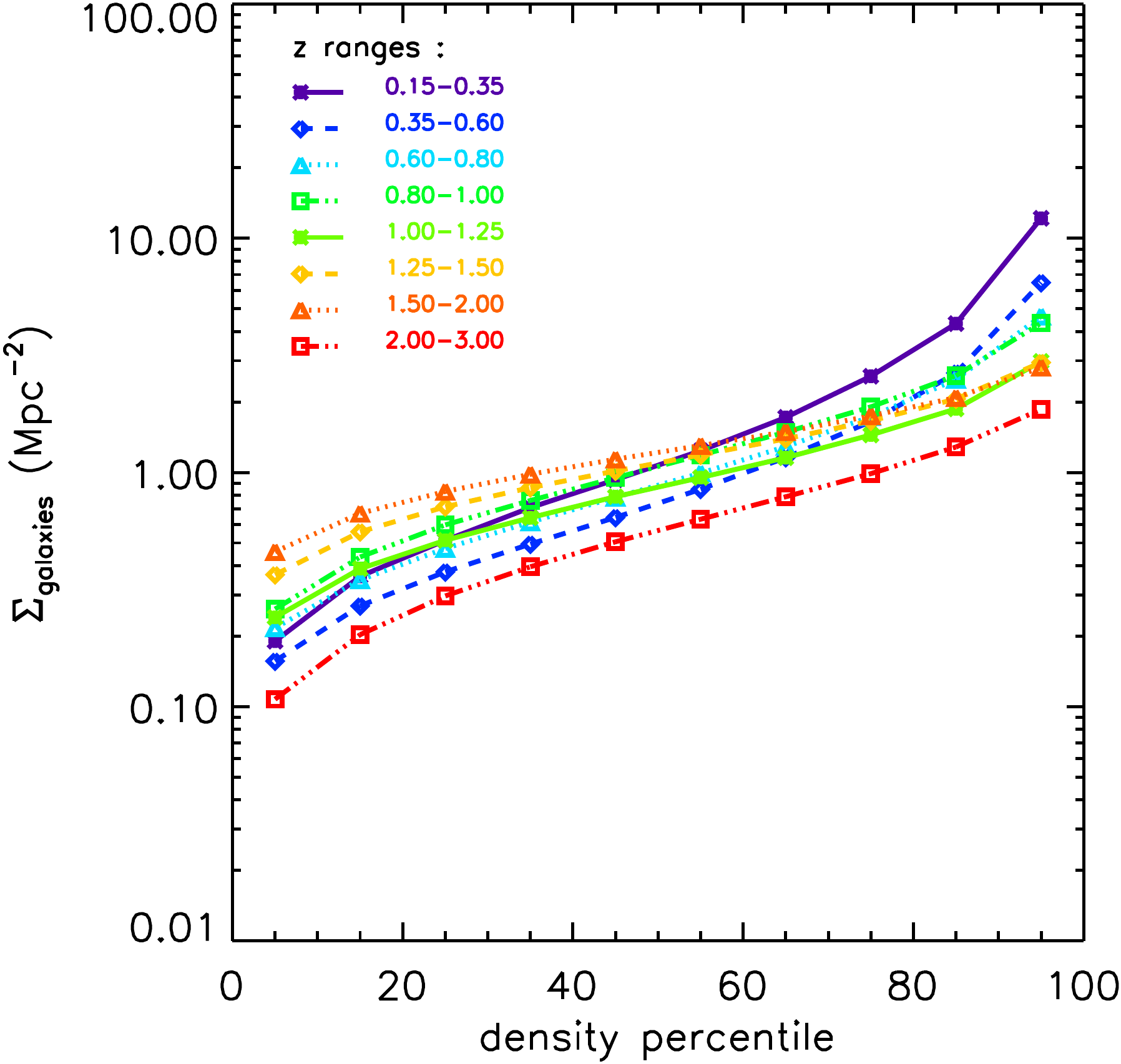}
\caption{The correspondence between the density percentiles (used in Fig. \ref{early_z} - \ref{sfrd}) and the absolute surface density per comoving Mpc$^2$ is shown for 7 redshift ranges in COSMOS.
}\label{density_per}
\end{figure}

\begin{figure}[ht]
\epsscale{1.}
%\plottwo{per_g_p_d_vor_raw_ov_p17-eps-converted-to.pdf}{per_g_p_d_vor_raw_ov_mil_p17-eps-converted-to.pdf}
\plotone{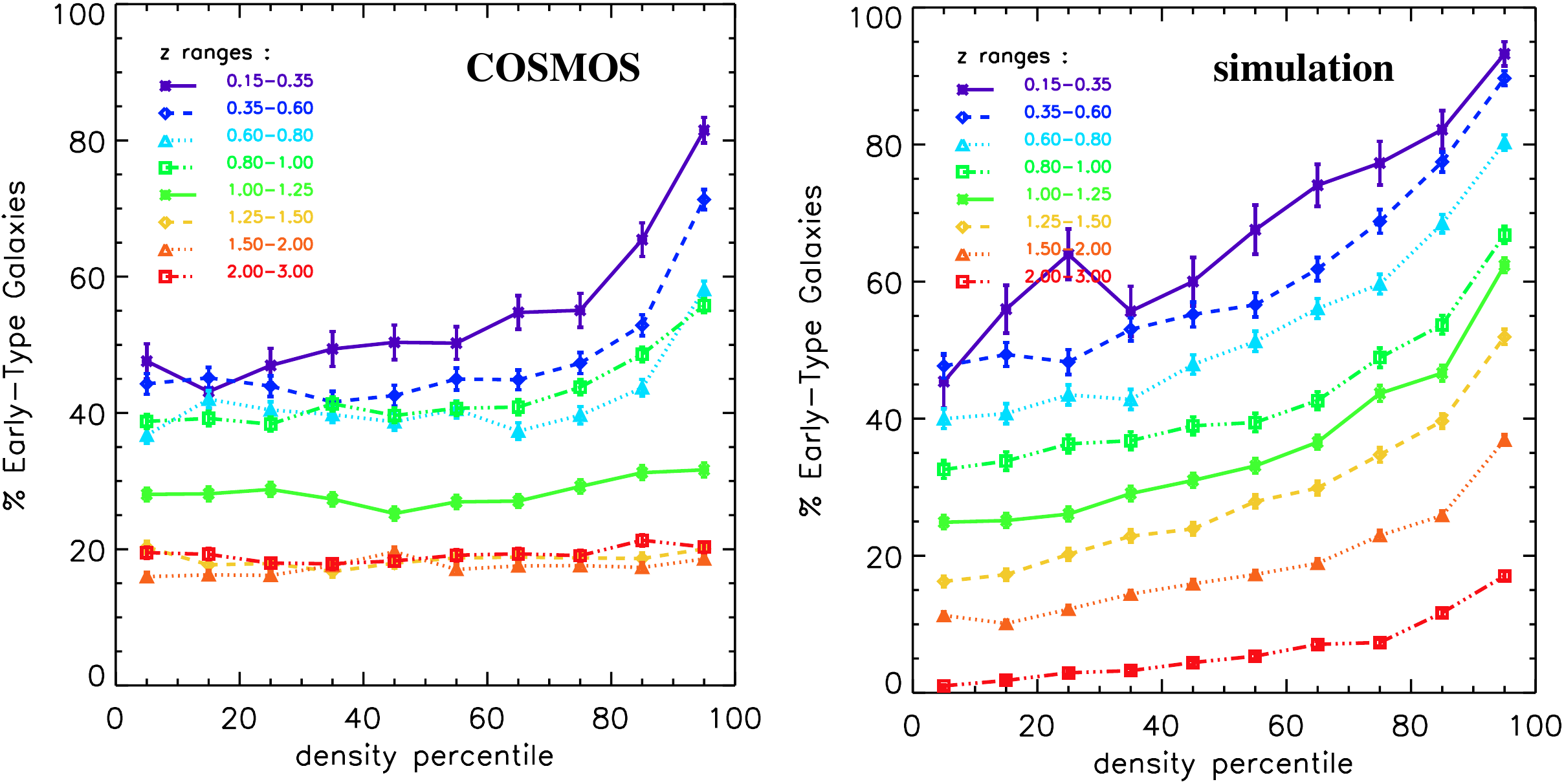}
\caption{The percentage of Early type galaxies (E-Sa) as a function of environmental 
density percentiles is shown for 8 redshift ranges for COSMOS galaxies (Left) and simulation (Right). At each redshift, the distribution of environmental densities is calculated and each galaxy's percentile within that distribution is determined. This effectively 
normalizes out the lower range of environmental densities at high redshift and the redshift dependence of the mean environmental density. The error bars show the dispersion in the median estimates for the sample 
in each bin; when the error bars are not shown, they are smaller than the symbol. The uncertainties shown in these and subsequent figures are calculated 
by bootstrap resampling. The translation between density percentile and actual surface density of galaxies is shown in Fig. \ref{density_per}.
}\label{early_z}
\end{figure}

Numerous studies have shown a strong dependence of the red galaxy fraction on environmental density at 
low redshift \citep[e.g. at $z < 0.1$][]{bal06}. Figures \ref{color_z} and \ref{early_z} clearly show a strong preference for the early type galaxies to 
inhabit the denser environments out to $z \sim 1.2$ although their total percentage decreases systematically with increasing z. Beyond $z \sim 1.2$, the early type galaxies are much less numerous and the strong environmental correlation disappears. \cite{iov10} analyzed the blue galaxy fraction in galaxy groups defined from the zCOSMOS spectroscopic sample and found a strongly increasing blue fraction at higher redshifts. 

To more clearly show the correlations with relative density as a function of redshift, we classify each galaxy by 
where it falls within the distribution of LSS densities at its redshift. At each redshift, the distribution of environmental densities is calculated and each galaxy's percentile within that distribution is determined. This effectively 
normalizes out the lower range of environmental densities at high redshift, and the redshift dependence of the mean environmental density. The translation between 
these density percentiles and absolute surface density is shown in Fig. \ref{density_per}.

Figure \ref{early_z} shows the variation in the percentage of early type galaxies (with SED corresponding to 
E-Sa galaxies) with density percentiles for 8 redshift ranges. This plot clearly shows the steep increase in the fraction of early type galaxies at z $< 1.2$ and 
the development of strong environmental dependence at the same time, starting at $z \sim 1.2$ in the observed galaxies. For the simulation galaxies, the environmental dependence
for the early type galaxies persists all the way out to z = 3, albeit with reduced strength (Fig. \ref{early_z} - right). The flattening of the density dependence in the early type fraction 
at the highest redshifts is likely due in part to the reduced dynamic range of environmental 
densities at high z (see Fig. \ref{density_per}) and the fact that at early epochs the evolution is driven by environment on smaller scales. Another notable difference between the 
COSMOS and simulation samples is the overall lower fraction of early type galaxies in the simulation at z $> 1.5$. In summary, the most notable difference between the simulation and the COSMOS galaxies is
that the simulation shows higher percentages of early type galaxies in the dense environments and smoother and 
more regular variations -- probably an expected  result of the strictly prescriptive semi-analytics.

In the following,
we refer to this transition in the density dependence for the observed galaxies as the 'Emergence of the Red Sequence'.
This is not to imply that red sequence galaxies do not exist at higher redshift, simply that they 
do not exhibit the clear density dependence seen at $z < 1.2$. 
The span of cosmic age over which this emergence takes place is only $\sim1$ Gyr. It is important to emphasizes that \emph {the simulation, which was subjected to the same redshift 
uncertainties, photometric selection and LSS mapping techniques, did in fact show environmental dependence all the way to z = 3, so the emergence of the environmental
dependence in the observed galaxies only at $z \sim 1.4$ is not due to any selection or measurement effect}.

\subsection{Star Formation Activity}\label{sfr_dis} 

\begin{figure}[ht]
\epsscale{1.}
\plottwo{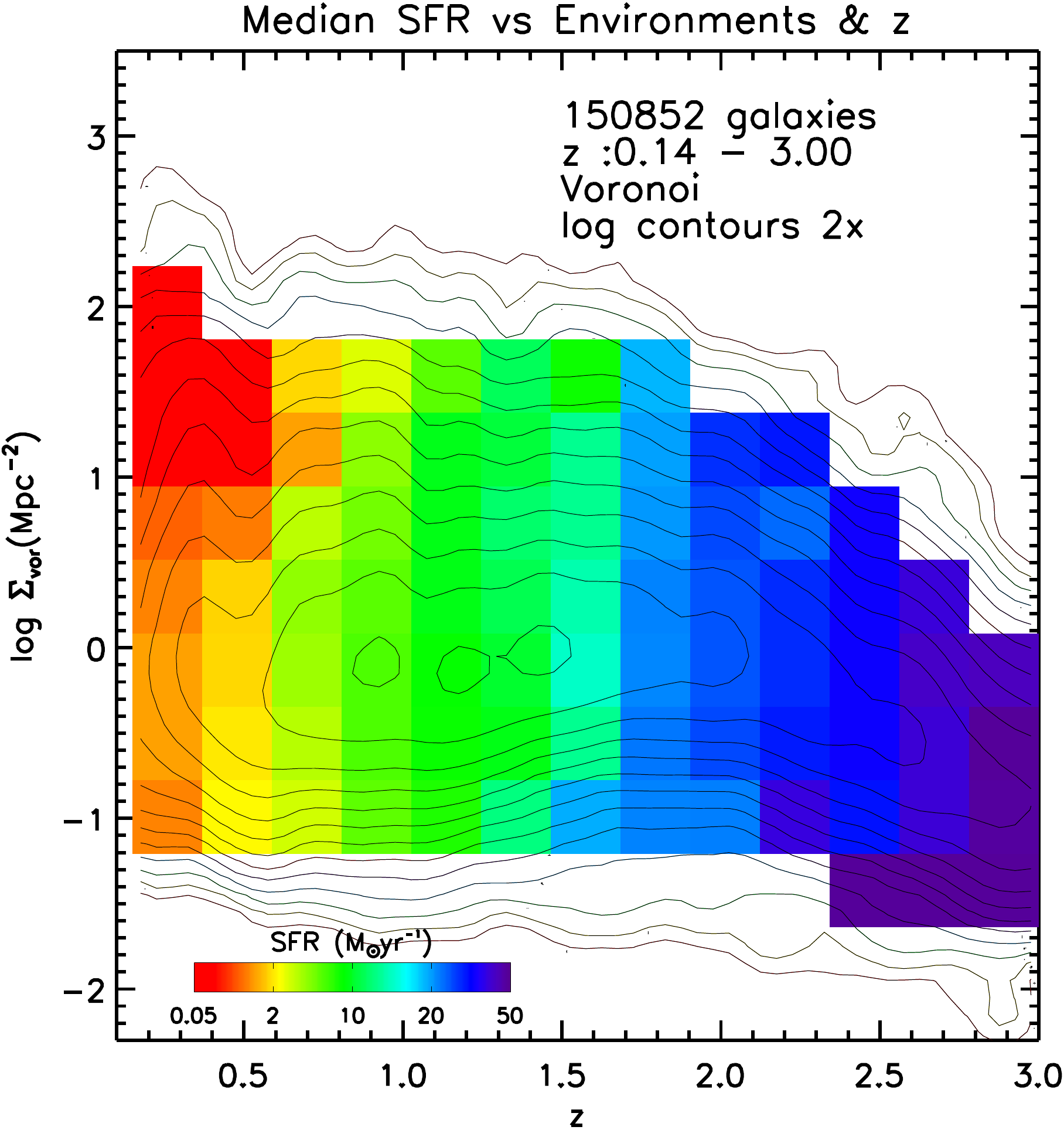}{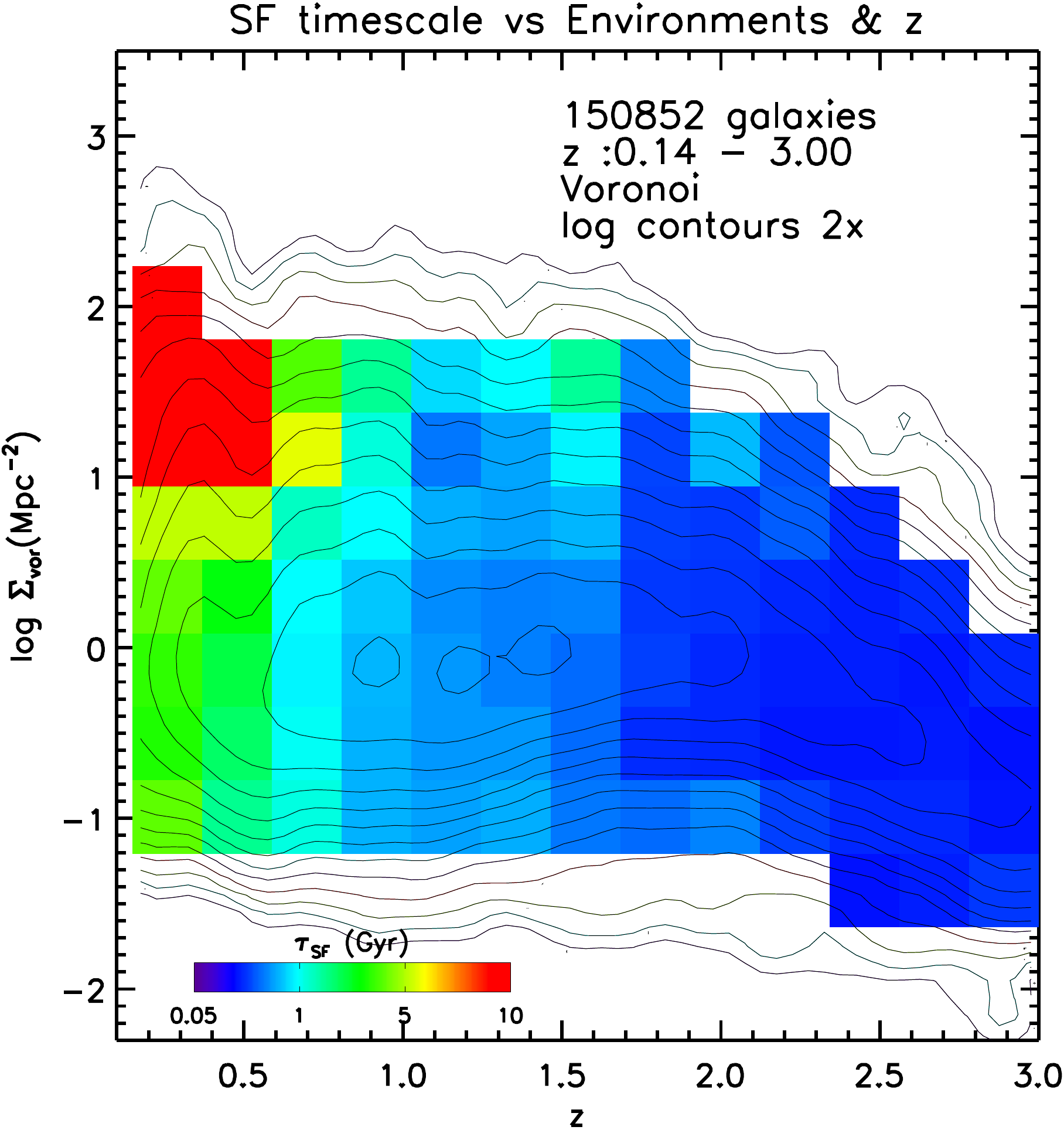}
\caption{The median SFR (Left) and star formation timescale ($\tau_{SF} = M_*/SFR$) (Right) for COSMOS galaxies are shown as a function of environmental density and redshift. }\label{sfr_density_z}
\end{figure}

The SFRs for each of the galaxies were estimated from the rest-frame NUV continuum of their 
SEDs and corrected for extinction, combined with SFR estimates from Spitzer 24$\mu$m data as described in \S \ref{sfr_section}. 
In Fig. \ref{sfr_density_z} the median SFRs and star formation timescales  ($\tau_{SF} = M_*/SFR$) are shown as a function of redshift and environmental density.
For both quantities, extremely strong environmental dependence is seen at low redshift with a factor of 10 change in both the SFR and timescale between 
the average at low z and that seen in the densest environments. As with the early type galaxy fraction, the environmental segregation falls off and disappears above z $\sim$1. In Fig. \ref{sfr_density_z}  
it can be seen that the median SFRs perhaps show some a very mild environmental dependence out to $z \sim 3$ but it certainly not as significant as the correlation at low z. 
At $z < 0.1$ using SDSS, \cite{kau04} found a strong dependence of the specific star formation rate ($sSFR = SFR/M_* = 1/\tau_{SF}$) with environment --
a factor 10 decrease in the sSFR going from low to high density environments.

\begin{figure}[ht]
\epsscale{1.}
%\plottwo{per_g_p_d_vor_raw_ov_p20-eps-converted-to.pdf}{per_g_p_d_vor_raw_ov_mil_p20-eps-converted-to.pdf}
\plotone{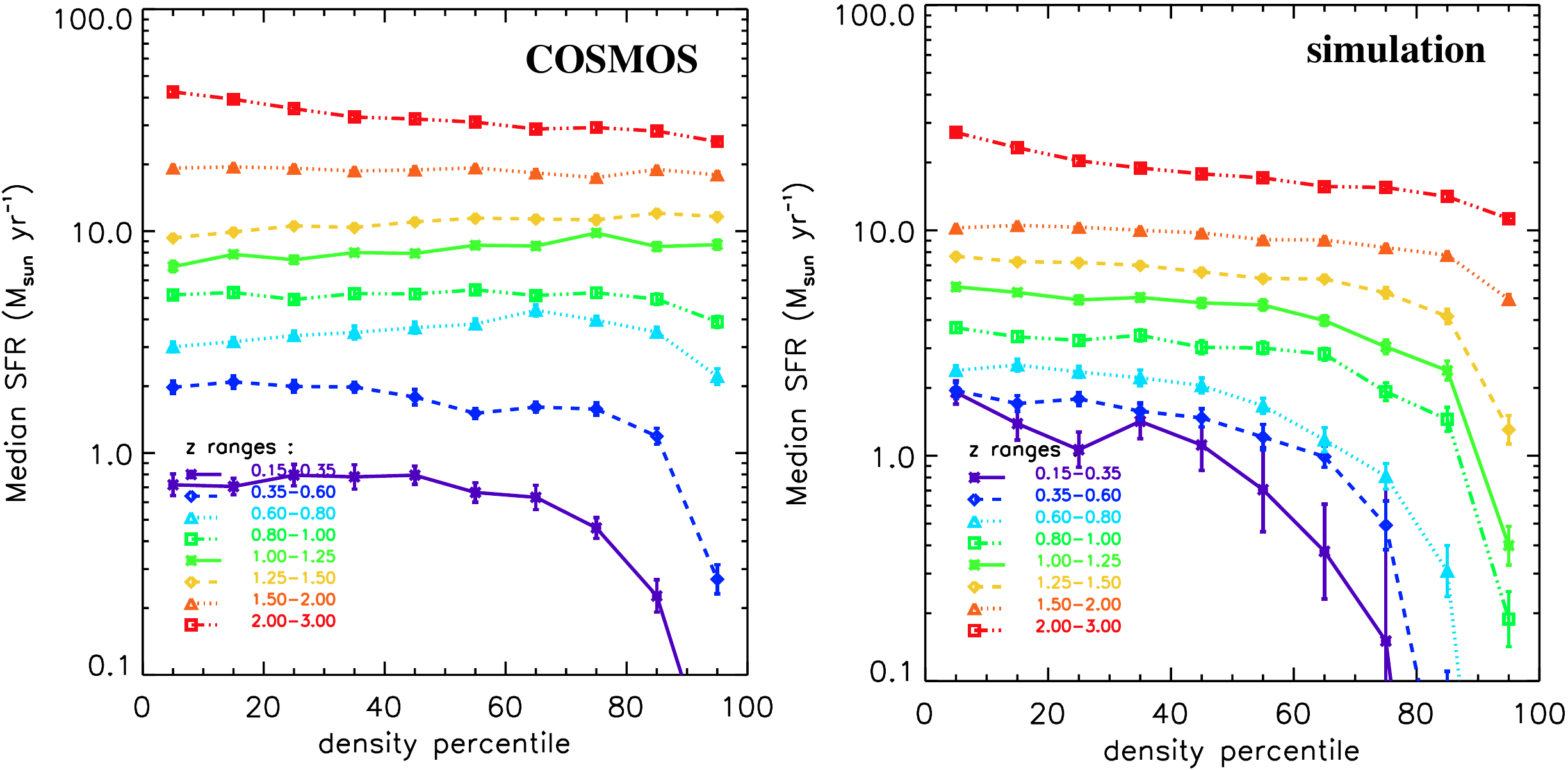}
\caption{The median SFR as a function of environmental 
density percentile for 8 redshift ranges for COSMOS galaxies (Left) and Millennium (Right).  The error bars show the dispersion in the median estimates for the sample 
in each bin; when the error bars are not shown, they are smaller than the symbol.
}\label{sfr_z}
\end{figure}

\begin{figure}[ht]
\epsscale{1.}
%\plottwo{per_g_p_d_vor_raw_ov_p21-eps-converted-to.pdf}{per_g_p_d_vor_raw_ov_mil_p21-eps-converted-to.pdf}
\plotone{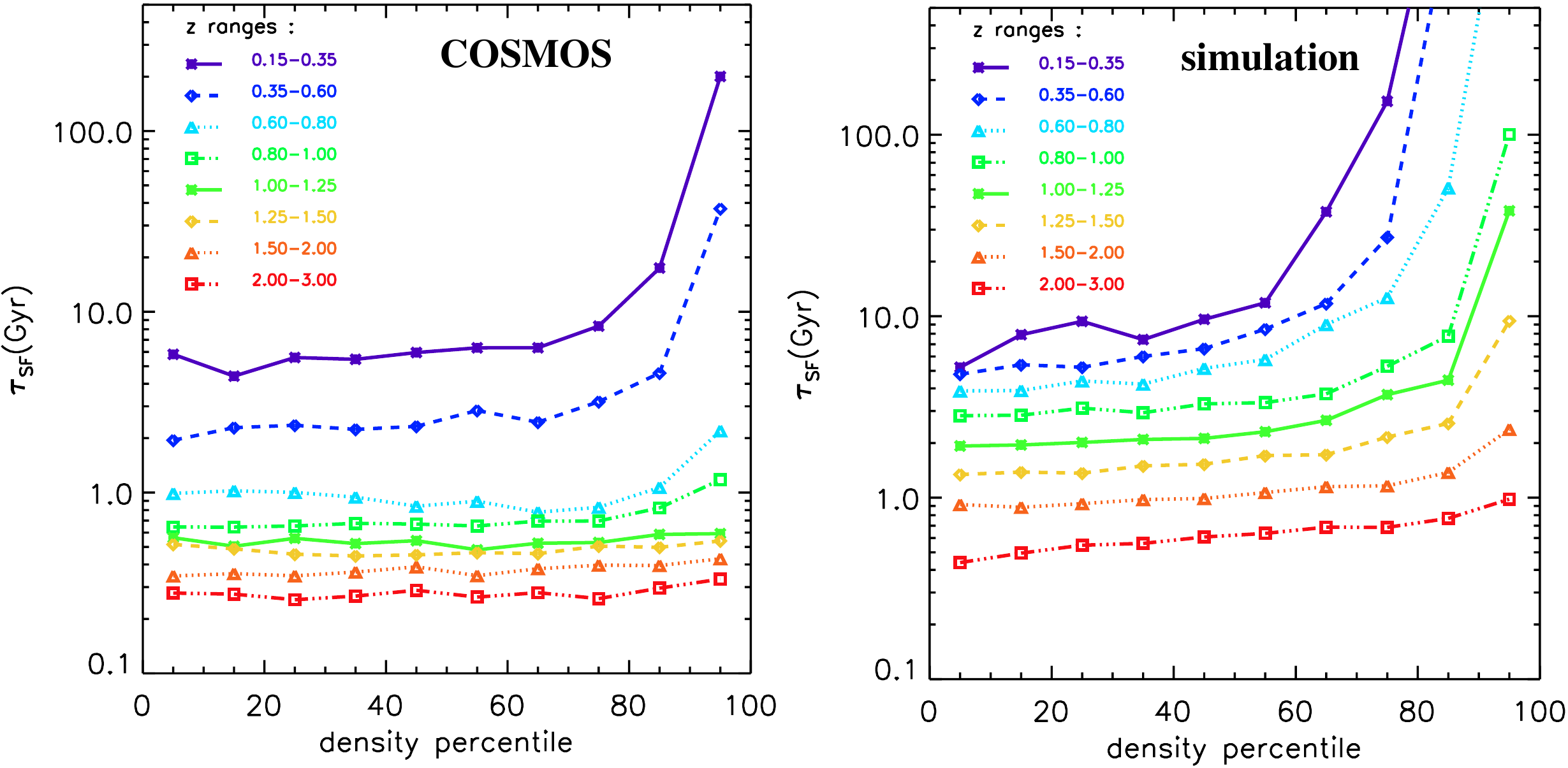}
\caption{The characteristic  SF timescale ($\tau_{SF} = M_*/SFR$) is shown for 8 redshift ranges for COSMOS galaxies (Left) and Millennium (Right).  The error bars show the dispersion in the median estimates for the sample 
in each bin; when the error bars are not shown, they are smaller than the symbol.
}\label{tau_sfr}
\end{figure}

In Figure \ref{sfr_z}, the median SFRs are shown for the COSMOS and simulation galaxies 
as a function of environmental density and redshift.  \emph{The observed galaxies exhibit a significantly stronger increase in SFRs with redshift 
than those in the simulation, but somewhat weaker environmental dependence at the lowest redshifts.} The COSMOS SFRs increase by a factor of $\sim40$
from z =0.1 to 2.5 while the galaxies in WMAP3YC show median SFRs up by a factor of $\sim25$ over the same range. 
Both the observed galaxies and those in the simulation also exhibit strong environmental dependence
out to z $\sim 1.0$ and 1.2 respectively. Figure \ref{tau_sfr} shows the variation in the 
characteristic star formation timescale (i.e. the median $\tau_{SF}$) -- 
this star formation timescale by an order of magnitude increase from z $= 3.0$ to 0.15 in less dense environments, 
and two orders of magnitude decrease in the denser environments.

In recent work, there has been major divergence regarding the dependence of the SFR in galaxies 
on their environment at $z \sim 0.8 - 1$. In the local universe, several investigations find the mean SFR of galaxies in 
dense environments to be much less than those of galaxies in lower density regions \citep{gom03,bal04,kau04}. 
\cite{elb07} and \cite{coo08} have suggested a reversal at $z = 0.8 - 1$ of the SFR-density relation (i.e. higher SFRs at higher densities); however,  \citep{pat09} found no such reversal for a cluster and its environment 
at $z = 0.1274$. We see no evidence of the claimed reversal in the density dependence using our  
sample of galaxies which is larger by a factor of 10-100 than those in the above studies and with consistent density estimators for 
the entire redshift range. (The basis of the reversal noted by \cite{coo08} is hard to assess since the effect shown in their  Fig. 12d is not clearly evident 
in Fig. 12b which plots the observed points from which Fig. 12d is derived.)  As noted by \cite{pat09}, the 
reversal claimed by \cite{elb07} actually occurs only in a narrow range of density and not at the very highest density. Using zCOSMOS data, \cite{cuc10,bol10} also see no reversal. Using [OII] emitters at $z \sim 1.2$ detected in narrow band imaging in COSMOS, \cite{ide12} found that the average SFR of star-forming galaxies 
was independent of both stellar mass and environmental density, consistent with our results at this and higher redshifts.

\subsection{Environmental Dependence of the Star Formation Rate Density} \label{discussion}

\begin{figure}[ht]
\epsscale{1.}
%\plottwo{per_g_p_d_vor_raw_ov_p15-eps-converted-to.pdf}{per_g_p_d_vor_raw_ov_mil_p15-eps-converted-to.pdf}
\plotone{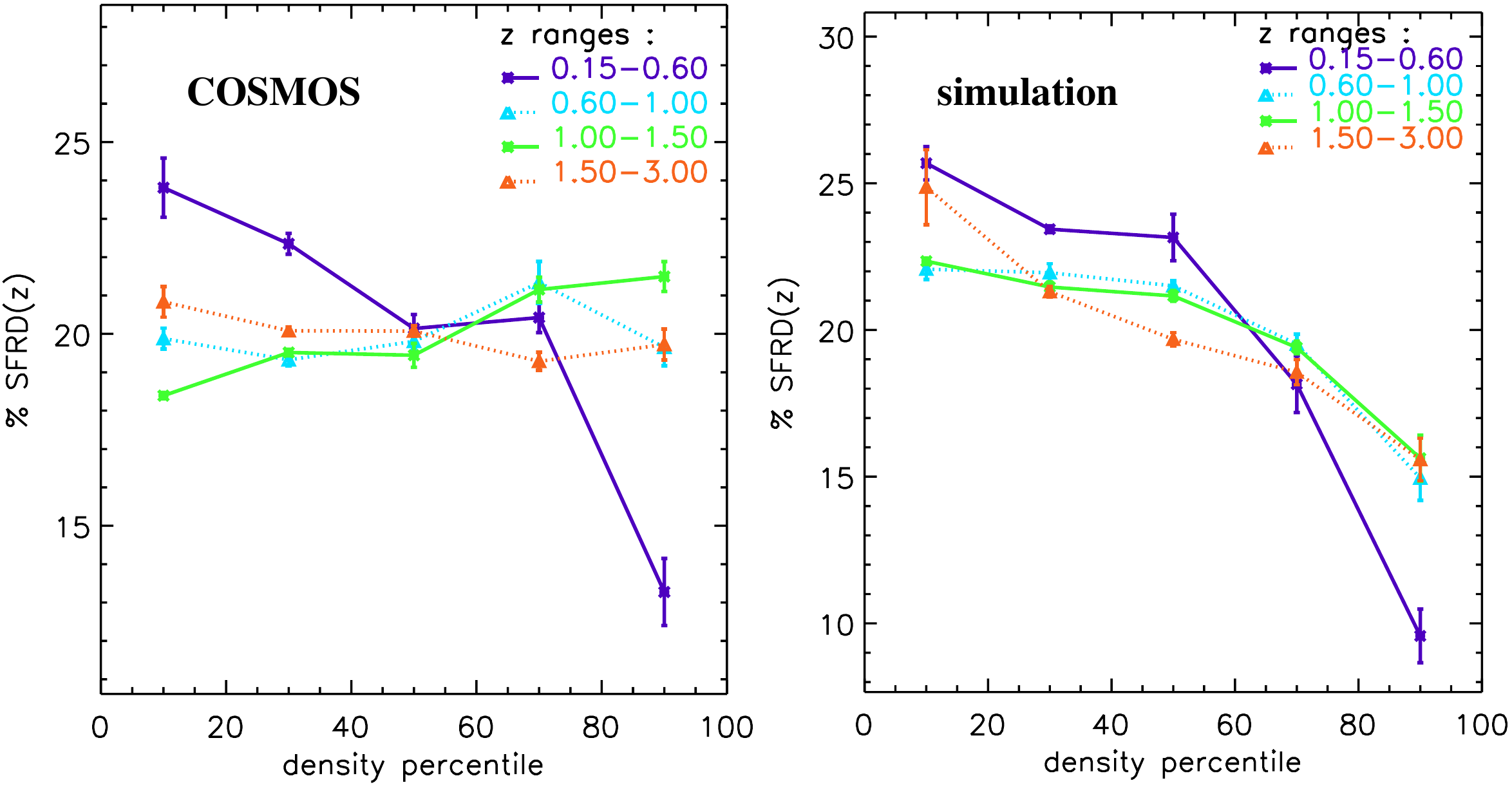}
\caption{The percentage of the star formation rate measured at each redshift attributed to different percentiles of the environmental density. The Left panel shows the COSMOS data; the Right shows the simulation SFRD.  At high redshift the SFRD is distributed equally across the density percentiles but at 
z $< 0.6$ the major contributions arise in the lower density environments. 
}\label{sfrd}
\end{figure}

It is now well established that the total SFR per unit of comoving volume or star formation rate density (SFRD) evolves strongly 
with cosmic time, decreasing by a factor of $\sim20$ from z = 2 to 0 \citep[see][and references cited there]{kar11}. Using the environmental 
densities derived here, it is possible to investigate how the SFRD at each epoch is distributed with environment. In Fig. \ref{sfrd}, the 
relative contributions to the total measured SFR(z) of galaxies in the different density percentiles are shown. Since there are, by construction, equal numbers 
of galaxies in each density percentile bin, this plot normalizes out the redshift variation of the number of galaxies in different 
density regimes. Fig. \ref{sfrd} shows that the SFRD is uniformly distributed amongst the density percentiles at all redshifts z $>$ 0.6, while below 
that redshift the SFRD shifts strongly to galaxies in lower density environments. Remarkably similar behavior is seen in the COSMOS (left panel) and simulation galaxies (right panel).

The preferential shift of the SFRD to lower density LSS is probably a result of two factors: 1) the galaxies in the high density regions evolved earlier 
 and 2) the shutdown of resupply of star forming gas in the dense environments (where the 
galaxy velocity dispersions are higher, and feedback could halt the diffuse gas accretion, see \S \ref{mat}). It is worth noting that 
since the mean stellar masses of galaxies in the dense environments are significantly higher, even above z = 0.6, the mass weighted 
SFRD would show even earlier environmental variation than the number-weighted SFRDs shown in Fig. \ref{sfrd}.

\subsection{Buildup of Stellar Mass in Passive and Star Forming Galaxies}\label{mat_dis} 

The SEDs of galaxies at $z < 3$ are separable into two distinct classes: the so-called Red Sequence (Early type) galaxies
with relatively low rates of on-going SF, hence called passive galaxies; and the Blue Cloud (Late type) galaxies with high SF rates. 
In color-magnitude diagrams, there is a much lower number of galaxies in the Green Valley between the Red Sequence
and Blue Cloud. Rather than the standard color-magnitude diagram, we show in Fig. \ref{mat1} this bifurcation of galaxy populations 
in more physical units: the maturity, $\mu = \tau_{SF} / \tau_{cosmic} = (M_*/SFR) /  \tau_{cosmic}$ where $\tau_{cosmic}$ is the age of the universe at each redshift. Since we are interested in the 
relative maturity of galaxy stellar populations over a range of redshift, we have normalized the SF timescales by the cosmic age 
at each redshift. The star formation timescale is estimated using the SFRs (from the UV continuum plus the IR, see \S \ref{sfr_section}) and stellar masses; hence, the maturity responds rapidly to changes in the SFRs.
 (The UV continuum at $\sim1800$\AA ~is produced by OB stars. For an instantaneous starburst with a Kroupa stellar IMF, the UV will fall by a factor of 10 within $30\times10^6$ yrs after the starburst ends \citep{sco11}.)
  Fig. \ref{mat1} shows the variation of the galaxy populations 
as a function of environmental density (left panels, low density and right panels, high density) and redshift (the rows) from  $z=0.15$ to $3.0$. 
The dashed line in the figures is the approximate color-dependent mass limit corresponding to the photometric selection function. 

\begin{figure}
%\figurenum{\ref{mat1}} 
\begin{minipage}[b]{1.0\linewidth}
\epsscale{1.0}
\plottwo{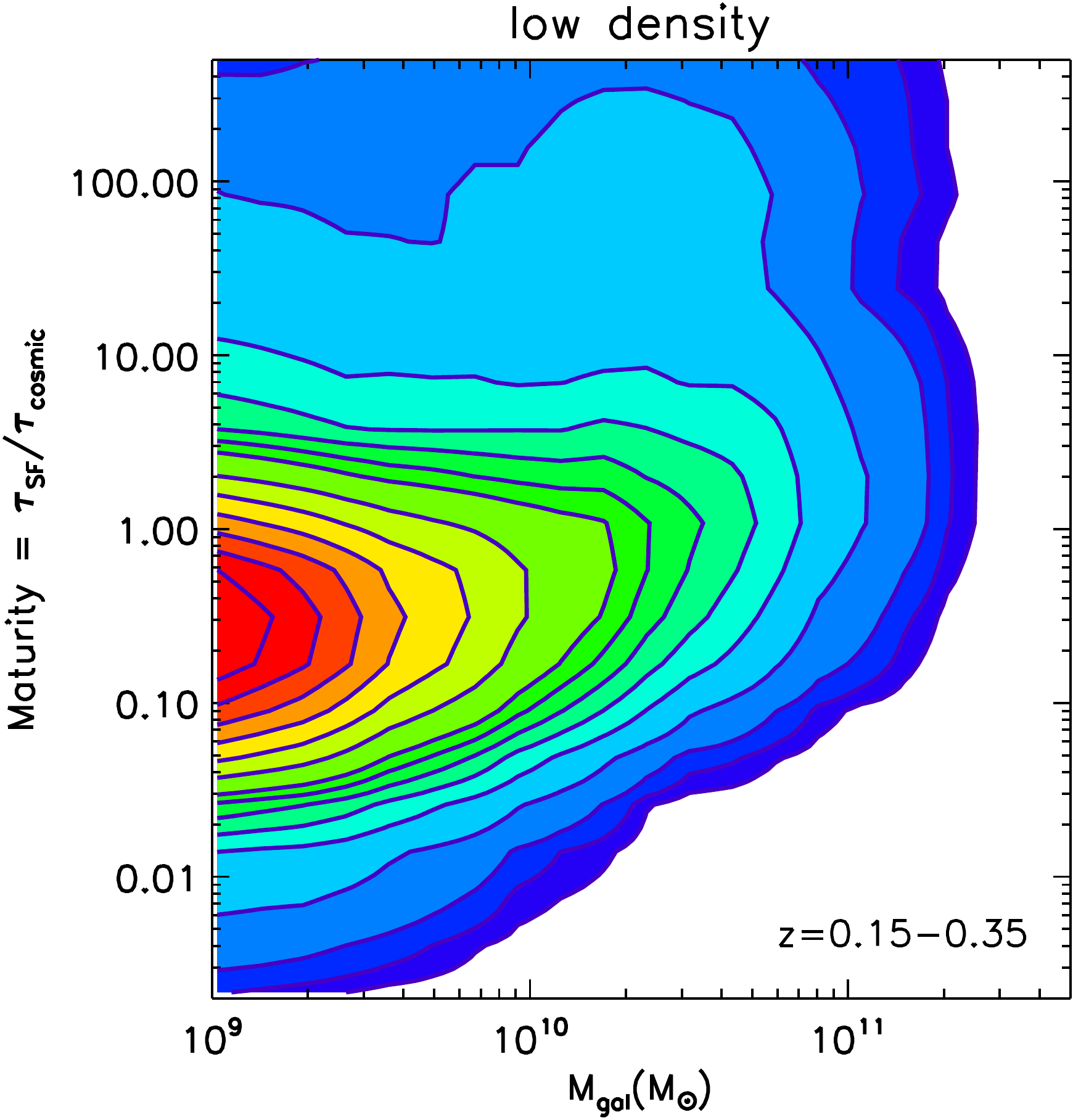}{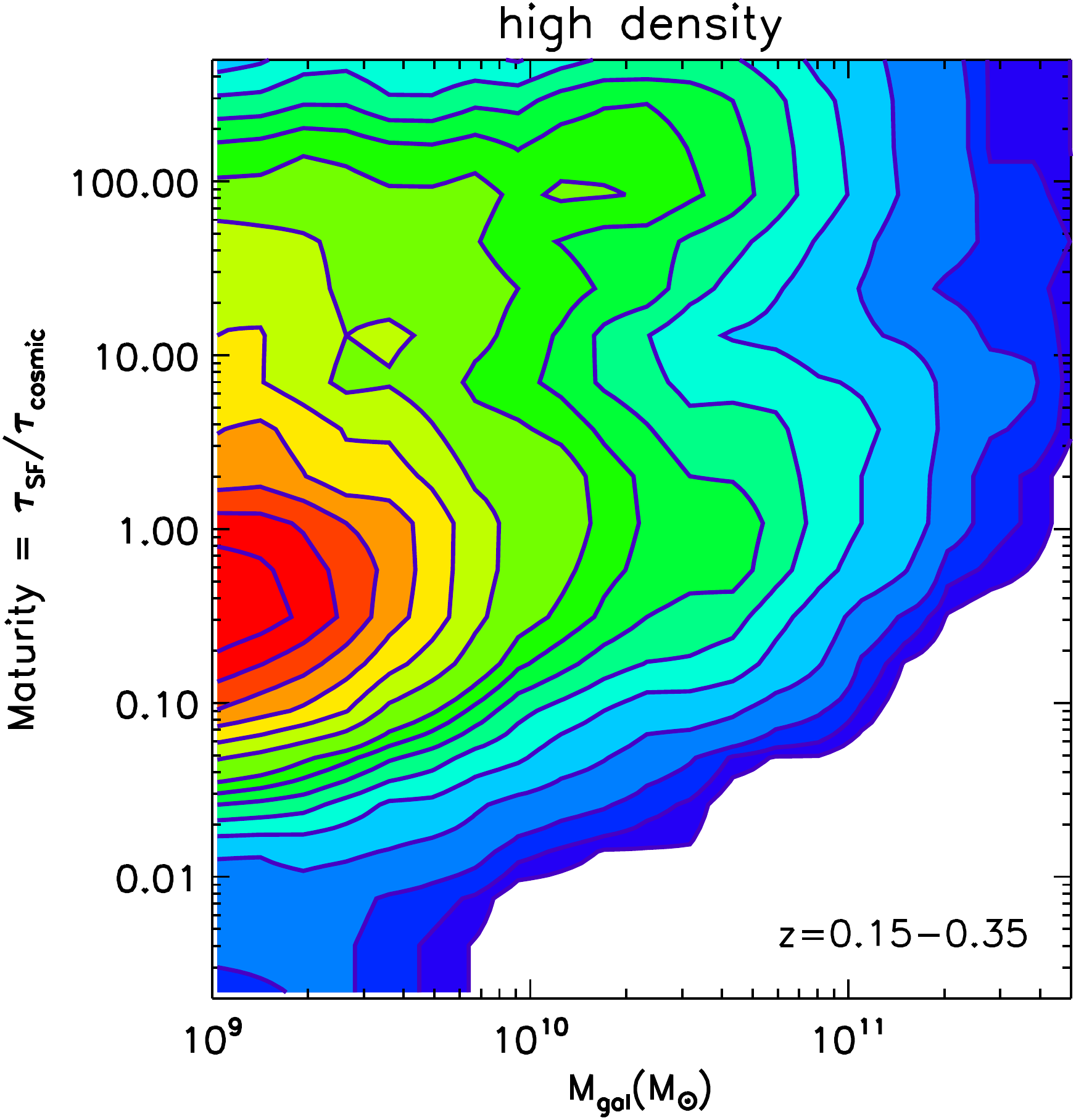}
\end{minipage}
\begin{minipage}[b]{1.0\linewidth}
\epsscale{1.0}
\plottwo{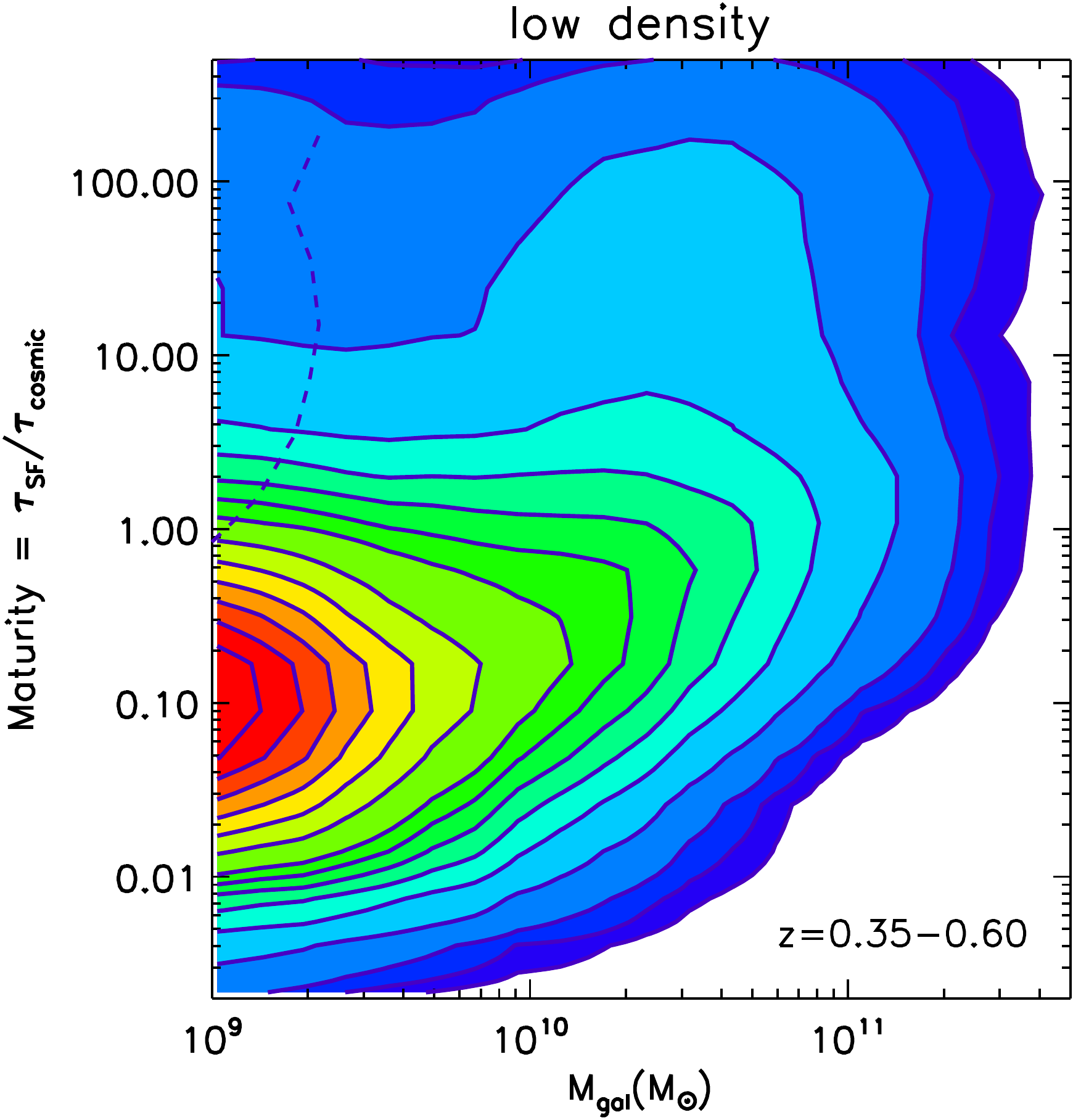}{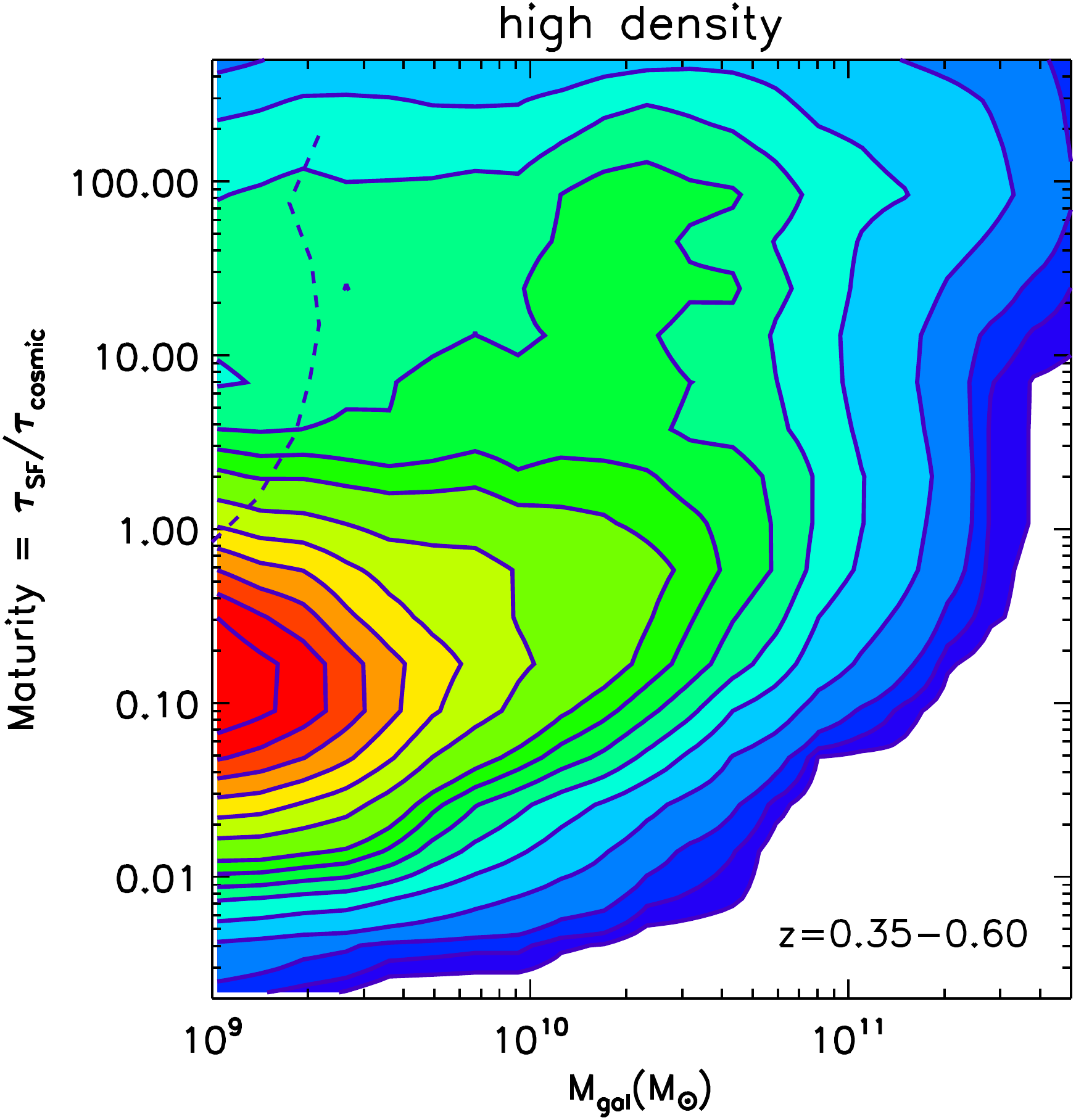}
\end{minipage}
\caption{The maturity  ($\tau_{SF} / \tau_{cosmic} $) is shown separately as a function of stellar 
mass for low (left panel of each pair) and high (right panel of each pair) density environments as a function of redshift (pairs of panels). The maturity is correlated with the rest frame colors of the galaxies (primarily the NUV for the SFR 
and the near infrared for the stellar mass). At low redshift, the maturity shows extremely
good separation of the blue cloud and red sequence galaxies. On each plot the dashed line indicates the mass limit of the selection function as a function of maturity. (These mass limits are the average for 
objects in each redshift range and hence, some galaxies at the low end of the redshift interval will 
appear to the left of the mass limit line.) Contours at :
0.05, 0.1 ,0.15 ,0.2, 0.25, 0.3, 0.4, 0.5, 0.6, 0.7, 0.8, 0.9 $\times$ peak. The high and low environmental density bins are such as to split the overall
galaxy sample in half at each redshift.
}\label{mat1} 
\end{figure}

\begin{figure}
\figurenum{\ref{mat1} b}
\begin{minipage}[b]{1.0\linewidth}
\epsscale{1.0}
\plottwo{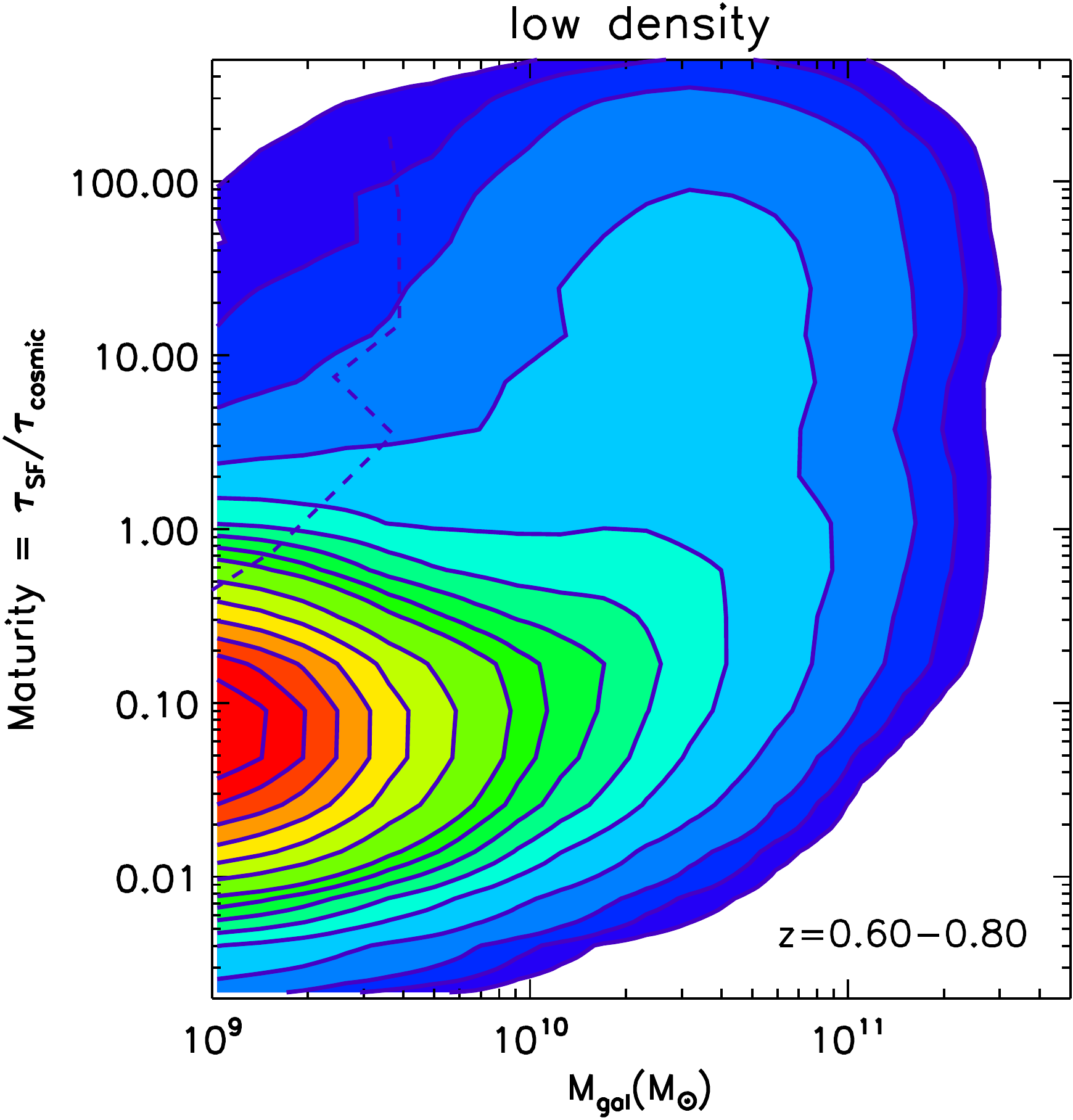}{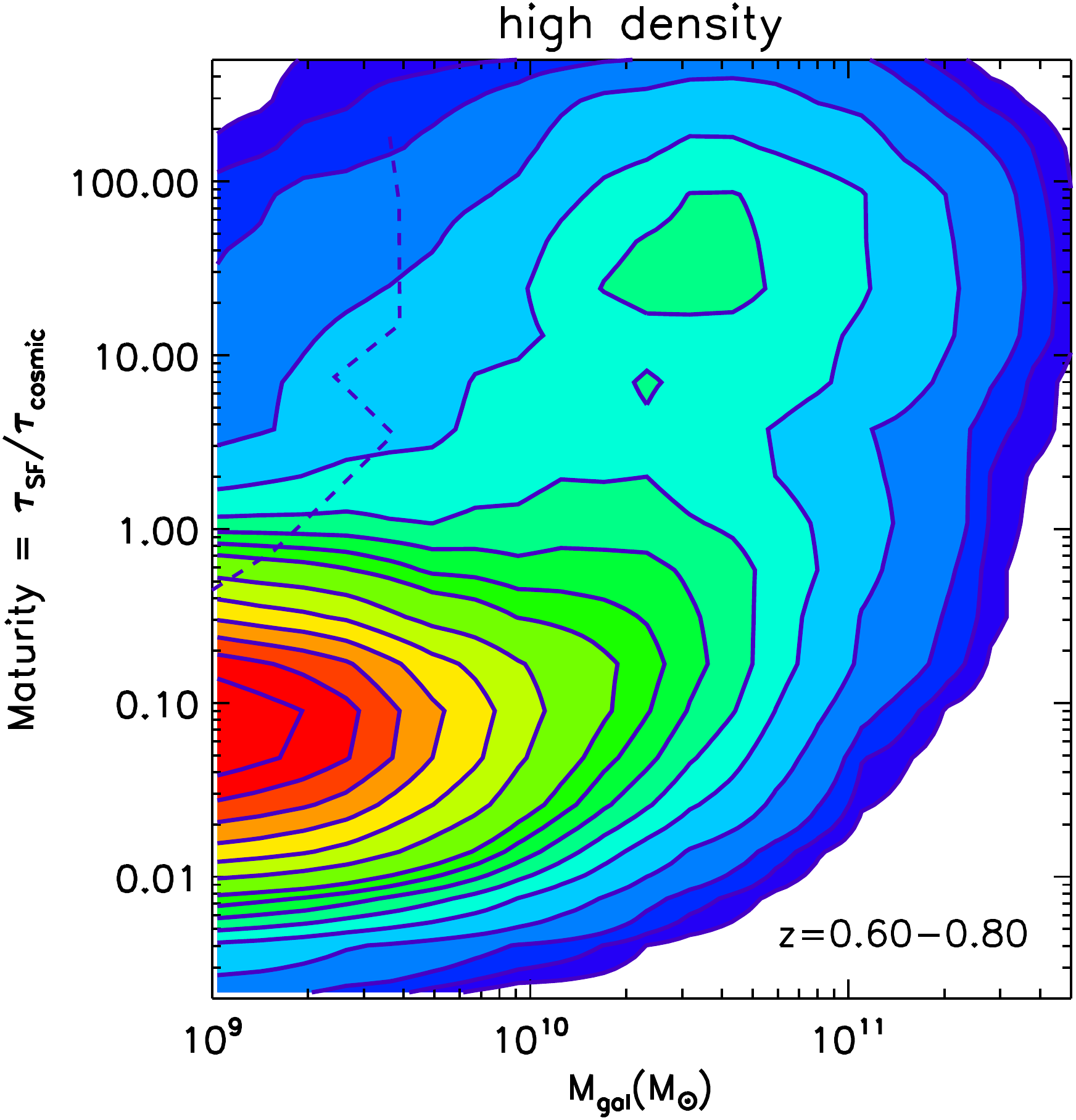}
\end{minipage}
\begin{minipage}[b]{1.0\linewidth}
\epsscale{1.0}
\plottwo{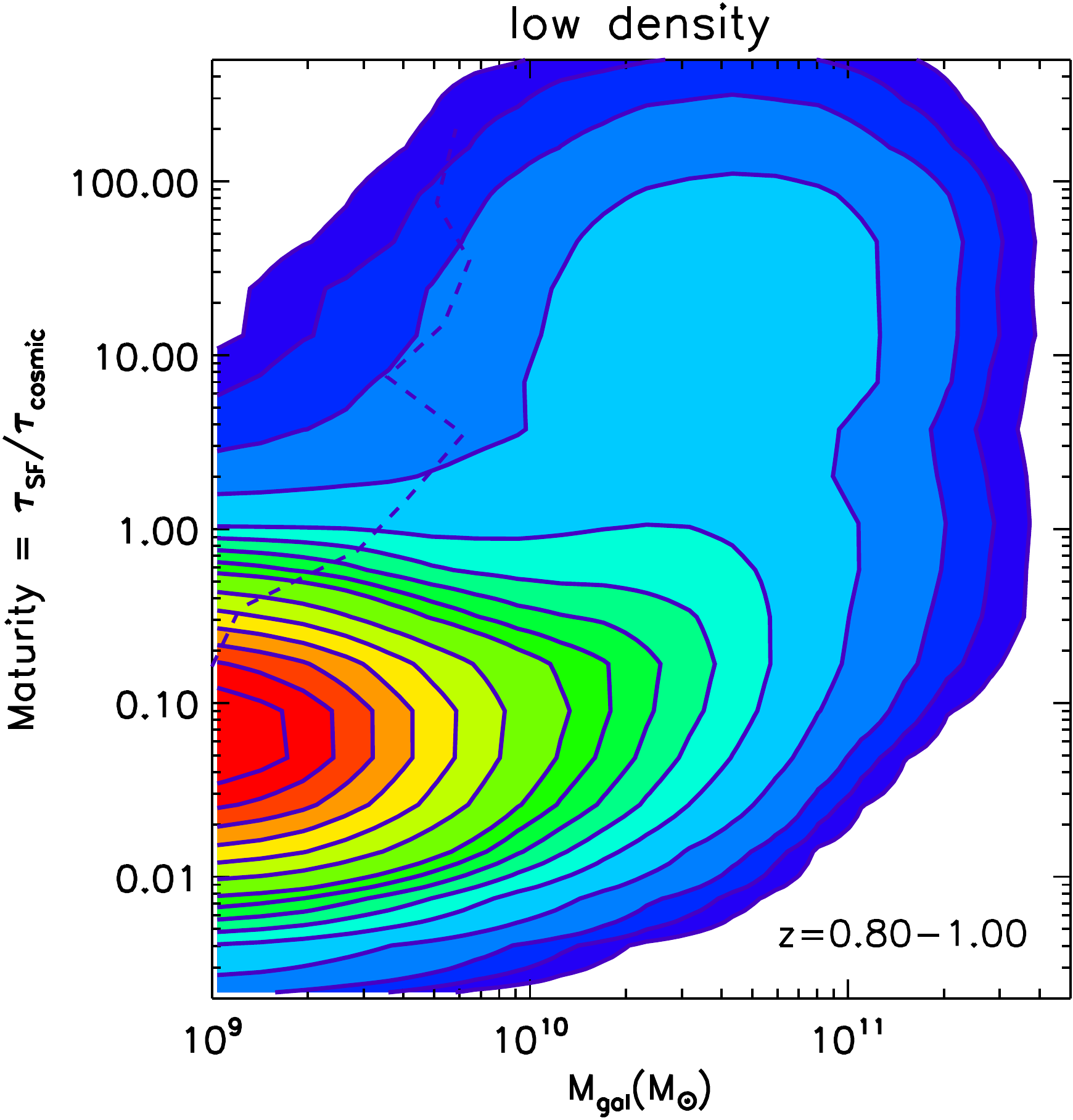}{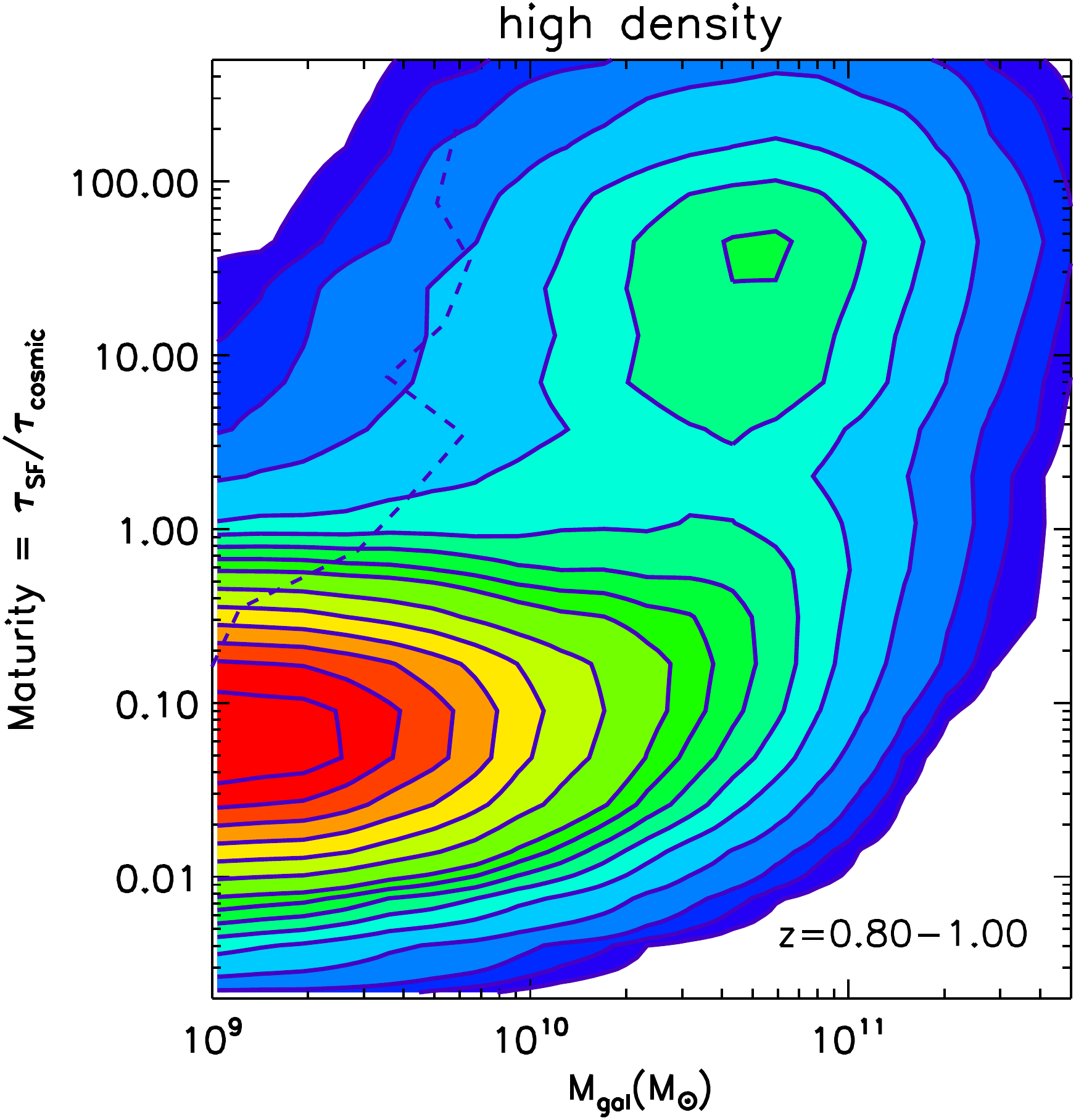}
\end{minipage}
\caption{continued
}
\end{figure}

\begin{figure}
\figurenum{\ref{mat1} c}
\begin{minipage}[b]{1.0\linewidth}
\epsscale{1.0}
\plottwo{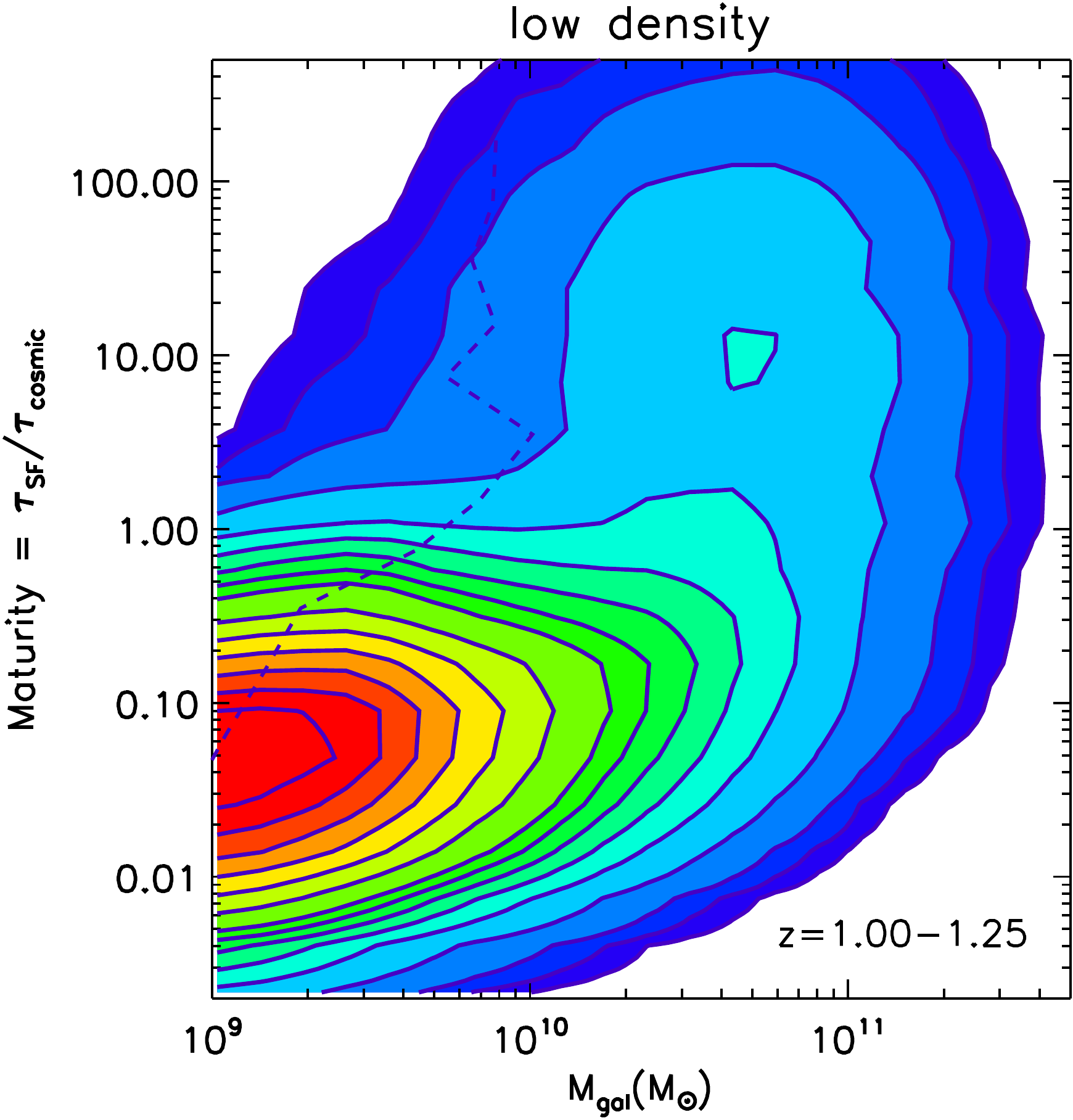}{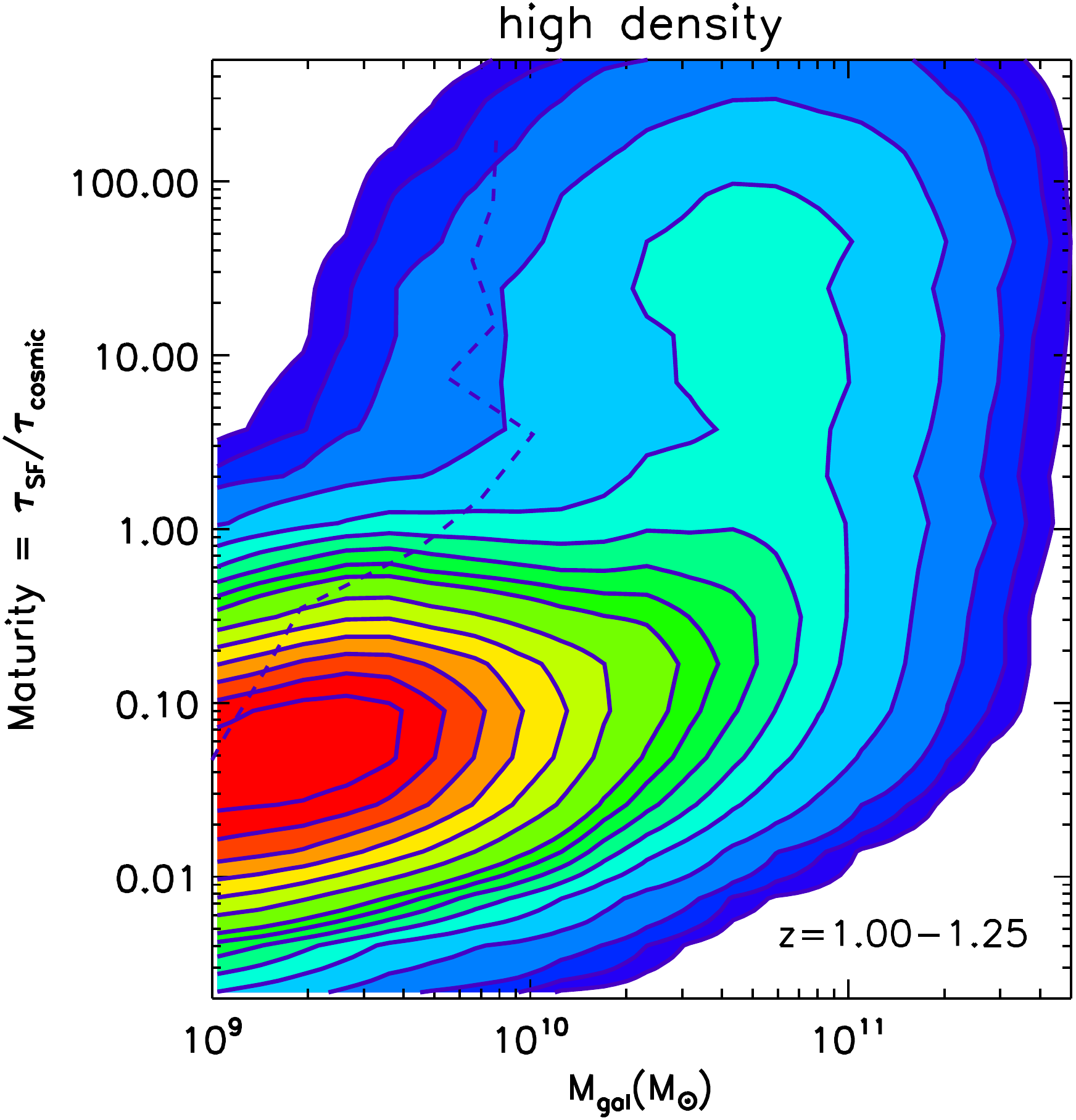}
\end{minipage}
\begin{minipage}[b]{1.0\linewidth}
\plottwo{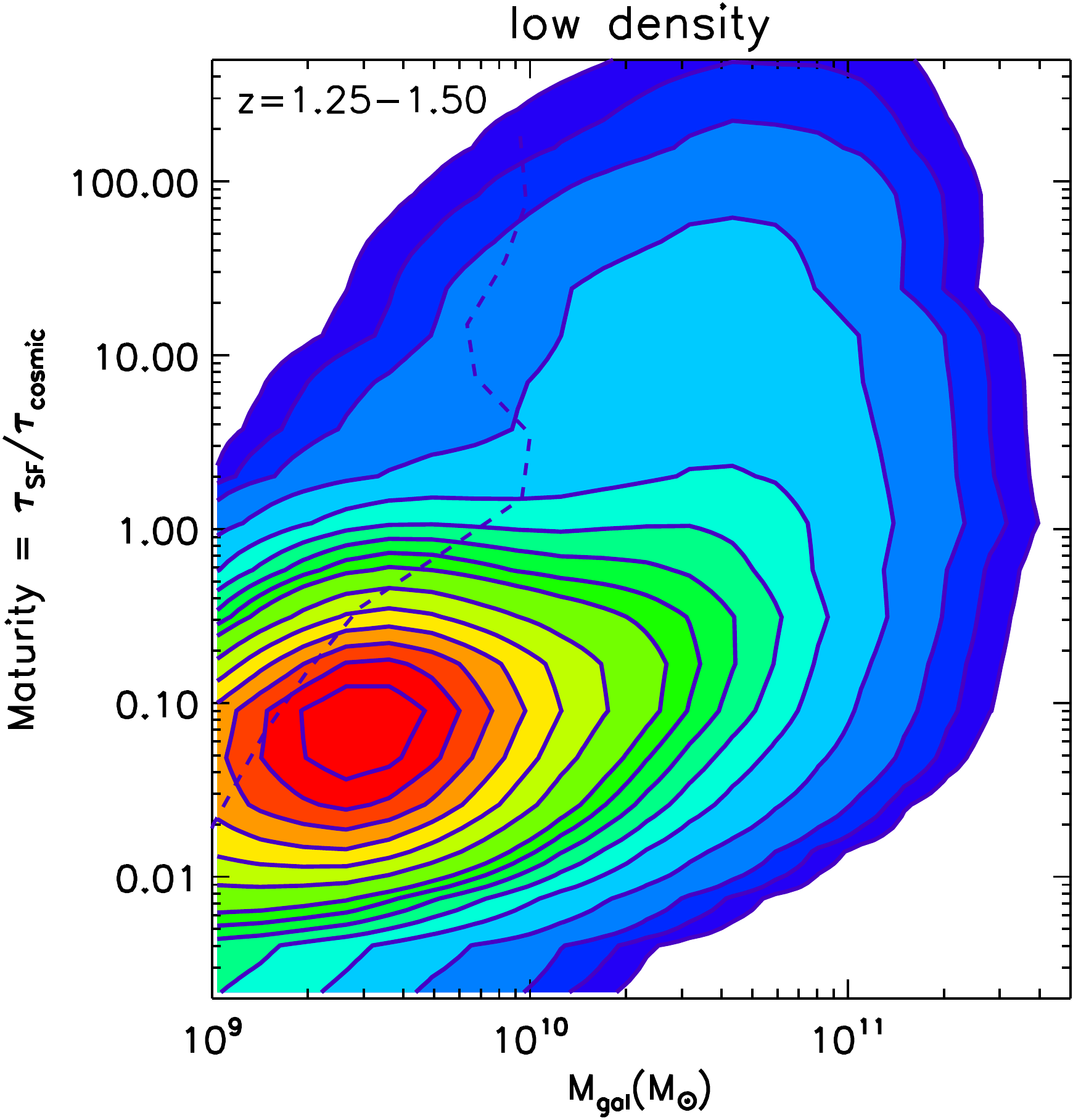}{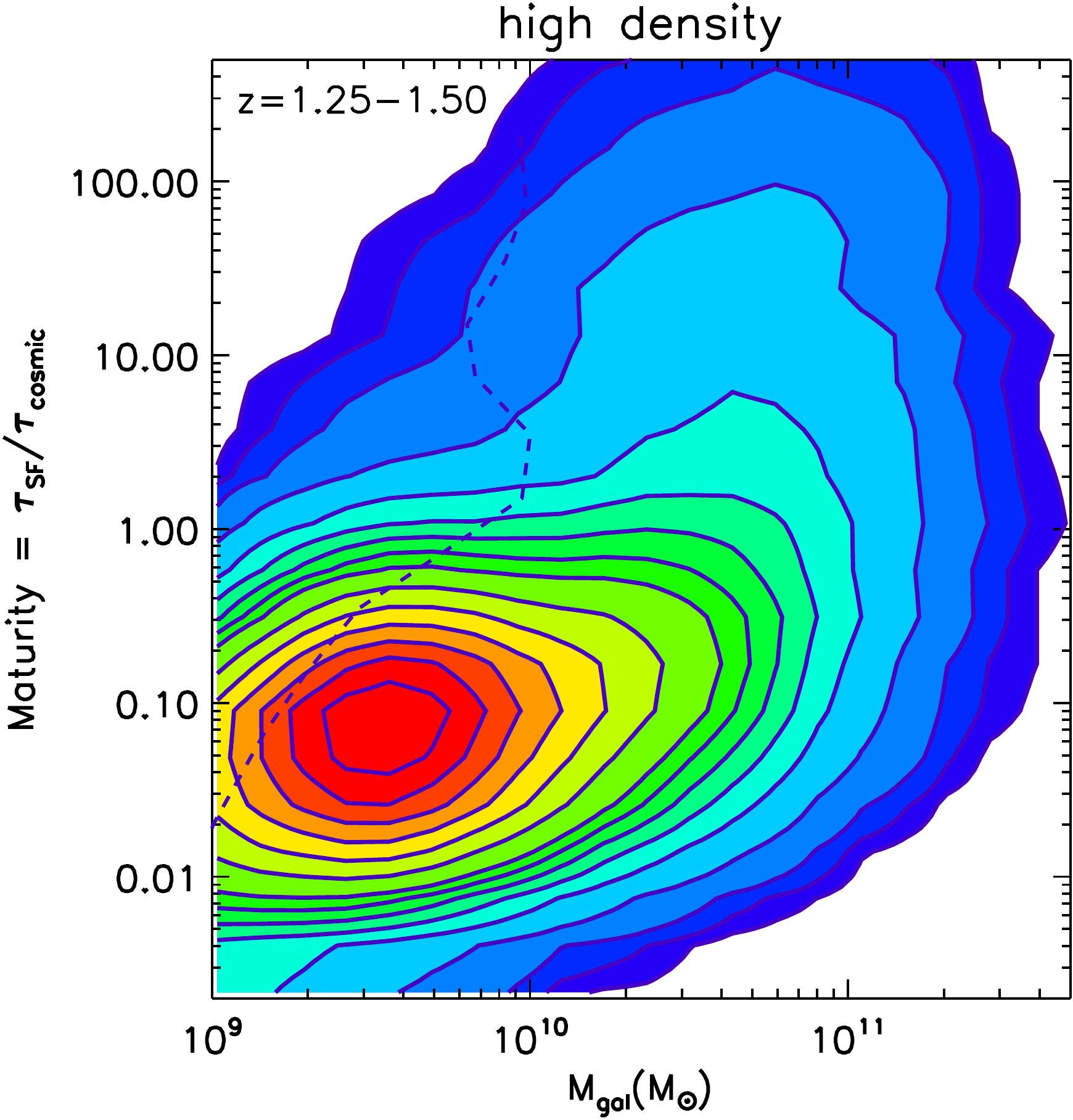}
\end{minipage}
%\begin{minipage}[b]{1.0\linewidth}
%\epsscale{0.5}
%\plottwo{m7l-eps-converted-to.pdf}{m7h-eps-converted-to.pdf}
%\plottwo{m8l-eps-converted-to.pdf}{m8h-eps-converted-to.pdf}
%\end{minipage}
\caption{continued
}
\end{figure}

At low z,  Fig. \ref{mat1} shows  very clear separation of the Red Sequence (Maturity $\mu > 10$) and Blue Cloud ($\mu \lesssim 2$) and an enhancement 
of the Red Sequence in the denser environments (right panel). In the Green Valley, between the Red Sequence and the Blue Cloud, the number density 
of galaxies (in the mass-maturity plane) is as low as 30\% of the peaks on either side. This enhancement of the Red Sequence in denser environments persists but with 
diminishing amplitude out to $z = 1$.  In these plots, the Red Sequence clearly extends to lower mass galaxies at decreasing redshift. In fact, 
comparing the plot for z = 0.15 - 0.35 and z = 0.35 - 0.60 (Fig. \ref{mat1} a), a major development is the appearance of the low mass
red galaxies at z = 0.15 - 0.35, which were not very apparent at z = 0.35 - 0.60. This strongly implies that such galaxies are the result of environmental 
quenching processes (such as ram pressure stripping or starvation of gas accretion) rather than dry merging since the lower mass red galaxies were not 
present in sufficient abundance at the earlier epoch. This corresponds to the environmental quenching as discussed by \cite{pen10}. 

The mass 
limit cutoff shown by the dashed line in each panel is at significantly lower mass than the mass at the peak of the Red Sequence, and therefore the 
disappearance of environmental dependence of the Red Sequence at $z > 1.1$ is not due simply to insufficient mass sensitivity for passive galaxies. 
Additionally, we note that such effects would not differentiate between low and high density environments. Thus, we conclude that environmental 
differentiation decreases at the higher redshifts. 

A more complete analysis of the galaxy mass function evolution is provided by \cite{ilb09}. 
The Red Sequence shows an obvious tilt in maturity as a function of stellar mass (Fig. \ref{mat1}-a), implying that the lower mass red galaxies 
were built up at later times than the high mass red galaxies. This behavior is seen in the COSMOS study of galaxy mass functions \citep{ilb10} 
and is commonly referred to as downsizing. Above $z = 1.25$, the minimum corresponding to the 
Green Valley disappears and the Red Sequence appears more as a plume extending out of the high mass end of the Blue Cloud (see Fig. \ref{mat1} c).  At $z > 1.1$, one can still see 
a very mild environmental segregation of the Red Sequence galaxies (i.e. a slightly higher density of such galaxies in the right panel of each redshift 
range).

\begin{figure}
\begin{minipage}[b]{1.0\linewidth}
\epsscale{1.}
\plottwo{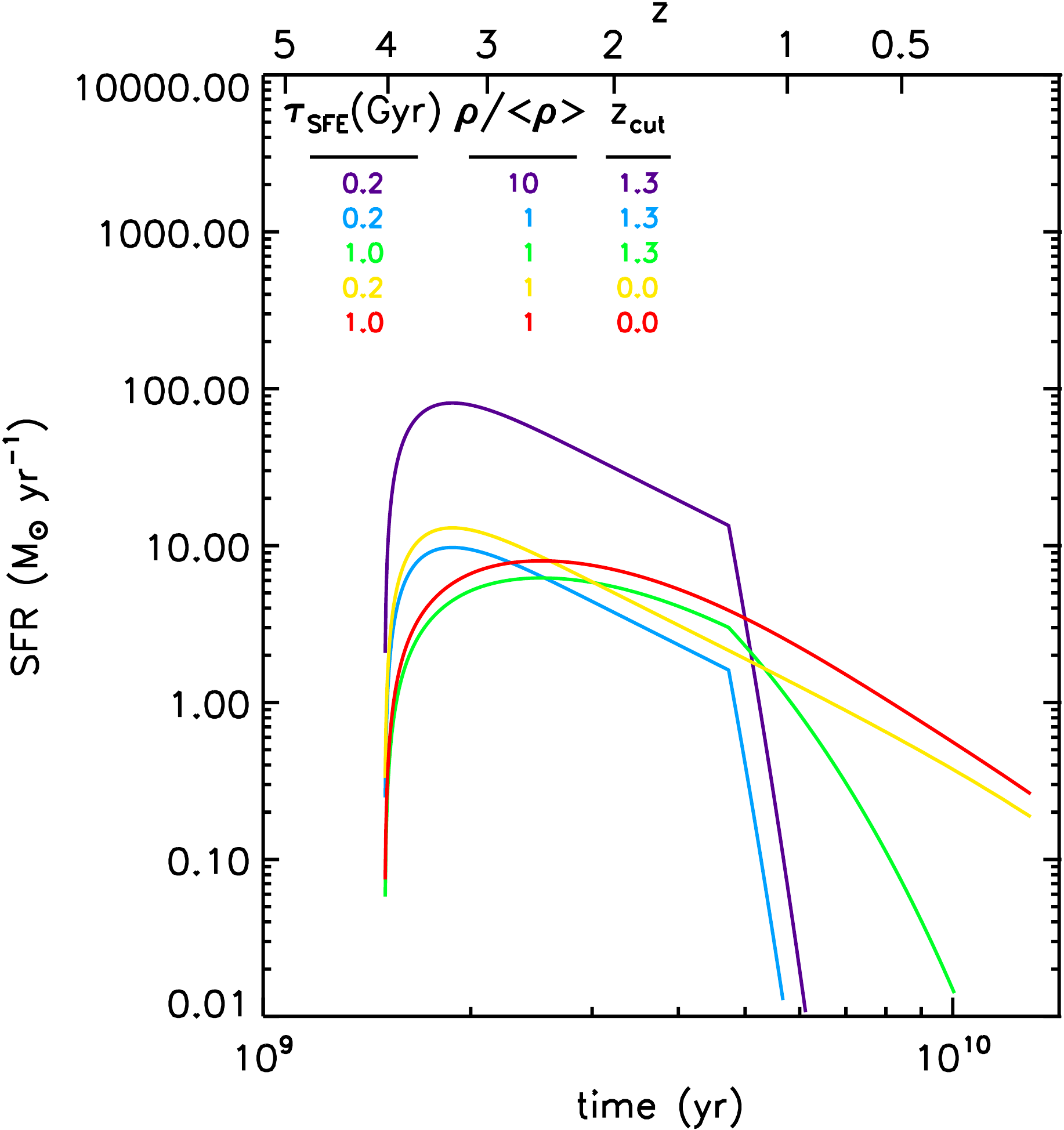}{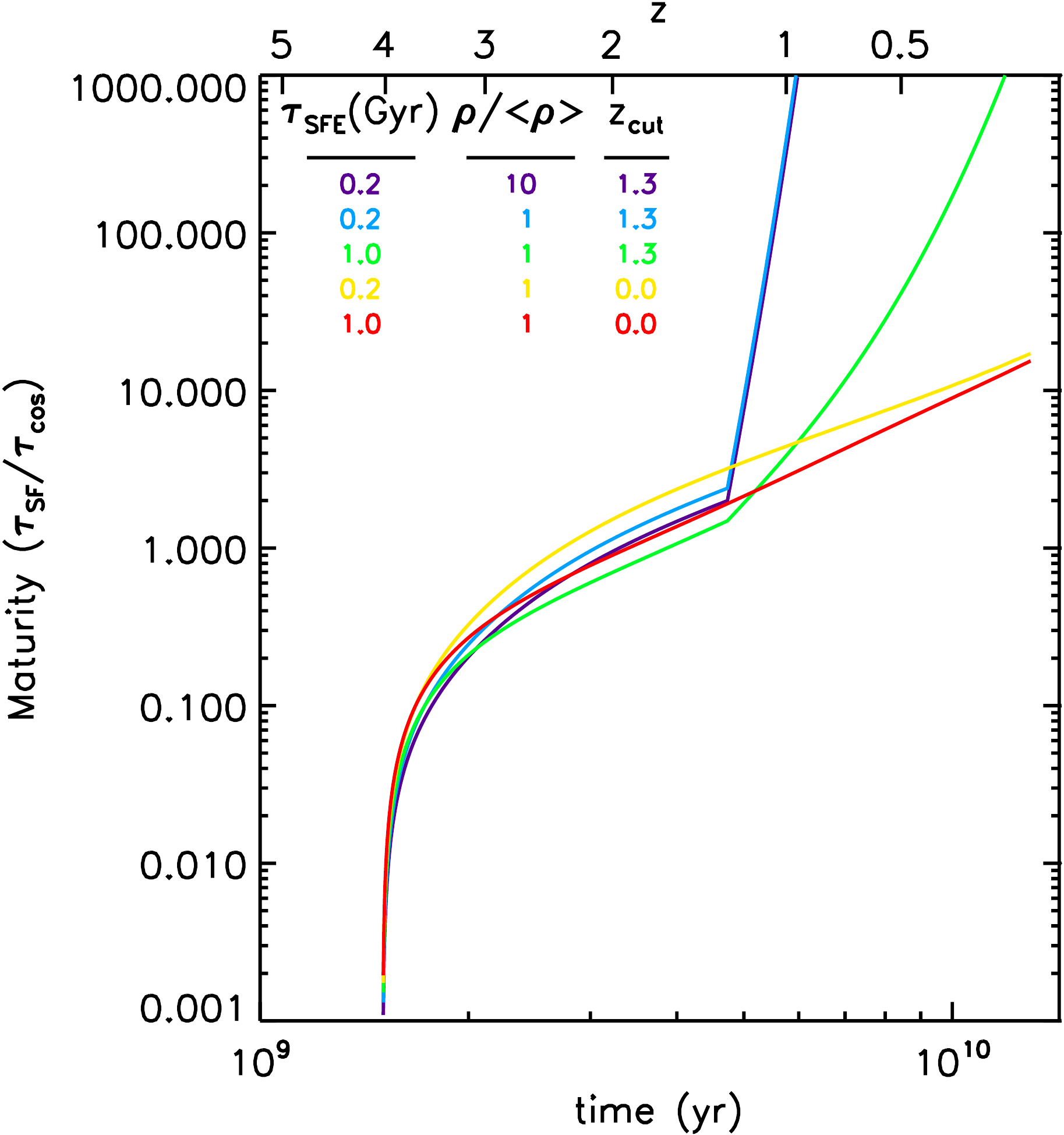}
\end{minipage}
\begin{minipage}[b]{1.0\linewidth}
\plottwo{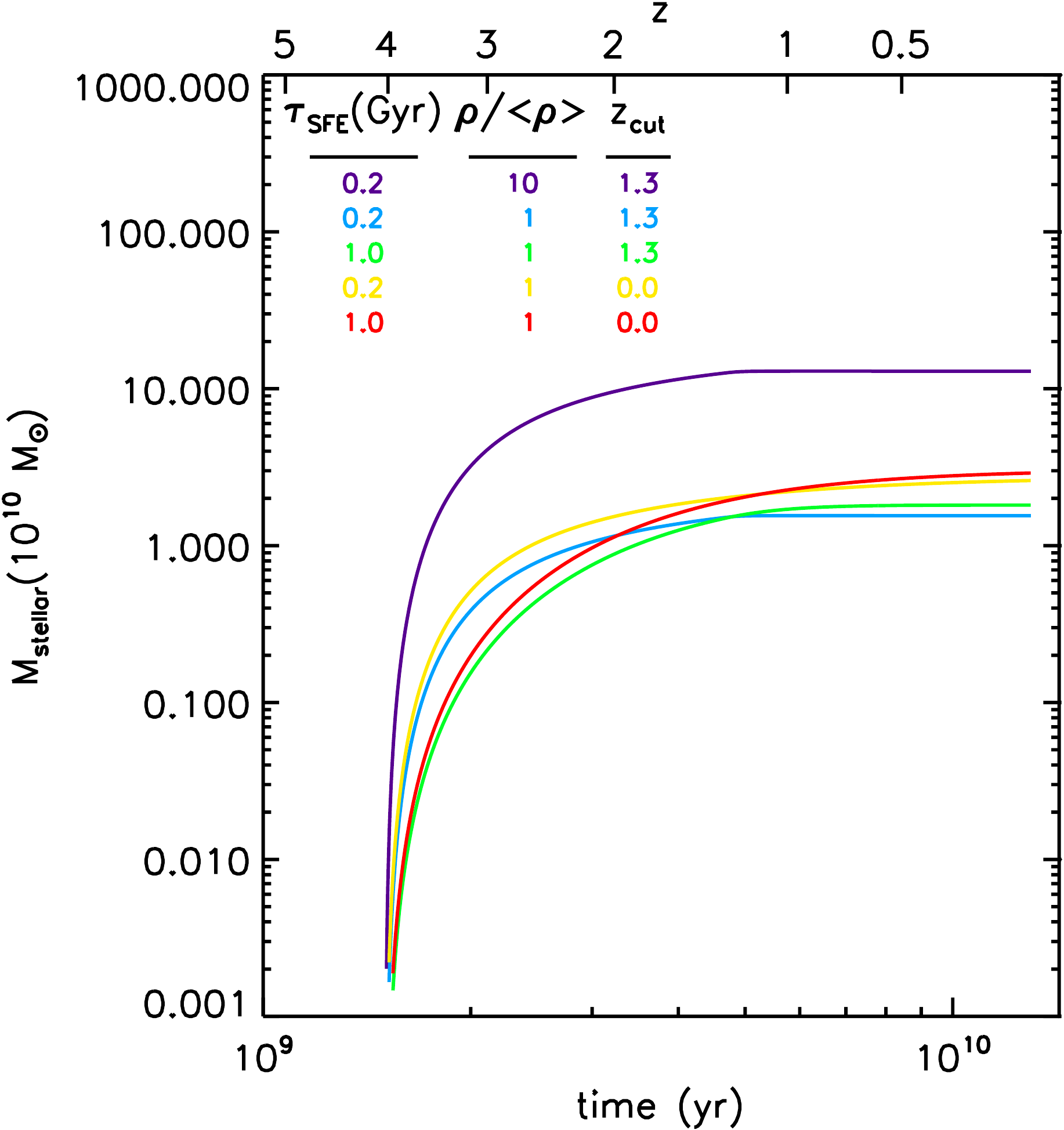}{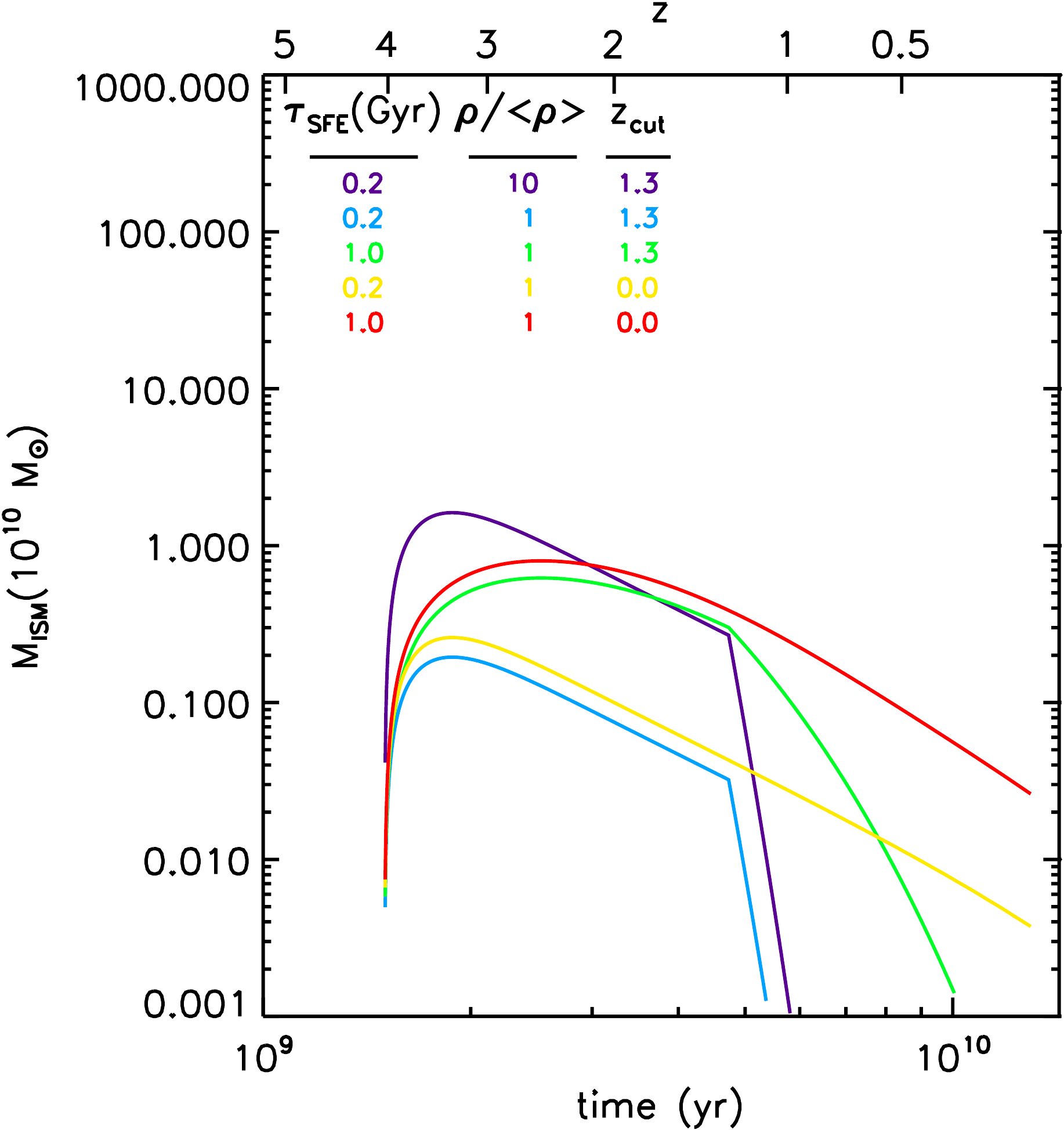}
\end{minipage}
\caption{The SFR (upper-left), maturity ($\tau_{SF} / \tau_{cosmic} $ -- upper-right), $M_{stellar}$ (lower-left) and $M_{ISM}$ (lower-right) are
shown for simple models with: 1) ISM buildup by accretion from the environment (with differing overdensity); 2) varying star formation efficiencies 
per unit mass of ISM and 3) varying the lower redshift cutoff for the accretion (see text for details). $\tau_{SFE} = 1$ Gyr corresponds to the typical 
efficiency observed in low-z spiral galaxies like the Milky Way; shorter timescales correspond to starburst activity. $\rho/<\rho> = 1$ corresponds to 
the galaxy being in the average environmental density at each redshift. The accumulated stellar mass and residual ISM mass are shown in the lower two panels.
}\label{model}
\end{figure}

\subsection{The Rapid Development of Passive Galaxies in Dense Environments}\label{mat} 

The mean cosmic ages at $z = 1$ and $1.3$ are 5.7 and 4.7 Gyr, so the abrupt development of environmental segregation for  
passive galaxies must take place in $\sim1$ Gyr. A number of mechanisms have been suggested for such environmental 
differentiation: galaxy-galaxy harassment, tidal and ram pressure stripping of the disk gas, shutoff of fresh gas accretion from the outer halo by ram pressure ('strangulation') and feedback from starburst and active nucleus activity. To explore the possible explanations for this rapid change, we have constructed a 
model for the bulk evolution of galaxies with the simple assumptions that they accrete interstellar/star forming gas from their 
local environment and form stars with a star formation efficiency similar to that seen at low-z. 

At low redshift, we may take the Milky Way 
as being typical of normal star forming galaxies. Here, the mass of ISM is $\sim3\times10^9$\msun and the SFR is $\sim3$\msun yr$^{-1}$, 
implying an e-folding timescale of $\sim10^9$ yrs for reducing the ISM and SFR. %This also provides a strong argument that the low-z 
%star forming galaxies must have significant on-going accretion since it is unlikely they will \emph{all} significantly halve their SF on a 1 Gyr timescale.
For the z = 1.3 star forming galaxies --  ISM consumption with an efficiency or timescale like that of local galaxies will only change the 
Maturity (M$_* / SFR /  \tau_{cosmic}$) by a factor of 2 on a Gyr timescale via star formation in the 
blue galaxies, changing the Maturity by a factor of a few within this time period.

An alternative mechanism to populate the Red Sequence might be the merging of lower mass Red Sequence galaxies in the dense environments at $z > 1.1$ (often referred to as dry merging); however, the overall mass and number of such pre-existing Red Sequence galaxies is insufficient even if the merger rate is sufficient. 
We are therefore forced to the conclusion that \emph{ there is rapid conversion of massive star forming 
galaxies to passive galaxies (with SFRs decreased by a factor of $>10$ within $\sim1$ Gyr) and the only way this can happen 
is by removal of the ISM}. 

One process which might remove the ISM rapidly and have the observed  
strong environmental dependence is ram pressure stripping of the ISM by cluster gas, deposited by prior star formation and AGN feedback processes.  
This strangulation process has been included in LSS evolution simulations by 
\cite{mcg09} and they predict that the environmental dependence of the passive galaxies should set in at $z \sim 1.4-1.5$ (i.e $\sim$1 Gyr earlier than 
seen here).

To illustrate the need for rapid depletion of the star forming gas, we have computed the evolution for an extremely  simple model in which 
galaxy ISM is supplied by accretion from the local environment and converted into stars at a rate or efficiency equal to that in the local 
universe for normal galaxies, as given above, (i.e. not undergoing a starburst). The ISM accretion or replenishment is taken to vary proportionately to the local environmental density ($\rho$) :

\begin{mathletters}
\begin{equation}
 \dot{M} = \dot{M_0} \delta \times \rho(z)   
\end{equation}
\end{mathletters}

\noindent where $\delta = \rho/<\rho(z)> $ is the local environmental overdensity and  $\rho(z) \propto (1+z)^3$ is the mean cosmic density, and $\dot{M_0}$
is a normalizing constant such that a significant star forming ISM has accumulated by z $\sim$ 4. This simplistic assumption is most reasonable for central galaxies but 
not so appropriate for satellite galaxies. The actual halo growth rate may vary as  much as $(1+z)^{2.2}$ 
\citep[e.g.][]{nei08} and if this were adopted it would make the ISM removal problem even more severe. The accretion may occur either as spherical or cold flow accretion.

The timescale for star formation in the accreted 
gas is taken to be 10$^9$ yr, i.e. similar to that computed above for the Milky Way \citep[also typical of low-z spiral galaxies - ][]{you91}. This 
adopted efficiency is similar to that implicit to the Kennicutt relation for typical spiral galaxies (but we omit the non-linear dependence on surface density 
for which one would need to know the size of the ISM disk). 

Using this simple model, we explore the 
effects of more rapid star formation or a denser environment (leading to higher accretion) on the evolution of the SFR and maturity parameter.
In Fig \ref{model}, the redshift evolution of the SFR, the maturity parameter ($\tau_{SF}/\tau_{cos}$),
of the stellar population, the accumulated stellar mass and ISM mass are shown for these evolutionary models. The \emph{red curve} illustrates the redshift 
evolution expected simply as a result of accretion from the environment and SF with the standard efficiency -- this model 
clearly cannot reproduce the rapid maturing (within $\sim 1$ Gyr) of the stellar population (upper right panel of Fig \ref{model}) as observed in the dense environments 
between z = 1.3 and 1.1. Similarly, decreasing the star formation timescale or gas consumption time ($\tau_{SFE}$) by a factor of 5 as might occur in a starbursting system (\emph{yellow curve} in Fig \ref{model}) simply shifts the peak 
star formation activity earlier, but does not significantly accelerate the change of the maturity parameter to high values. In this case, the accretion of fresh ISM continues 
and the associated star formation keeps the maturity low.  To model the effect of exhaustion of the existing ISM supply by star formation when accretion processes are abruptly terminated, 
we ran models with a cutoff redshift $z_{cut}$ = 1.3 (green curve); this clearly accelerates the maturation of the galaxies although still not as rapidly (within $\sim 1$ Gyr)
as required by the observations for the dense environments. Here the SF slowly dies out with an e-fold timescale of 1 Gyr but is not abruptly terminated. This model corresponds to those of 
\cite{bou10} which have an abrupt accretion cutoff once the halo mass exceeds $\sim10^{12}$\msun. Instead, \emph{one needs to actually strip the existing ISM from the galaxies or accelerate the 
star formation process (decreased SF timescale) as shown in the blue and purple curves and halt accretion of fresh ISM}. The model with $\rho/<\rho> = 10$ 
(purple curve) is included simply to illustrate that when the accretion rate is scaled up by a factor of 10, the temporal 
variation remains unchanged (although of course the SFRs and final mass of stars is 10 times larger). 
 
In summary, the rapid maturing of galaxies in dense environments seen here at z $\sim 1.2$, requires both termination of the 
fresh resupply of ISM and an elevated rate of depletion of the existing ISM, either through stripping from the galaxy or enhanced star efficiency. 
The termination of accretion within dense environments might be caused by the higher virial velocities of galaxies in the dense environments 
and disconnection of the galaxies from the filamentary/cold accretion flows found in lower density environments. For exhaustion of the existing ISM, ram 
pressure stripping seems more likely than enhanced star formation rates, since the latter may happen as a result of interactions with characteristic timescales of $10^8$ yrs in some 
galaxies, but is unlikely to occur for a significant fraction of the  galaxies in the dense environments within $\sim1$ Gyr. The IGM accumulated in the densest LSS from galactic mass-loss SF and AGN winds, combined with the nascent inter-cluster gas would be the agent for the ram pressure. \cite{kau04} have argued similarly, based on 
very detailed analysis of star formation histories and structural characteristics of galaxies as a function of environment in SDSS at $z < 0.1$. They point out that since the structural properties are not so environmentally dependent, it is unlikely that galaxy interactions and merging are driving the decrease in the sSFR and increase in the red fraction in dense environments. \cite{pen10} find that the quenching of SF activity can be empirically modeled as \emph{separable} stellar mass and environmental
density dependent terms, but do not identify the physical mechanisms associated with each.  \cite{pen10} argue that the environmental quenching occurs only below $M_* \sim 10^{10}$\msun, i.e.  satellite galaxies and that the central galaxies show no effect.

\section{Conclusion}

New high accuracy photometric redshifts for a sample of 155,954 galaxies at z = 0.15 to 3.0 have been used to map the cosmic large 
scale structure in the 2 square degree COSMOS survey field. Approximately 260 significantly overdense structures are detected, including 
high density, circularly symmetric structures, and elongated filamentary structures extending up to 15 Mpc. The density distributions and their 
evolution with redshift are in good agreement with semi-analytic models based on the $\Lambda$CDM simulation.

We have also presented preliminary analysis of the correlation of galaxy properties with their large scale structure densities. 
At z $< 1.2$, the red and low SFR galaxies are strongly correlated with the higher density environments and 
this environmental segregation increases systematically to lower redshift. Above z = 1.2, the environmental correlations 
are greatly reduced in both the observations and the simulation. The contributions to the overall SFRD are uniformly spread across all environments 
down to z $\sim 0.6$ while at lower redshift, the SFRD shifts to lower density environments. 

The density maps presented here for  the COSMOS field are available for general use at the IPAC/IRSA COSMOS 
archive at :  \href{http://irsa.ipac.caltech.edu/data/COSMOS/ancillary/densities/}{{\bf COSMOS LSS download}} which is :

\noindent {\bf http://irsa.ipac.caltech.edu/data/COSMOS/ancillary/densities/}.  These files include: images of all redshifts slices 
 in Figures \ref{vor}, \ref{vor_den} and \ref{ad}; animations for these figures, 
 and 3D FITS files.

\acknowledgments
 We thank the referee for helpful suggestions which have very much improved this work
 and we thank Zara Scoville for proofreading of the manuscript.
  Support
 for this work was provided by NASA through Contract Number 12712786
 issued by JPL.   Additional information on the COSMOS 
 survey is available from the main COSMOS web site at 
{\bf$<$http://www.astro.caltech.edu/cosmos$>$}. It is a pleasure to
 acknowledge the excellent services provided by the NASA IPAC/IRSA
 staff in providing online archive and server capabilities for the
 COSMOS datasets. The environmental densities and mpeg versions with all 127 redshift slices of the maps shown in Fig. \ref{vor}-\ref{ad} will be available in FITS format from the NASA IPAC/IRSA archive at
{\bf$<$ http://irsa.ipac.caltech.edu/data/COSMOS/$>$}. KS acknowledges support from the National Radio Astronomy Observatory
which is a facility of the National Science Foundation operated under cooperative agreement by Associated
Universities, Inc. GQ acknowledges support from the National basic research program of China (program 973 under grant No. 2009CB24901), the Young Researcher Grant of National
Astronomical Observatories, CAS, the NSFC grants program (No. 11143005) and the Partner Group program of the Max Planck Society.
 
 {\it Facilities:} \facility{HST (ACS)}, \facility{Subaru (SCAM)}, \facility{Spitzer (IRAC)}.



\eject
\appendix

\section{Bright Star Masking}\label{appendix}

\begin{figure}[ht]
\epsscale{0.7}
\plotone{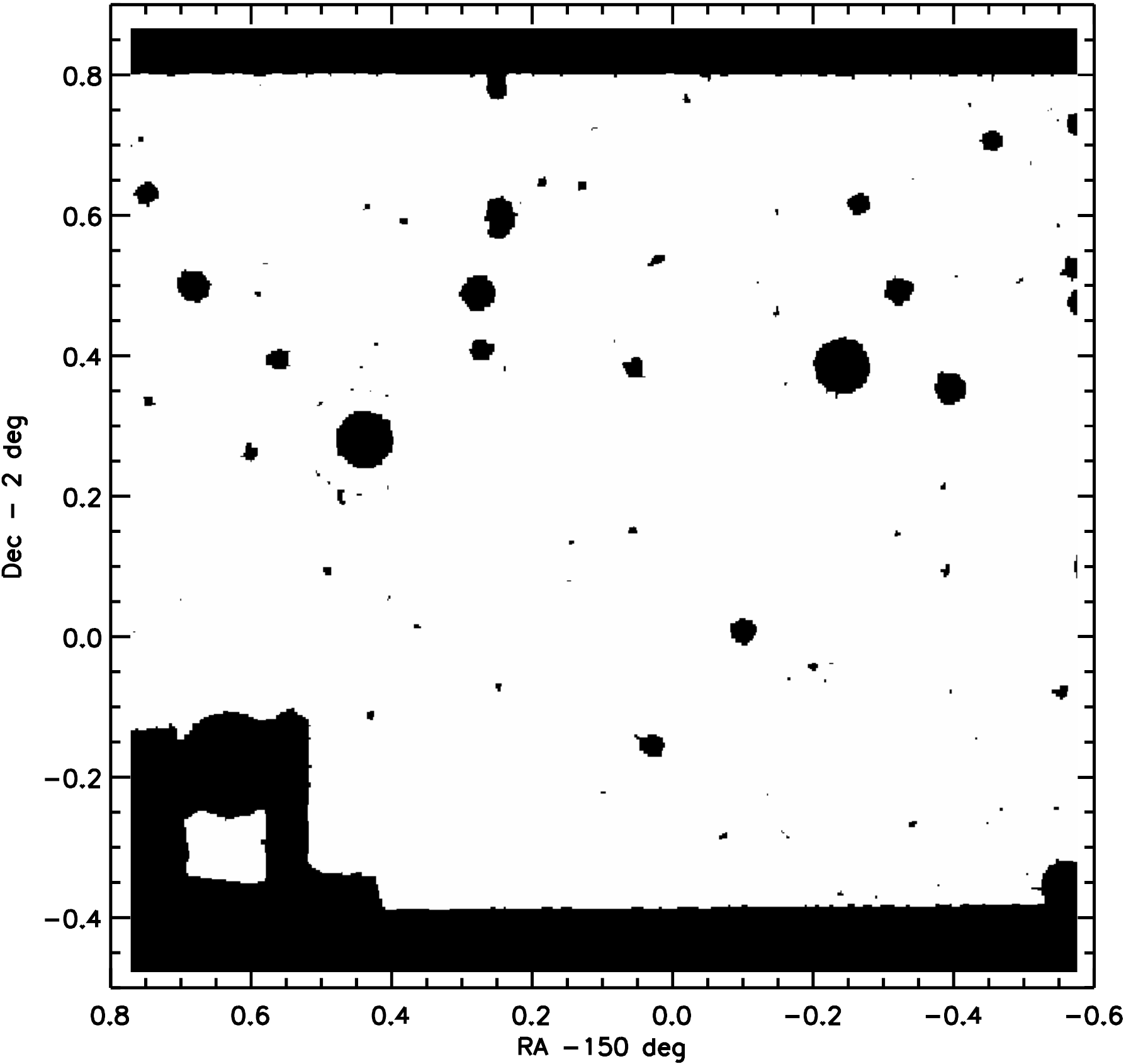}
\caption{Masked areas around bright stars are shown in black (as generated from COSMOS I and B band masks). In these regions, accurate photometry is precluded and all galaxies in these regions
are excluded from the LSS mapping and analysis. They appear as blank areas in the LSS maps at all redshifts.}\label{mask}
\end{figure}

For completeness, we show in Fig \ref{mask} the areas in the COSMOS field where bright stars precluded accurate photometry 
and hence where galaxies are not included in the photometric redshift catalog. These regions will also be devoid of LSS.

\eject

\bibliography{scoville_lss}

\end{document}